\documentclass[amssymb,prd,superscriptaddress,aps,nofootinbib,twocolumn,preprintnumbers,longbibliography]{revtex4-1}
\usepackage[utf8]{inputenc}
\usepackage{amsmath}
\usepackage{amssymb}
\usepackage{xcolor} 
\usepackage{hyperref}
\usepackage{graphicx, epsfig, amssymb} 
\usepackage{ulem}

\def\be{\begin{equation}}
\def\ee{\end{equation}}
\newcommand{\bseq}{\begin{subequations}}
\newcommand{\eseq}{\end{subequations}}

\newcommand{\di}{\mathrm d}

\newcommand{\V}{\vec{V}}

\allowdisplaybreaks 

\begin{document}
\preprint{CERN-TH-2019-165, INR-TH-2019-018, KCL-PH-TH/2019-77}

\title{Secular effects of Ultralight Dark Matter on Binary Pulsars}
% {Binary Pulsars as Resonant Detectors of Ultralight Dark Matter}

   \author{Diego Blas}\email{diego.blas@cern.ch}
   \affiliation{Theoretical Particle Physics and Cosmology Group, Department of Physics,\\
King's College London, Strand, London WC2R 2LS, UK}

   \author{Diana L\'opez Nacir}\email{dnacir@df.uba.ar}
 \affiliation{CONICET, Universidad de Buenos Aires, Facultad de Ciencias Exactas y Naturales,
  Instituto de F\'isica de Buenos Aires (IFIBA), Buenos Aires C1428EGA, Argentina}

   \author{Sergey Sibiryakov}\email{sergey.sibiryakov@cern.ch}
 \affiliation{Theoretical Physics Department, 
CERN, CH-1211 Gen\`eve 23, Switzerland}
 \affiliation{Institute of Physics, LPPC, Ecole Polytechnique 
F\'ed\'erale de Lausanne, CH-1015, Lausanne, Switzerland}
 \affiliation{Institute for Nuclear Research of the Russian Academy
of Sciences, 60th October Anniversary Prospect, 7a, 117312  Moscow, Russia}

\begin{abstract}

Dark matter (DM) can  consist of very light bosons behaving as a
classical scalar field that experiences  coherent oscillations. 
The presence of this DM field would perturb the 
dynamics of celestial bodies, either because the (oscillating) DM
stress tensor modifies the gravitational potentials of the galaxy, or
if DM is directly coupled to the constituents of the body.  
We study secular variations of the orbital parameters of binary
systems induced by such perturbations. Two classes of effects are
identified. Effects of the first class appear if the frequency of DM
oscillations is in resonance with the orbital motion; these exist for
general DM couplings including the case of purely gravitational
interaction. Effects of the second class arise if DM is coupled
quadratically to the masses of the binary system members and do not
require any resonant condition. 
The exquisite precision of binary pulsar timing 
can be used to constrain these effects.
 Current observations are not sensitive to oscillations in the
 galactic gravitational field, though 
 a discovery of pulsars in regions of high DM density may
 improve the situation.  
For DM with direct coupling to ordinary matter, 
the current timing data are already competitive with other
existing constraints in the range of DM masses $\sim
10^{-22}-10^{-18}\,{\rm eV}$. Future 
observations are expected to 
increase the sensitivity and probe new regions of parameters.     
 
\end{abstract}

\maketitle

%%%%%%%%%%%%%%%%%%%%%%%%%%%%%%%
\section{Introduction}
%%%%%%%%%%%%%%%%%%%%%%%%%%%%%%%

Dark matter (DM), a key ingredient in the success of the standard
cosmological model, remains a mysterious component of our Universe.
Its properties are still largely unknown, despite 
an impressive progress in characterizing and testing the signatures
of different DM  candidates. 

In this paper we consider the possibility
that DM is composed of ultralight scalar particles. Natural candidates
of this kind are axion-like and dilaton-like particles, whose
theoretical and phenomenological motivation can be found in
\cite{Marsh:2015xka,Damour:2010rp,Arvanitaki:2014faa}.  
Due to huge phase space occupation numbers, DM in these models is
described as a classical scalar field $\Phi$
\cite{Hu:2000ke,Amendola:2005ad,Hui:2016ltb}.

Several observables have been identified to probe different mass
ranges of ultralight dark matter (ULDM). Current constraints from
observations of the cosmic microwave background and large scale
structure on linear scales robustly exclude $m_{\Phi}\lesssim
10^{-24}$ eV~\cite{Hlozek:2014lca}  (with a factor of $10$
improvement in $m_\Phi$ expected in the future
\cite{Hlozek:2016lzm}). 
The analysis of the Lyman-$\alpha$ forest
\cite{Irsic:2017yje,Armengaud:2017nkf,Kobayashi:2017jcf} and galactic
dynamics \cite{Bar:2018acw,Bar:2019bqz,Safarzadeh:2019sre} further
excludes $m_\Phi\lesssim 10^{-22}$\,eV and shows a tension between the
data and predictions of ULDM models with
$m_\Phi\sim 10^{-22}- 10^{-21}$\,eV. Complementary constraints in
this mass range can come from pulsar-timing arrays
\cite{Khmelnitsky:2013lxt,Porayko:2014rfa}. Yet higher masses can be
probed by the dynamics of  star clusters
\cite{Marsh:2018zyw} or the spectral shape of 21-cm absorption feature 
\cite{Schneider:2018xba,Lidz:2018fqo}  (see \cite{Boyarsky:2019fgp}
for the discussion of the robustness of 21-cm bounds).
Finally, ULDM masses up to $m_\Phi\sim 10^{-18}$\,eV might be explored
using 21-cm intensity mapping \cite{Marsh:2015daa}. 
Independently of their role as DM, the existence of ultralight bosons
has potentially observable signatures due to superradiance instability of
rotating  black
holes
\cite{Arvanitaki:2010sy,Arvanitaki:2014wva,Arvanitaki:2016qwi,Brito:2017zvb,
Davoudiasl:2019nlo}.
These test may probe (or rule out)
the existence of scalar degrees of freedom with masses in the range
$m_{\Phi}\sim 10^{-21} - 10^{-10}\,\rm{eV}$. 

The previous bounds are based only on the gravitational interaction of
ULDM and hence apply to any ULDM model. 
But DM candidates may also be {\it directly} coupled to standard model
(SM) fields through non-gravitational interactions. An interesting
possibility is a universal dilaton-like coupling between ULDM field
$\Phi$ and ordinary matter which respects the 
weak equivalence principle (WEP). Still, this interaction 
violates the strong equivalence principle (SEP)
with non-universal {\it effective couplings} to $\Phi$ being generated
for objects with large gravitational self-energy
\cite{Damour:1992we}.  
Powerful constraints on the SEP violation and other consequences of
universally coupled ultralight scalar fields (non necessarily DM)
come from the tests
within the Solar System 
\cite{Williams:2012nc,2018NatCo...9..289G,Talmadge:1988qz,
Adelberger:2003zx,Iorio:2005fk,
Bertotti:2003rm}. 
In addition, for ULDM case, one can infer constraints on the direct
coupling from the bounds on the stochastic gravity wave background 
\cite{Porayko:2014rfa,2003ApJ...599..806A} (see the discussion in
\cite{Blas:2016ddr}). 

In our previous work \cite{Blas:2016ddr} we showed that competitive
constraints on ULDM in the mass range $10^{-22}\,\text{eV}\lesssim
m_\Phi \lesssim 10^{-18}\,\text{eV}$ are obtained from timing of
millisecond binary pulsars. That work focused on ULDM universally
coupled to the binary system members and studied the secular change in
the orbital period induced by this coupling. The study was generalized
in \cite{LopezNacir:2018epg,Armaleo:2019gil} to the case of ultralight vector
and tensor DM. Recently
Ref.~\cite{Rozner:2019gba} considered instantaneous variations
of the binary orbit due to ULDM. These are, however, strongly
suppressed compared to the secular effects.

In this paper we extend the analysis of \cite{Blas:2016ddr} by including the 
possibility of {\it non-universal} couplings  and studying the secular effect
of ULDM on all orbital parameters appearing in the 
pulsar timing model 
\cite{Teukolsky1976,1986AIHS...44..263D,Edwards:2006zg}.   
It is worth stressing that the couplings of ULDM to the two components
of the binary pulsar system are naturally expected to be different
even if at the fundamental level ULDM couples to the SM fields
universally. This is a consequence of the SEP violation and large
gravitational binding energy of the pulsar constituting a few tens of
per cent of its mass. Irrespectively of whether the companion is a
normal star, a white dwarf or another neutron star (NS) with a different
mass, its binding energy will generically be different.

We will see that non-universality of ULDM--pulsar coupling leads to a
rich phenomenology that can be tested by current and future
observations. In particular, in contrast to the universal coupling
case, we find a non-zero drift of the orbital period for binaries with
circular orbits. Given that the orbits of most of the binary pulsars
are close to circular\footnote{This is due to the circularizing effect
  of accretion from the companion star at an earlier stage
  of the binary system evolution \cite{Lorimer:2008se}.} 
\cite{Manchester:2004bp,pscat}, 
this greatly increases the number of
systems that can be used in the analysis.

The paper is organized as follows. In Sec.~\ref{sec:DMfield} we
summarize the relevant properties of the ULDM field in the Milky Way
halo. The effective Lagrangian describing interaction of a binary
system with ULDM is presented in Sec.~\ref{sec:Lag}. We first deal
with the case of pure gravitational interaction (Sec.~\ref{sec:Lagpg})
and then include the direct coupling (Sec.~\ref{DCcase}). 
Section~\ref{sec:osc} contains the derivation of the key equations
governing the secular evolution of the binary system. The case of only
gravitational interaction is considered in Sec.~\ref{pureGrav}. We
allow for direct coupling in Sec.~\ref{directCoup}, studying first the
ULDM-induced motion of the binary barycenter (Sec.~\ref{sec:BB}), next
turning to the evolution of the orbital elements for linearly
(Sec.~\ref{sec:bina}) and quadratically (Sec.~\ref{sec:quadra})
coupled ULDM. 
In  Sec.~\ref{sec:bounds} we provide numerical
estimates for the size of possible effects and compare them to the
current sensitivity. We first consider in Sec.~\ref{sec:resonant} the
effects that appear when the frequency of ULDM oscillations is in
resonance with the orbital motion of the binary. We separately discuss
the cases of pure gravitational interaction
(Sec.~\ref{sec:gravbounds}), universal direct coupling
(Sec.~\ref{sec:UCbounds}) and non-universal coupling
(Sec.~\ref{sec:genericbounds}). In Sec.~\ref{sec:detuning} we study
the  timing constraints when the ULDM mass deviates from the resonant value.  In Sec.~\ref{sec:nonresonant} we explore
non-resonant secular effects that appear for quadratically coupled
ULDM and lead to constraints in a wide range of ULDM masses.   
Section~\ref{sec:concl} is devoted to
conclusions and outlook. 
Two Appendices are added to make the paper
self-contained. Appendix~\ref{osculating} summarizes the Keplerian
dynamics and the formalism of osculating
orbits. Appendix~\ref{AppendixLPE2} contains the osculating orbit
equations for a binary interacting with ULDM prior to extraction of
secular contributions.

%%%%%%%%%%%%%%%%%%%%%%%%%%%%%%%
\section{Ultralight dark matter in the halo}\label{sec:DMfield} 
%%%%%%%%%%%%%%%%%%%%%%%%%%%%%%%

We assume that DM is described by a scalar field $\Phi$ minimally
coupled to gravity. 
For gravity, we use the standard Einstein-Hilbert action, while 
for DM the action reads\footnote{We use the signature $(-,+,+,+)$ and
  work in the natural units $\hbar=c=1$.}, 
\begin{equation}
S_{\rm DM}=\frac{1}{2}\int
\di^4x\sqrt{-g}\left[-g^{\mu\nu}\partial_{\mu}\Phi
\partial_{\nu}\Phi-m_{\Phi}^2\Phi^2\right]\,.\label{eq:PhiS}
\end{equation}
We are interested in the mass range  
$m_\Phi \sim 10^{-22}- 10^{-18}$\,eV.  
The  occupation numbers for modes within the halo of a typical galaxy 
are sufficiently high that  the classical  field
theory description is applicable. 
In the non-relativistic limit, the field oscillates in time with
frequency set by the mass, whereas its spatial gradients are
suppressed. The rapid oscillations are conventionally factored out and
the field is written as,
\begin{equation}
 \Phi=e^{-im_\Phi t}\Psi+e^{im_\Phi t}\Psi^*\,, \label{eq:DMfield}
\end{equation}
 where  
\begin{equation}
\Psi= \frac{1}{2} \Phi_0(\vec{x},t)\, e^{-i \Xi(\vec{x},t)}\,. \label{eq:Psi}
\end{equation}
One can  compute the stress energy tensor $T_{\mu\nu}$ of the scalar
field and   take the average $\langle  ...\rangle$ over the ``fast''
oscillations  of frequency $ m_{\Phi}$.  
At  leading order in gradients of $\Phi_0$ and $\Xi$ one finds,
\begin{equation}
\langle T_{00}\rangle\simeq
T_{00}\simeq\frac{m_\Phi^2\Phi_0^2}{2}~,~~~~
\langle T_{0i}\rangle\simeq
\frac{m_\Phi\Phi_0^2}{2}\partial_i\Xi\,.
\end{equation}
This leads to the identification of 
\begin{equation}
\rho_{\rm DM}\equiv \frac{m_\Phi^2\Phi_0^2}{2}~,~~~~
\vec{V}\equiv -\frac{\nabla \Xi}{m_\Phi} \label{rhoV}
\end{equation}
as the DM density and velocity respectively. The spatial stress tensor
reads,
\begin{equation}
T_{ij}=p_{\rm DM} g_{ij}+\ldots \label{Tij}\,,
\end{equation}
where
\begin{equation}
p_{\rm DM}= -\rho_{\rm DM}\cos(2m_\Phi t+2\Upsilon)~,~~~~
\Upsilon=\Xi\big|_{\vec{x}=t=0} \label{pDM}\,,
\end{equation}
and dots stand for terms suppressed by gradients. We see that the ULDM
field is characterized by a large oscillating pressure. This, however,
vanishes upon averaging, so that at large time and length scales the field
behaves as a pressureless fluid, similar to the standard cold DM. On
the other hand, the gradient terms neglected in (\ref{Tij}) survive
the averaging and lead to deviations from cold DM behavior at scales
comparable to the de Broglie wavelength of the field
$\lambda_{\rm dB}\equiv 2\pi/(m_{\Phi}V)$. 
The description of self-gravitating ULDM in the non-relativistic
regime which is valid at all scales is provided by the
Schr\"odinger--Poisson system of equations for the complex amplitude
$\Psi$ and the Newtonian potential $\phi$, see e.g. \cite{Hui:2016ltb}
for review.

In a virialized halo one expects the DM field to consist of a
superposition of waves with random phases. The amplitudes of the modes
are expected to follow approximately the Maxwell distribution with the
dispersion $V_0$ determined by the typical velocity at a given
location in the halo. Such superposition produces in space an
interference pattern of wavepackets with characteristic size 
$\lambda_{\rm dB}/2$. Inside an individual wavepacket the field performs
coherent oscillations with the period $t_{\rm osc}=2\pi/m_\Phi$ and
slowly changing amplitude and phase. Each coherence patch is characterized by a
local DM density and velocity, which exhibit large (order-one)
fluctuations between the patches. These expectations are born out by
numerical
simulations~\cite{Schive:2014dra,Schive:2014hza,Schwabe:2016rze,Veltmaat:2018dfz}
that show a granular structure of DM distribution in halos
characteristic of wave interference. Due to motion of wavepackets, the
field at a fixed point in space looses coherence after time $t_{\rm
  coh}\sim \lambda_{\rm dB}/(2V)$. The local DM density averaged over
time scales much bigger than $t_{\rm coh}$ reproduces the smooth halo
profile, whereas the local velocity averages to zero.

Let us compare the scales introduced above to the typical
characteristics of binary pulsar systems. The ULDM oscillation period
is 
\begin{equation}
t_{\rm osc}\simeq 1.15\, {\text{hours}} \left(\frac{10^{-18}
    \mbox{eV}}{  m_{\Phi}}\right)\,, 
\end{equation} 
which for $m_\Phi\sim 10^{-22}-10^{-18}$\,eV is comparable to pulsar
orbital periods. This allows for a resonant interaction between ULDM
and a binary. The de Broglie wavelength of the ULDM field is set by
the velocity dispersion $V_0$ of the halo which can be estimated from
the virial velocity. For the Milky Way we assume $V_0\sim 10^{-3}$
\cite{Lisanti:2016jxe}, so that
\begin{equation}
\lambda_{\rm {dB}}\sim 1.3\times 10^{12}\,\mbox{km}\left(\frac{10^{-3}}{V_0}\right) \left(\frac{10^{-18} \mbox{eV}}{ m_{\Phi}} \right)\,. \label{eq:dB}
\end{equation} 
This is typically much larger than  the size of the orbit of the
binary systems we consider, with semi-major axes always satisfying $a
\lesssim 10^8\, {\rm km}$.  
Also the coherence time 
\be
\label{eq:tcoh}
t_{\rm coh}\sim  65\,\mbox{years}\left(\frac{10^{-3}}{V_0}\right)^2
\left(\frac{10^{-18} \mbox{eV}}{ m_{\Phi}} \right),\, 
\ee
is larger or comparable to the observation time. Therefore we can
assume that the binary system is located entirely inside a single ULDM
coherence patch and the field oscillations remain coherent on the
relevant time scales.

It will be convenient to work in the binary rest frame. We will need the field together with its 
spatial gradients in the vicinity of the binary and will neglect
higher order spatial derivatives. Thus we write,
\be
\Phi(\vec{x},t)=\frac{\sqrt{2\rho_{\rm DM}}}{m_\Phi} 
e^{-\vec{S}\cdot\vec{x}}
\cos\big(m_\Phi (t- \V \cdot \vec{x})+ \Upsilon\big)\,,\label{eq:phidm}
\ee 
where $\vec{V}$ now denotes the relative velocity of DM with respect
to the binary barycenter (BB) and we have introduced the vector
\begin{equation}
\vec{S}\equiv -\nabla \log\Phi_0 \label{Sdef}
\end{equation}
to characterize the spatial variation of the field amplitude. The
granular structure of the ULDM halo described above implies that the
absolute value of $\vec{S}$ is of order $2\lambda_{\rm dB}^{-1}$. We
will consider $\rho_{\rm DM}$, $\vec{S}$, $\vec{V}$ and $\Upsilon$ in 
(\ref{eq:phidm}) as constant over the binary system
size. 

Most of the millisecond binary pulsars observed to date are located
in $\sim 2$\,kpc neighbourhood of the Solar System. The Milky Way halo
models give DM density in this region 
$\bar\rho_{\rm DM} \sim 0.3\div 0.6\, {\rm GeV/cm}^3$
\cite{Piffl:2014mfa,McKee:2015hwa,Evans:2018bqy}. 
In our numerical estimates we will use a conservative proxy
$\bar\rho_{\rm DM} = 0.3\, {\rm GeV/cm}^3$. It is worth mentioning,
however, that the actual ULDM density in the vicinity of the binary
may differ by a factor of a few from $\bar\rho_{\rm DM}$ due to large
fluctuations between the coherence patches.

Throughout our
analysis we will neglect any effect that  the binary system may
produce on the ULDM distribution  (that is, the {\it back-reaction}
effect). To estimate when this approximation is valid, 
we follow Ref.~\cite{Hui:2016ltb} and treat the ULDM coherence patches
as quasiparticles of mass 
$M_{\rm qp}\sim \rho_{\rm DM}(\lambda_{\rm dB}/2)^3\sim 
\pi^3 \rho_{\rm DM}/(m_\Phi V_0)^3$.
The interaction between the binary and ULDM may be viewed as
scattering of quasi-particles that lasts for the coherence time
$t_{\rm coh}$. It leads to the change of quasiparticle momentum and
energy, $\Delta p\simeq -\dot p_{\rm b} t_{\rm coh}$
and $\Delta {\cal E}\simeq -\dot {\cal E}_{\rm b} t_{\rm
  coh}$, where $\dot p_{\rm b}$, $\dot {\cal E}_{\rm b}$ are the time
derivatives 
of the binary momentum and energy. The back-reaction is
negligible as long as $\Delta p$ and $\Delta {\cal E}$ are small
compared to the typical quasiparticle momentum 
$p_0\sim M_{\rm qp}V_0$ and its total energy (including the rest mass)
${\cal E}_0\sim M_{\rm qp}$. In this way we arrive at the conditions
\begin{align}\label{BackRe}
\frac{m_\Phi^2 \dot p_{\rm b}}{\pi^2 \rho_{\rm DM}}\ll 1\;,
\qquad\qquad
\frac{m_\Phi^2 \dot {\cal E}_{\rm b}V_0}{\pi^2 \rho_{\rm DM}}\ll 1\;.
\end{align}
These will be verified {\it a posteriori} upon estimating $\dot p_{\rm
  b}$, $\dot {\cal E}_{\rm b}$ in Sec.~\ref{sec:bounds}. Note that
they are easier to satisfy for lighter ULDM, which is one of the
reasons to restrict our study to the range $m_\Phi\lesssim
10^{-18}\,{\rm eV}$.

%%%%%%%%%%%%%%%%%%%%%%%%%%%%%%%
\section{Interaction of dark matter with compact bodies}\label{sec:Lag}
%%%%%%%%%%%%%%%%%%%%%%%%%%%%%%%

To model the binary system, we treat the pulsar and its companion as 
point particles. 
These will always interact with DM gravitationally. In addition,
there may be an interaction between the bodies and $\Phi$ from a
direct coupling. In all cases, in the limit of point particles
the effective action for the bodies in the system will be  
\be
S_{\rm B}=-\sum_{A=1,2}\int \di \tau_A \, M_A(\Phi) \,, \label{eq:sources}  
\ee 
where $\tau_A$ is the proper time and  $M_A(\Phi)$ is a function of
the DM field \cite{Damour:1992we}.  

In order to characterize the main potentially observable effects that
the DM field produces on a binary system, it will be enough to work
at  leading (Newtonian) order in the velocities of  the members of the
system, and at first order in the corrections due to the presence of
$\Phi$ (in an analogous way as  customarily done for gravitational
waves).  In the cases where the description of the orbital motion
requires the inclusion of  relativistic corrections, our results can
be  generalized in a systematic way by means of the post-Newtonian
formalism \cite{will1993theory}.

%%%%%%%%%%%%%%%%%%%%%%%%%%%%%%%
\subsection{Pure gravitational interaction: oscillations in the
  spacetime curvature}\label{sec:Lagpg} 
%%%%%%%%%%%%%%%%%%%%%%%%%%%%%%%

Let us start with the case when there is no direct coupling between
DM and the bodies, so that their masses are independent of $\Phi$,
$M_A(\Phi)\equiv M_A$. 
Following \cite{Khmelnitsky:2013lxt,Blas:2016ddr} we begin by
computing the perturbation of the metric produced by the oscillating
wave (\ref{eq:phidm}) in the Newton gauge,
\begin{equation}
\label{eq:FRW}
\di s^2=-(1+2\phi) \di t^2+(1-2\psi)\delta_{ij}\di x^i\di x^j\,.
\end{equation}
We are interested in resonant effects due to the oscillating pressure
(\ref{pDM}). Thus we can neglect the gradient contributions into the
DM energy-momentum tensor which are suppressed by the DM velocity. The
linearized Einstein equations read,
\bseq 
\begin{align}
&\nabla^2\psi=4\pi G\rho_{\rm DM}\,,~~~~\nabla\dot\psi=0\,,\\
&2\ddot\psi\delta_{ij}+\nabla^2(\phi\!-\!\psi)\delta_{ij}
-\nabla_i\nabla_j(\phi\!-\!\psi)=8\pi G p_{\rm DM}\delta_{ij}\,,
\label{Einstij}
\end{align}
\eseq
where $G$ is the Newton's gravitational constant. Their solution
splits into the time-independent part and an oscillatory homogeneous
contribution\footnote{More precisely, the oscillatory contribution has
a weak dependence on the spatial coordinates suppressed by the DM
density gradients, which we neglect in our derivation.}, 
\be
\psi=\bar\psi(\vec{x})+\tilde\psi(t)\;,
\ee
and similarly for $\phi$. The time-independent part locally has the form,
\be
\bar\phi=\bar\psi=A_i x^i+B_{ij}x^ix^j\,,~~~~~B_{ii}=2\pi
  G\rho_{\rm DM}\,,
\ee 
where the constant vector $A_i$ and tensor $B_{ij}$ correspond to the
acceleration and tidal forces produced by the galactic gravitational
field. These are standard effects in general relativity (GR) 
\cite{Damour:1991rd}
and are taken into account in the analysis of pulsar
timing\footnote{We note that in the case of ULDM, large density
  fluctuations between coherence patches can lead to an additional
  acceleration of the binary as a whole, on top of the acceleration in
the large-scale galactic field. This effect, however, appears
degenerate with the gravitational pull of other celestial bodies.}.
In what follows we focus on the effects due to the oscillating nature
of the ULDM field, and thus omit the time-independent part of the
gravitational potentials. 

From (\ref{Einstij}) we find that the
oscillating part of the potential $\psi$ satisfies
\begin{equation}
\ddot{\tilde\psi}=-4\pi G\rho_{\rm DM}\cos(2m_\Phi t+2\Upsilon)\,. \label{eq:ddpsi}
\end{equation}   
As will be clear in a moment, this expression is sufficient to study
the dynamics of the binary system within our approximations. 
In particular, we will not need the Newtonian potential $\tilde\phi$.

The equations of motion for the binary are conveniently formulated in
the Fermi normal coordinates $\{t_{F},\vec{\xi}\}$
associated to the binary barycenter (BB) rest frame. This corresponds
to the geodesic motion 
of a free particle in the metric
(\ref{eq:FRW}) with the initial conditions $\vec{x}=\dot{\vec{x}}=0$ at
$t=0$. 
The explicit change of coordinates is: 
\begin{subequations}\label{FNC}
\begin{align}
t_{F}(t,\vec{x})&=t +\int_{0}^{t}\di t' 
\tilde\phi(t')-\frac{\dot{\tilde\psi}(t)}{2}x^2\,,\\
\vec{\xi}(t,\vec{x})&=\vec{x}\big(1-\tilde\psi(t)\big)\,.
\end{align} \end{subequations}
Including  the
Newtonian potential $\phi_N$ generated by the orbiting bodies, the metric in the new coordinates reads
\begin{equation} 
\di s^2=-(1+\ddot{\tilde\psi}\xi^2+2 \phi_N) \di t_{F}^2+
\delta_{ij}(1-2\phi_N) \di \xi^i \di \xi^j\,, \label{eq:BBframe}
\end{equation}
which is independent of  $\tilde\phi(t)$.
The dynamics of the binary system is obtained from  the  geodesic
deviation equations, where one adds the modifications to the 
curvature $\delta R^{\mu}_{\nu\rho\sigma}$ produced by the DM 
field~\cite{1978ApJ...223..285M}. Defining
  $\vec{r}\equiv \vec{\xi}_1-\vec{\xi}_2$, one gets an extra acceleration
  $\delta \ddot{r}^i=-\delta R^i_{0j0}r^j$.
Since the perturbation of the Riemann tensor is gauge independent
at the linearized level, we can evaluate it in the Newton gauge
(\ref{eq:FRW}). In this way we arrive at
\begin{equation}
\ddot{r}^i + \frac{GM_Tr^i }{r^3}= - \ddot{\tilde\psi}\,  r^i\,, 
\label{eq:bin_gra}
\end{equation}  
where dot stands for the derivative with respect to $t_{F}$ and
$M_T\equiv M_1+M_2$.  

Alternatively, from the non-relativistic limit of
Eq.~(\ref{eq:sources}) one can derive the Lagrangian  
 \begin{eqnarray}
 \label{eq:L_grav}
L&=&M_1\left(-1+\frac{{\dot\xi}^2_1}{2}\right)
+M_2\left(-1+\frac{{\dot\xi}^2_2}{2}\right)\,\,\,\,\,\,\,\,\nonumber
\label{eq:Lag}\\
&&~~~~~+\frac{G M_1 M_2}{|\vec{\xi}_1-\vec{\xi}_2|}
-\frac{\ddot{\tilde\psi}}{2}\big(M_1 \xi_1^2+M_2 \xi_2^2\big)\,,
\end{eqnarray}
whose variation reproduces Eq.~\eqref{eq:bin_gra}. Notice that the
orbital dynamics decouples from the center-of-mass motion. The BB
coordinate 
\be
\vec{R}=\frac{M_1\vec{\xi}_1+M_2\vec{\xi}_2}{M_T}
\ee
obeys the equation $\ddot{\vec{R}}=-\ddot{\tilde\psi}\vec{R}$, which
admits the trivial solution $\vec{R}=0$. In other words, BB indeed
follows the geodesic which we have chosen as the origin of our Fermi
frame. We are going to see that the situation is different if ULDM
couples non-universally to the bodies in the binary.

%%%%%%%%%%%%%%%%%%%%%%%%%%%%%%%
\subsection{Direct coupling of ULDM to matter: 
violation of the equivalence principle}\label{DCcase}
%%%%%%%%%%%%%%%%%%%%%%%%%%%%%%%
 
We now include the possibility of direct coupling encapsulated by the
dependence of the masses of the bodies on the ULDM field. We assume
that the change of the masses due to this coupling is small, so that
it can be Taylor expanded in the amplitude of the field. We analyze
the cases of linear and quadratic couplings,
\be
\label{massdep}
M_A(\Phi)=
 \bar M_A(1+\alpha_A \Phi)~~~\text {or}~~~
M_A(\Phi)=\bar M_A\bigg(1+\frac{\beta_A}{2} \Phi^2\bigg),
\ee
where $\bar M_A$ is the mass of the body in the absence of the
field. 

Part of the {\it sensitivities} $\alpha_A$, $\beta_A$ arises from the
coupling of the field $\Phi$  with the SM fundamental fields. 
This coupling may be universal (that is, given by the same interaction
with the same  coupling constant for all fields)  
or not.  For instance, any direct  coupling of $\Phi$ with the Higgs
field will produce a non-universal coupling to the masses of
electrons, protons and neutrons (the main sources of mass in the
stars), see e.g.~\cite{Graham:2015ifn}. 

The microscopic couplings are typically inversely proportional to some
high-energy scale $\Lambda$, which sets the cutoff of the effective
low-energy theory. Generically, one expects on dimensional grounds
$\alpha_A\sim \Lambda^{-1}$, $\beta_A\sim \Lambda^{-2}$. In this case,
the linear coupling will dominate the low-energy phenomenology and the
quadratic interaction can be neglected. However, the linear coupling
may be forbidden if the underlying theory possesses an appropriate
symmetry, for example, is invariant under $\Phi\mapsto -\Phi$. Then the dominant
term in the interaction between $\Phi$ and SM will be quadratic. We study
the two types of couplings in (\ref{massdep}) separately, as
representatives of these two classes of theories. 
 
For compact objects, such as NSs,
the mass has also a contribution coming from the self-gravity of the
body. 
This contribution couples to $\Phi$ differently from matter and thus 
generally violates the SEP even if the WEP is
satisfied. To be more specific, consider the case when $\Phi$ couples
to the SM fields universally through a Jordan--Fierz metric 
$\tilde g_{\mu\nu}={\cal A}^2(\Phi)g_{\mu\nu}$ and define  
\be
\alpha_A(\Phi)=\frac{\di \log {\cal A}(\Phi)}{\di \Phi}\bigg|_{\Phi=0}
\,,~~~\beta_A(\Phi)= \frac{\di^2 \log {\cal A}(\Phi)}{\di \Phi^2}
\bigg|_{\Phi=0}\,. 
\label{eq:alpha}
\ee  
If the self-gravity of the body is sufficiently weak, the linear
sensitivity can be expanded\footnote{The calculation in
  \cite{Damour:1992we} assumes a constant background field far from
  the body, rather than
  an oscillating wave. Since in our case the time scale of
  the oscillations is large compared to the scales relevant
  for the inner structure of the body and the corresponding near field
  physics, the evolution of the field can be considered as
  adiabatic.} as \cite{Damour:1992we}, 
\begin{equation}
\alpha_A=\alpha \,(1+c_1s_A+c_2s_A^2+\ldots)\,,\label{LOcomp}
\end{equation} 
where  $c_1<0$, $c_2$, etc. are order-one coefficients and $s_A\simeq
GM_A/r_A$ is the ratio of the gravitational binding energy to the rest
mass of the body (here $r_A$ is the size of the body). In case
$\alpha=0$ a similar expression holds for the quadratic sensitivity
$\beta_A$ upon replacing $\alpha$ by $\beta$. The binding energy is
negligible for ordinary stars and white dwarfs, whereas for neutron
stars the typical values are $s_A\simeq 0.1\div 0.2$ depending on the
NS mass  \cite{Damour:1992we}. This implies that the two bodies comprising a binary pulsar
system will typically have a $10\%$ difference in their sensitivities, 
\begin{equation}\label{PerUniv}
\frac{\alpha_1-\alpha_2}{\alpha_1+\alpha_2}\sim 0.1\,
\end{equation}
and similarly for the case of quadratic coupling\footnote{The
  sensitivities vanish in the extreme case of a black hole due to the
  no-hair theorems \cite{Damour:1992we}. Thus, the asymmetry
  (\ref{PerUniv}) is expected to be of order one in the case of a
  NS--black hole binary.}.

The preceding logic does not take into account non-perturbative
effects that can further boost the difference in the sensitivities of
compact bodies. A well-known example of such effect is spontaneous
scalarization \cite{Damour:1993hw,Damour:1996ke} which typically
occurs for sufficiently large negative $\beta$. In this case a
NS develops a non-zero scalar charge $\alpha_A\neq 0$ even if the
fundamental linear coupling of the SM fields to $\Phi$ vanishes,
$\alpha=0$. The physical origin of this effect is easy to
understand. The coupling of the scalar field to nuclear matter inside
the star gives a contribution to the effective scalar mass,
\be
m_{\rm eff}^2=m^2_\Phi-\beta T_{{\rm m}\,\mu}^{~\;\mu}\;,
\ee
where $T_{{\rm m}\,\mu}^{~\;\mu}=-\rho_{\rm m}+3p_{\rm m}$ is the
trace of the matter energy-momentum tensor and $\rho_{\rm m}$, $p_{\rm
m}$ are the density and pressure in the NS interior. For 
$\rho_{\rm m}>3p_{\rm m}$ and large negative $\beta$ this mass becomes
tachyonic, implying an instability of the trivial field configuration
$\Phi=0$ inside the NS. A necessary and sufficient condition for the
development of the instability is that the corresponding `Compton
wavelength' should be smaller than the size of the body, 
$1/\sqrt{|m_{\rm eff}^2|}\lesssim r_A$. In this case the NS generates
a non-trivial scalar profile which corresponds to an effective scalar
charge\footnote{It has been further shown that in a binary NS--NS
system a non-trivial scalar profile due to scalarization of one of the
binary members can lead to induced scalarization of the other with the
transition to a phase with non-zero scalar charge occurring 
dynamically \cite{Barausse:2012da,Palenzuela:2013hsa}.}. 

As it is clear from the above reasoning, spontaneous scalarization can
happen also for $\beta>0$ if the nuclear equation of state is such
that the pressure in the NS interior is large, $p_{\rm m}>\rho_{\rm m}/3$
\cite{Horbatsch:2010hj,Mendes:2014ufa,Palenzuela:2015ima,
  Mendes:2016fby}. If instead $p_{\rm m}$ is less than $\rho_{\rm
  m}/3$ positive values of $\beta$ will increase the mass of the
scalar field inside the NS. This expels the scalar field from the star
interior and reduces the sensitivities \cite{Damour:1993hw}. If $m_{\rm
  eff}^2\gg 1/r^2_A$ only the outer layer of the NS of thickness $\sim
m_{\rm eff}^{-1}$ will interact with the scalar field.

The upshot of this discussion is that non-perturbative phenomena can
dramatically modify the interaction of the scalar field $\Phi$ with
NSs, compared to its bare coupling to the fundamental SM
fields. This strengthens the role of pulsar timing as a probe of
scalar-tensor theories complementary to the 
laboratory tests or tests performed in a weak gravitational field
\cite{Damour:1996ke,Damour:1998jk,Shao:2017gwu}.
A caveat is that the non-perturbative effects, and hence the bounds on
the fundamental parameters $\alpha$, $\beta$ inferred from them,
strongly depend on the assumed equation of state of nuclear matter
\cite{He:2014yqa,Shao:2017gwu}. 
For this reason,  we will adopt a model-independent perspective and
search for constraints on the {\it effective} parameters $\alpha_A$,
$\beta_A$ 
rather than on  fundamental model parameters
(cf. \cite{Horbatsch:2011nh}).  
Finally,  if non-perturbative effects are
absent, one can   assume the typical  values given by
Eqs.~(\ref{LOcomp}), (\ref{PerUniv}). 
 
The effective Lagrangian describing the non-relativistic motion of a
binary system interacting with ULDM in Fermi coordinates has the form
(\ref{eq:L_grav}) with the masses promoted to the field-dependent
expressions (\ref{massdep}). In deriving the equations of motion we
keep only terms that are first-order in the sensitivities,
leading-order in the non-relativistic approximation and up to linear
in the spatial gradient of $\Phi$. Combining the equations for
$\vec{\xi}_1$, $\vec{\xi}_2$ and using the expression (\ref{eq:phidm})
for the ULDM field we obtain in the case of linear coupling,
\bseq
\label{eq:Rcmr12}
\begin{align}
 \ddot{\vec{R}}=&-\alpha_T\dot\Phi\dot{\vec{R}} 
+\alpha_{T}  (\dot{\Phi}\, \V+ \Phi\,\vec{S} )\nonumber\\
&- \Delta\alpha\dot\Phi\, \frac{\bar\mu\,\dot{\vec{r}}}{\bar M_T} 
+\Delta\alpha\Phi\,\frac{G \bar\mu\,\vec{r} }{r^3}
-\ddot{\tilde\psi}\vec{R}\;,\label{eq:Rcm}
\\
\ddot{\vec{r}}=&- (1+\alpha_{T} \Phi)\frac{G\bar M_T \vec{r} }{r^3}
-\alpha_{\mu}\dot{\Phi}\,\dot{\vec{r}}
\nonumber\\\label{eq:r12}
 &+ \Delta\alpha (\dot{\Phi}\, \V+ \Phi\,\vec{S})
-\Delta\alpha\dot\Phi\,\dot{\vec{R}}
- \ddot{\tilde\psi}\,  \vec{r}
\,,\,
\end{align}  
\eseq
where the BB position is defined using the unperturbed masses,
\be
\label{BBnew}
\vec{R}=\frac{\bar M_1\vec{\xi}_1+\bar M_2\vec{\xi}_2}{\bar M_T}\;,
\ee
and $\Phi$, $\dot\Phi$ are evaluated at $\vec{\xi}=0$. We have
introduced the following notations,
\bseq
\label{alphanots}
\begin{align}
&\Delta\alpha=\alpha_1-\alpha_2\;,
&\alpha_T=\frac{\alpha_1\bar M_1+\alpha_2\bar M_2}{\bar M_T}\;,\\
&\bar\mu=\frac{\bar M_1\bar M_2}{\bar M_T}\;,
&\alpha_\mu=\frac{\alpha_1\bar M_2+\alpha_2\bar M_1}{\bar M_T}\;.
\end{align}
\eseq
Two comments are in order. First, we observe that the orbital dynamics
and the BB motion are coupled whenever the interaction with the scalar
field is non-universal, $\Delta\alpha\neq 0$. Second, $\vec{R}=0$ is
no longer a solution of the equations of motion due to an extra
(non-universal) force exerted by the ULDM field on the system as a
whole. In other words, the BB does not move along a geodesic and
cannot be chosen as the origin of the Fermi normal coordinates. The
Fermi system constructed in the previous subsection and used to derive
(\ref{eq:L_grav}) should be understood as associated to the {\it
  unperturbed} motion of the BB obtained if one neglects all terms
proportional to $\alpha_A$. The values of $\vec{R}$ and
$\dot{\vec{R}}$ induced in this frame are ${\cal O}(\alpha)$. Thus,
the terms containing these quantities on the r.h.s. of
(\ref{eq:Rcmr12}) are ${\cal O}(\alpha^2)$ and will be neglected in
what follows. We will discuss below to what extent the modification of the BB
trajectory at order ${\cal O}(\alpha)$ implied by (\ref{eq:Rcm}) may  be relevant for observations.

For the case of quadratic coupling the equations read,
\bseq
\label{eq:Rcmr12beta}
\begin{align}
 \ddot{\vec{R}}=&\beta_{T} \Phi (\dot{\Phi}\, \V+ \Phi\,\vec{S} )
- \Delta\beta\Phi\dot\Phi\, \frac{\bar\mu\,\dot{\vec{r}}}{\bar M_T} 
+\frac{\Delta\beta\Phi^2}{2}\,\frac{G \bar\mu\,\vec{r} }{r^3}\;,\label{eq:Rcmbeta}
\\
\ddot{\vec{r}}=&-\left(1+\frac{\beta_{T} \Phi^2}{2}\right)
\frac{G\bar M_T \vec{r} }{r^3}
-\beta_{\mu}\Phi\dot{\Phi}\,\dot{\vec{r}}
\nonumber\\\label{eq:r12beta}
 &+ \Delta\beta\Phi (\dot{\Phi}\, \V+ \Phi\,\vec{S})
- \ddot{\tilde\psi}\,  \vec{r}
\,,\,
\end{align}  
\eseq
where we have already neglected the terms proportional to $\vec{R}$,
$\dot{\vec{R}}$. The coefficients $\beta_T$, $\beta_\mu$,
$\Delta\beta$ are defined similarly to (\ref{alphanots}). 
Note that these equations can be obtained from (\ref{eq:Rcmr12}) by
the substitution $\alpha_A\mapsto\beta_A$, $\Phi\mapsto\Phi^2/2$,
$\vec{S}\mapsto2\vec{S}$. 
To simplify
the notations, we will omit the overbar on
the values of the masses at $\Phi=0$ in the rest of the paper.

The equations (\ref{eq:Rcmr12}), (\ref{eq:Rcmr12beta}) describe the
perturbation of the Keplerian motion due to the interaction with the
ULDM field. The orbital motion will be also affected by the GR effects,
such as the periastron precession and emission of gravitational
waves. As we mentioned above, as long as both GR and ULDM corrections are small, they sum up
linearly and can be treated independently. Thus, we are not going to
consider the GR corrections explicitly in our analysis, focusing on
the new effects due to ULDM. When comparing our results to the data,
we will use the residuals derived after subtracting the known GR
contributions.

%%%%%%%%%%%%%%%%%%%%%%%%%%%%%%%
\section{The perturbed  Keplerian problem}\label{sec:osc}
%%%%%%%%%%%%%%%%%%%%%%%%%%%%%%%

We now derive the modifications to Keplerian orbits from the terms in
equations \eqref{eq:r12} or \eqref{eq:r12beta} depending on $\Phi$.   
We treat the new terms as perturbations and work  in the framework of
osculating Keplerian orbits \cite{danby1970fundamentals,poisson2014gravity}.
Our calculation is
analogous to the case of a gravitational wave perturbation of
Ref.~\cite{Turner:1979yn}. 

\begin{figure} [htbp]
        \center{\includegraphics[width=8cm]{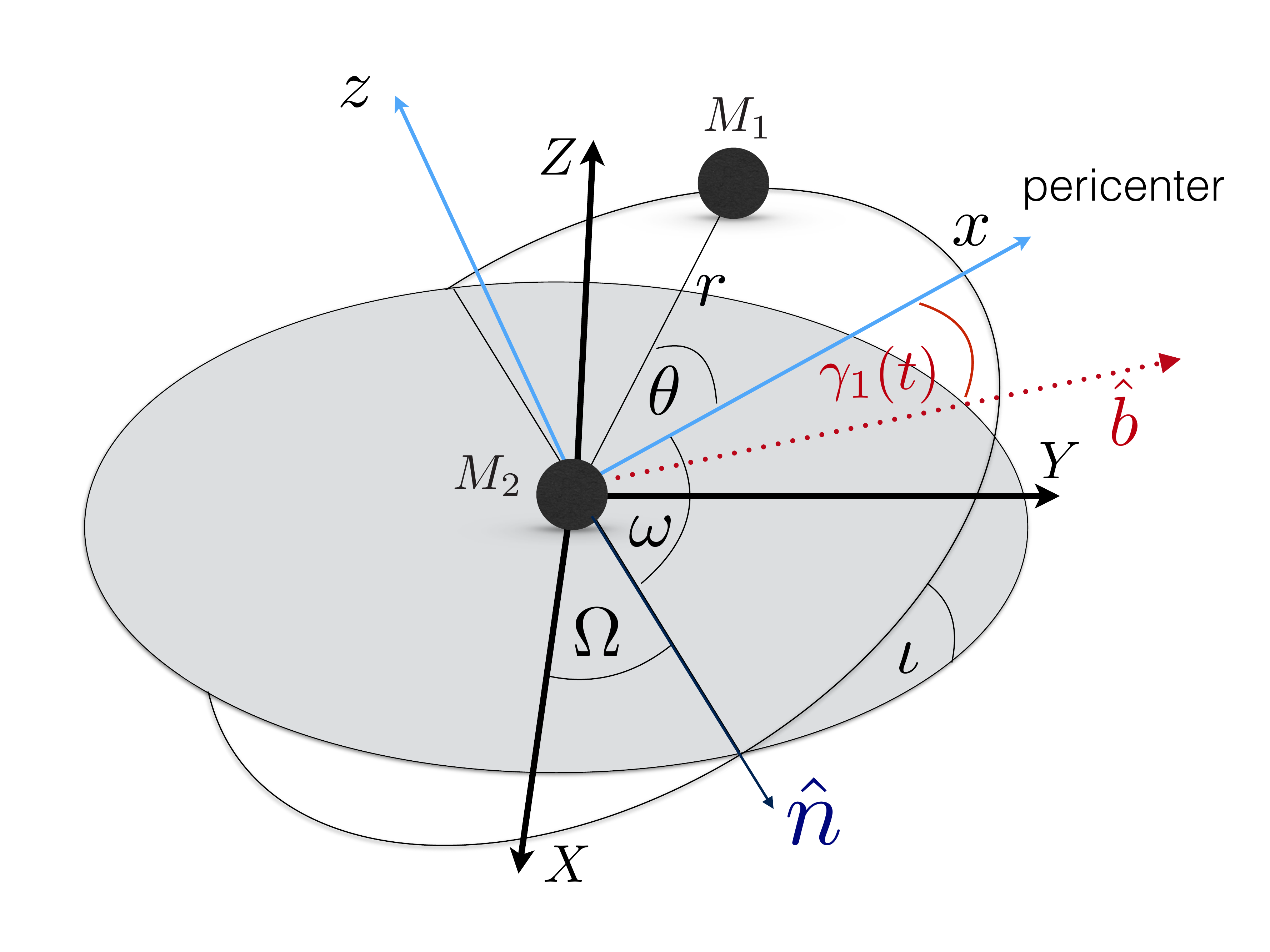}}
 \caption{Orientation of the binary system orbit in the fundamental
   reference frame $(X,Y,Z)$. The orbital frame 
$(x, y, z)$ is centered at the body with mass $M_2$. It has $x$ and
$y$ axes in the plane of the orbit, with the $x$-axis pointing towards
the pericenter. The angle between the axis $X$ and the unit vector
$\hat n$ pointing towards the ascending node is denoted by $\Omega$,
the inclinations between the planes $xy$ and $XY$ is denoted by
$\iota$, and the angle between $\hat n$ and the $x$-axis is
$\omega$. $\theta$ is the angle between the direction towards the body
$M_1$ and the $x$-axis. We also show the vector $\hat b$ that will be
used in Sec.~\ref{directCoup}. The observer looks at the system from
$Z\to -\infty$.}
\label{orbits}
\end{figure}

We start with unperturbed Keplerian motion and introduce Cartesian
coordinate system $(x,y,z)$ centered on the body with mass $M_2$ with 
the axes
$x$ and $y$ lying in the orbital plane and the unit vector $\hat x$
pointing towards the pericenter. We also use cylindrical coordinates
$(r,\theta,z)$, so that the position of the body $M_1$ is given by
$\vec{r}=r\cos{\theta}\hat{x}+ r \sin\theta\hat{y}$. The system
$(x,y,z)$ is rotated with respect to the observer reference frame
$(X,Y,Z)$ by the Euler angles $\Omega$ (longitude of the
  ascending node), $\iota$ (inclination) and $\omega$ 
(longitude of the pericenter), see Fig.~\ref{orbits}.
The motion within the orbital plane is characterized by the orbit's
semimajor axis $a$, its eccentricity $e$ and the time of the
pericenter passage $t_0$, see Appendix~\ref{osculating} for more
details. 

A perturbing force $\vec F$ leads to deviations from the Kepler
motion. The idea of the osculating orbits method is to find at every
given instance a Keplerian orbit that touches the actual
trajectory. The parameters of the latter orbit --- its {\it orbital
  elements} --- evolve with time according to the equations
\cite{Turner:1979yn} (see Appendix~\ref{osculating} for an outline of
the derivation),
\begin{subequations}\label{Lagrange}
\begin{align}
\label{Lagrange1}
\dot a &=
\frac{2}{\omega_b}\left\{\frac{F_{r}
    e}{\sqrt{1-e^2}}\sin \theta+\frac{F_{\theta}}{r}a\sqrt{1-e^2}\right\}\,, \,\,\,\\
\label{Lagrange2}
\dot e&= \frac{\sqrt{1-e^2}}{a\omega_b}\left\{F_{r} \sin
  \theta+F_{\theta}\bigg(\cos 
  \theta+\frac{1}{e}-\frac{r}{ae}\bigg)\right\}\,, \\
\label{Lagrange3}
\dot \Omega&= \frac{F_{z}r
  \sin(\theta+\omega)}{a^2\omega_b\sqrt{1-e^2}\sin\iota}\,, \\
\label{Lagrange4}
\dot\iota&=\frac{F_{z}r \cos(\theta+\omega)}{a^2\omega_b\sqrt{1-e^2}}\,, \\
\dot \varpi&= \frac{\sqrt{1-e^2}}{a e\omega_b}
\left\{-F_{r} \cos \theta+F_{\theta}\sin \theta\left[1+\frac{r}{a(1-e^2)}\right]\right\}\nonumber\\
&+ 2\sin^2\left(\frac{\iota}{2}\right)\dot\Omega\,,
\label{Lagrange5}  \\
\dot \epsilon_1&=
-\frac{2rF_{r}}{a^2\omega_b}+\left[1\!-\!\sqrt{1-e^2}\right]
\dot\varpi + 2\sqrt{1-e^2}  \sin^2\left(\frac{\iota}{2}\right)
\dot\Omega\,,
\label{Lagrange6}
\end{align}
\end{subequations}
where
\be
\label{orbfreq}
\omega_b=\sqrt{\frac{GM_T}{a^3}},
\ee
is the binary orbital frequency and we have decomposed the perturbing force
in the cylindric coordinates,
\be
\label{Fdecomp}
\vec{F}=F_r\hat{r}+F_{\theta}\hat{\theta}+F_z\hat{z}\;.
\ee
In Eqs.~(\ref{Lagrange5}), (\ref{Lagrange6}) we have introduced the
argument of the pericenter,
\be
\varpi=\omega+\Omega\,,
\ee
and the parameter
\be
\epsilon_1=\omega_b(t-t_0)+\varpi-\int \di t \,\omega_b\,,
\ee 
related to the pericenter passage time $t_0$.

Working at leading order in $\vec F$ we can evaluate the r.h.s. 
of Eqs.~\eqref{Lagrange} on the unperturbed Keplerian orbit. We will
 be interested in 
secular effects accumulating over many orbits.
%\footnote{As shown in
%\cite{Turner:1979yn} for the tensorial gravitational wave, the
%secular effect are gauge (coordinate)-invariant.}. 
Oscillations of ULDM will also produce non-secular effects that can,
in principle, be used to further probe the existence of $\Phi$ 
\cite{Rozner:2019gba}. 
However, one expects non-secular perturbations to be suppressed
compared to the secular ones.

Equations (\ref{Lagrange}) are to be supplemented
by the equations \eqref{eq:Rcm} or \eqref{eq:Rcmbeta} 
describing acceleration of the BB. We now consider separately
 the cases of purely gravitational interaction ($\alpha_A=\beta_A=0$) and
 direct coupling ($\alpha_A,\beta_A\neq 0$).

%%%%%%%%%%%%%%%%%%%%%
\subsection{Interaction only through gravity}\label{pureGrav}
%%%%%%%%%%%%%%%%%%%%%

We consider first the case of no direct coupling $\alpha_{A}=\beta_A=0$. 
As discussed in Sec.~\ref{sec:Lagpg}, in this case the BB follows a
geodesic in the galactic gravitational field (including the Newtonian
potential from DM). 

For the orbital dynamics, we read out the components of the perturbing
force from Eq.~(\ref{eq:bin_gra}): 
$F_{r}= -\ddot{\tilde\psi}r$, $F_{\theta}=F_{z}=0$. Vanishing of $F_z$
implies that the orbital elements $\Omega$ and $\iota$ describing the
orientation of the orbital plane with respect to the reference
coordinate system will remain constant. This can be traced back to the
absence of a preferred direction. The non-trivial orbital equations
reduce to,  
\begin{subequations}\begin{align}
\frac{\dot{a}}{a}&=-\frac{2e}{\omega_b\sqrt{1-e^2}}\ddot{\tilde\psi}\,
\frac{r}{a}\sin\theta\,,
\label{eqa_uni}\\
\dot{e}&=-\frac{\sqrt{1-e^2}}{
  \omega_b}\ddot{\tilde\psi}\,\frac{r}{a}\sin\theta\,,
\label{eqe_uni}\\
\dot{\varpi}&=\frac{\sqrt{1-e^2}}{
  e\omega_b}\ddot{\tilde\psi}\,\frac{r}{a}\cos\theta\,,
\label{eqw_uni}\\
\dot\epsilon_1&=\frac{2}{\omega_b}\left(\frac{r}{a}\right)^2
\ddot{\tilde\psi}+[1-\sqrt{1-e^2}]\dot{\varpi}\,.
\end{align}
\end{subequations} 
Note that Eqs. (\ref{eqe_uni}) and (\ref{eqa_uni}) imply the
conservation of the combination $a(1-e^2)$ which is connected
to the angular momentum $L$ of the osculating orbit, see Eq.~(\ref{orbelLA1}).  

Using Eq.~(\ref{eq:ddpsi}) and the 
Fourier decomposition of  the variables that describe the unperturbed
Keplerian motion (see Eqs.~(\ref{KeplFour}) in
Appendix~\ref{osculating}), 
we see that there is a secular effect   
 when a resonant condition 
$m_{\Phi}\approx N\omega_b/2$, $N \in \mathbb{N}$
 is satisfied. More precisely, let us define,
\be
\label{domegag}
\delta\omega_2=2m_{\Phi}-N\omega_b,
\ee
which we will assume to be small, $\delta\omega_2\ll 2 m_{\Phi}$.
One computes the secular contributions into the temporal derivatives of
the orbital parameters by averaging  over
time intervals $\Delta t$ satisfying  
$P_b/N\ll\Delta t\ll 2\pi/\delta\omega_2$, where we have introduced
the orbital period,
\be
\label{Pb}
P_b=2\pi\sqrt{\frac{a^3}{GM_T}}\,.
\ee
The result is,
\begin{subequations}\label{LPE1}
\begin{align}
\left\langle\frac{\dot{a}}{a}\right\rangle&=-4  G\rho_{\rm DM}P_b\frac{J_N(Ne)}{N}\sin\gamma_{2}(t)\,,\\
\left\langle\dot{e}\right\rangle&=-2  G\rho_{\rm DM}P_b(1-e^2)\frac{J_N(Ne)}{eN}\sin\gamma_{2}(t)\,,\\
\langle\dot{\varpi}\rangle&=-2 G\rho_{\rm DM}P_b\sqrt{1-e^2}\frac{J'_N(Ne)}{eN}\cos\gamma_{2}(t)\,,\\
\langle\dot{\epsilon}_1\rangle&=8 G\rho_{\rm
  DM}P_b\frac{J_N(Ne)}{N^2}\cos\gamma_{2}(t)
+[1-\sqrt{1-e^2}]\langle\dot{\varpi}\rangle,
\end{align}
\end{subequations} 
where $J_N(x)$ is the Bessel function, $J_N'(x)$
is its derivative with respect to the argument and 
\begin{equation}
\label{gammag}
\gamma_{2}(t)=\delta\omega_2 \,(t-t_0)+2 m_{\Phi}\,t_0+2 \Upsilon\,,
\end{equation}
is the combination describing the phase difference between the
ULDM-induced metric oscillations and the binary orbital motion. 

Expanding $J_N(N e)$ and $J_N'(N e)$ for small values of $e$,
 \begin{subequations}\label{Jexpa}\begin{align}
J_N(N
e)&=\frac{1}{N!}\left(\frac{Ne}{2}\right)^N\big(1+\mathcal{O}(e^2)\big)\,,\\ 
 J'_N(N e)&=\frac{1}{2(N-1)!}\left(\frac{Ne}{2}\right)^{N-1}
\big(1+\mathcal{O}(e^2)\big)\,,
 \end{align}
  \end{subequations}
one observes
that the  effect on the  orbital period  increases with $e$, and
vanishes as $e\to 0$. Therefore,  we expect to obtain the strongest
constraints from  measurements of $\langle \dot{P_b}\rangle$ for
highly eccentric systems. 
The pericenter advance $\langle \dot\varpi\rangle$ is, on the contrary,
enhanced for orbits with low eccentricity, notably for $N=1$. However,
measurements of $\langle \dot\varpi\rangle$ for nearly circular orbits
suffer from large uncertainties. Therefore, mildly eccentric orbits
are potentially the best candidates to constraint this effect. 

The numerical estimates of the effects (\ref{LPE1}) and their
comparison with the precision of pulsar timing 
measurements will be discussed in 
Sec.~\ref{sec:gravbounds}.

%%%%%%%%%%%%%%%%%%%%%
\subsection{Direct coupling}\label{directCoup}
%%%%%%%%%%%%%%%%%%%%%
 
We now turn on the direct coupling between ULDM and the binary. We
will assume that it generates stronger effects than the purely
gravitational interaction of the previous subsection, which will
be ignored. 

%%%%%%%%%%%%%%%%%%%%%%%%%%%%%%%
 \subsubsection{Motion of the binary barycenter}\label{sec:BB}
%%%%%%%%%%%%%%%%%%%%%%%%%%%%%%%

We start by analyzing the effects on the motion of the BB. Consider
first the linear coupling case. At  leading order in $\alpha_A$ one
can replace the dynamical variables on the r.h.s. of (\ref{eq:Rcm}) by
their unperturbed values. Then, using the Keplerian expression for
relative acceleration of the binary, the BB equation can be cast in
the form,
\begin{equation}\ddot{\vec{R}}=\alpha_{T}
  (\dot{\Phi}\V+\Phi{\vec{S}})
-\Delta\alpha\, \frac{\mu}{M_T} \,
\frac{\di(\Phi\,\dot{\vec{r}})}{\di t}.\label{eq2:r12}
\end{equation}
We are interested in the secular contribution that survives after
averaging over the orbital motion and the oscillations of the field
$\Phi$. Only the last term produces such contribution provided the
system is close to the resonance. The 
latter corresponds to the condition,
\be
\label{reslin}
\delta\omega_1\equiv m_{\Phi}-N\omega_b\ll m_\Phi~,~~~~~~~~N\in\mathbb{N}\;.
\ee
Using the expression (\ref{eq:phidm}) for the ULDM field and
Eqs.~(\ref{KeplFour}) for the Fourier decomposition of the Keplerian
variables, one averages over time intervals $\Delta t$, such that 
$P_b/N\ll\Delta t\ll 2\pi/\delta\omega_1$. This yields,
%\begin{widetext}
\be
\label{linearBB}
\begin{split}
\langle\ddot{\vec{R}}\rangle=
-\sqrt{2\rho_{\rm DM}} \Delta\alpha& \frac{\mu}{M_T} a\delta\omega_1
\bigg[\frac{J'_N(Ne)}{N}\cos\gamma_1(t)\,\hat x\\
&
\hspace{-.2cm}-\sqrt{1-e^2}\frac{J_N(Ne)}{Ne}\sin\gamma_1(t)\,\hat y\bigg],
\end{split}
\ee
%\end{widetext}
%\vspace{.1cm}
where
\begin{equation}\label{gammal}
\gamma_{1}(t)=\delta\omega_1 \,(t-t_0)+  m_{\Phi}\,t_0+  \Upsilon\;.
\end{equation} 
We observe that a non-universal coupling leads to a secular
acceleration of the BB along a direction in the orbital plane. If
$N=1$, the effect survives for nearly circular orbits,
\be
\label{linearBBcirc}
\langle\ddot{\vec{R}}\rangle=
-\sqrt{2\rho_{\rm DM}} \Delta\alpha \frac{\mu}{2M_T}
a\delta\omega_1\hat b(t)\,\quad\text{for}~N=1,~e \ll 1\;.
\ee  
We have introduced here the time-dependent unit vector (see
Fig.~\ref{orbits})\footnote{\label{foot:circ}
This can be equivalently written in the form,
\be
\label{bdef1}
\hat b(t)=\hat n\cos\big(\gamma_1(t)-\omega\big)
-(\hat z\times\hat n)\sin\big(\gamma_1(t)-\omega\big)\;,
\ee
which shows that for $N=1$ resonances 
$\hat b(t)$ is regular in the limit of strictly circular orbits. 
In this limit the pericenter passage time
$t_0$ entering into the definition of $\gamma_1(t)$, as well as 
the pericenter longitude
$\omega$, are not separately well-defined.
On the other hand, the combination $\omega-\omega_b
t_0$ is regular and equals to the angle at $t=0$ between the 
radial vector $\vec{r}$ and the direction towards the
ascending node of the orbit. Thus, the quantities that remain regular
for circular orbits can depend only on the difference 
$\gamma_1(t)-N\omega$. 
This is indeed the case of Eq.~(\ref{bdef1}) (for $N=1$).
} 
\be
\label{bdef}
\hat b(t)=\hat x\cos\gamma_1(t)-\hat y\sin\gamma_1(t)\;.
\ee
Note that the secular BB acceleration is suppressed by the small
frequency difference $\delta\omega_1$. In particular, it vanishes
exactly at resonance. As will be discussed in Sec.~\ref{sec:bounds}, 
this makes the BB motion less sensitive to the
ULDM effects than the orbital dynamics.

%We will discuss the prospects for observing 
%this effect in Sec.~\ref{sec:bounds}.

For quadratic coupling one uses Eq.~(\ref{eq:Rcmbeta}). The square of
the field amplitude at the binary system location entering the
r.h.s. can be written as,
\be
\label{phi2}
\frac{\Phi^2}{2}=\frac{\rho_{\rm DM}}{2m_\Phi^2}\big(
1+\cos(2m_\Phi t+2\Upsilon)\big)\;.
\ee
The oscillating term leads to the same phenomenology as in the case of
linear coupling. The resonance condition now coincides with that for
pure gravitational interaction, see Eq.~(\ref{domegag}), and the
corresponding secular BB acceleration is obtained from
(\ref{linearBB}) by the replacement,
\be
\label{lintoquad}
\Delta\alpha\mapsto\Delta\beta\frac{\sqrt{2\rho_{\rm
      DM}}}{N\omega_b}\,,~~~
\delta\omega_1\mapsto
\delta\omega_2\,,~~~
\gamma_1(t)\mapsto\gamma_2(t)\;.
\ee 
\vspace{.1cm}

An important difference is introduced by the constant
term in (\ref{phi2}), which gives rise to a secular BB acceleration
irrespective of whether the system is in resonance or not,
\be
\label{quadraticBB}
\langle\ddot{\vec{R}}\rangle_{\rm non-res}=\beta_T\frac{\rho_{\rm
    DM}}{m_\Phi^2}\vec{S}\;.
\ee
In particular, the acceleration will be present even for the ULDM mass
$m_\Phi$ well above the orbital frequency. 
In fact, it will be present even for solitary pulsars and stars\footnote{This effect is also felt by the Solar System barycenter. However, 
there will be a relative acceleration with respect to the typical pulsar, since the systems are located  in different coherent patches, cf. \eqref{eq:dB}}. The effect
has a clear physical
interpretation. Recall that the 
masses of the binary constituents depend on the local DM density, whereas
$\vec{S}$  points towards decreasing  DM density. For
$\beta_T>0$ ($\beta_T<0$) the system tends to minimize its mass by
moving towards a region of lower (higher) $\rho_{\rm DM}$.

%%%%%%%%%%%%%%%%%%%%%%%%%%%%%%%
\subsubsection{Evolution of the orbital elements: Linear coupling}\label{sec:bina}
%%%%%%%%%%%%%%%%%%%%%%%%%%%%%%%
 
We first derive the orbital
equations in the linear coupling case and then discuss the
modifications for quadratic coupling. 
From Eq.~(\ref{eq:r12}) we obtain the components of the perturbing
force,  
\begin{subequations}\label{forceDC}
\begin{align}
 F_{r}&=-\alpha_{T}
 \frac{GM_T}{r^2}\Phi-\alpha_{\mu}\dot{r}\,\dot{\Phi}+\Delta\alpha
 (\dot{\Phi} V_r+\Phi S_r)\,,\\ 
 F_{\theta}&= -\alpha_{\mu}r\dot{\theta}\,\dot{\Phi}+\Delta\alpha
 (\dot{\Phi} V_{\theta}+\Phi S_{\theta})\,,\\ 
 F_{z}&= \Delta\alpha(\dot{\Phi} V_{z}+\Phi S_{z})\,.
\end{align}
\end{subequations}
These are to be inserted into 
Eqs.~(\ref{Lagrange}). The result is quite lengthy and not
particularly illuminating; for
completeness we give it in Appendix~\ref{AppendixLPE2}.
What we are
interested in are the secular contributions which appear 
when the system satisfies the resonant condition (\ref{reslin}). 
Averaging over multiple orbits we obtain the following equations,
\begin{widetext}
\begin{subequations}\label{LPE2}
\begin{align}
\left\langle\frac{\dot{a}}{a}\right\rangle
=&2\sqrt{2\rho_{\rm DM}} \left\{(\alpha_{T} +2\alpha_{\mu})J_N(Ne)\sin\gamma_1
+\Delta\alpha\left[J'_N(Ne)\, U_x(\gamma_1) 
+ \frac{\sqrt{1-e^2}}{e}J_N(Ne)\, {U}'_y(\gamma_1)\right]\right\}\,,\\
\langle \dot{e}\rangle=&\sqrt{2\rho_{\rm DM}}\frac{(1-e^2)}{e}
\left\{(\alpha_{T} +2\alpha_{\mu}) J_N(Ne)\sin\gamma_1
+\Delta\alpha
\left[\left(J'_N(Ne)-\frac{J_N(Ne)}{Ne}\right) U_x(\gamma_1) \right.\right.\nonumber\\
 &\qquad\qquad\qquad\qquad
\left.\left.-\left(\frac{1}{\sqrt{1-e^2}}\frac{J'_N(Ne)}{N}
-\frac{\sqrt{1-e^2}}{e}J_N(Ne)
     \right){U}'_y(\gamma_1)\right]\right\}\,,\\
\langle\dot{\Omega}\rangle=&\sqrt{2\rho_{\rm
    DM}}\frac{\Delta\alpha}{\sin\iota}
\left[-\frac{J_N(Ne)}{Ne}\cos\omega\, {U}_z(\gamma_1)
+\frac{1}{\sqrt{1-e^2}}\frac{J'_N(Ne)}{N}\sin\omega\, 
{U}'_z(\gamma_1)\right]\,,\\
\langle\dot{\iota}\rangle=& \sqrt{2\rho_{\rm DM}}\Delta\alpha
\left[\frac{J_N(Ne)}{Ne}\sin\omega\, U_z(\gamma_1)
+\frac{1}{\sqrt{1-e^2}}\frac{J'_N(Ne)}{N} \cos\omega\,{U}'_z(\gamma_1)
\right]\,,\\
\langle\dot{\varpi}\rangle=&\sqrt{2\rho_{\rm DM}}\frac{\sqrt{1-e^2}}{e}\left\{ 
(\alpha_{T} +2\alpha_{\mu})J'_N(Ne)\cos\gamma_1-\Delta\alpha
\left[\left(\frac{\sqrt{1-e^2}}{e}J_N'(Ne)
-\frac{1}{e^2\sqrt{1-e^2}}\frac{J_N(Ne)}{N}\right)U_{y}(\gamma_1)\right.\right.
\nonumber\\
&\qquad\qquad\qquad\qquad
\left.\left.+\left(\frac{J_N'(Ne)}{Ne}-\frac{(1-e^2)}{e^2}J_N(Ne)\right)
{U}'_x (\gamma_1)\right]\right\}
+2\sin^2\left(\frac{\iota}{2}\right)\langle\dot{\Omega}\rangle\,,\\ 
\langle \dot{\epsilon}_1\rangle=&
2\sqrt{2\rho_{\rm DM}}\left[(\alpha_{T}-\alpha_{\mu})
  \frac{J_N(Ne)}{N}
\cos\gamma_1+\Delta\alpha\left(\frac{\sqrt{1-e^2}}{e}\frac{J_N(Ne)}{N}
U_y(\gamma_1)- \frac{J_N'(Ne)}{N}{U}'_x(\gamma_1)
\right)\right]
\nonumber\\
&
+[1-\sqrt{1-e^2}]\langle\dot{\varpi}\rangle+
2\sqrt{1-e^2}
\sin^2\left(\frac{\iota}{2}\right)\langle\dot{\Omega}\rangle\,.
\end{align}
\end{subequations} 
\end{widetext}
Here we have introduced the time-dependent vectors,
\bseq
\begin{align}
\label{defL}
\vec{U}(\gamma_1)=& \frac{1}{a\,\omega_b}\bigg(\V\cos\gamma_{1}+
\frac{\vec{S}}{m_\Phi}\sin\gamma_{1}\bigg),\\
\label{defLpr}
\vec{U}'(\gamma_1)=& \frac{1}{a\,\omega_b}\bigg(-\V\sin\gamma_{1}+
\frac{\vec{S}}{m_\Phi}\cos\gamma_{1}\bigg).
\end{align} 
\eseq
In deriving (\ref{LPE2}) we used the Fourier decomposition of the
Keplerian motion, see Eqs.~(\ref{KeplFour}). 
Note that for $|\vec{S}|\sim m_\Phi |\vec{V}|$, which is typical of
the ULDM halo, the magnitude of $\vec{U}$, $\vec{U}'$ is
given by the value of $|\vec{V}|$
in units of the orbital velocity,  
\begin{equation}\label{aomega0}
~~a\,\omega_b \sim 1.5\times 10^{-4}   \left(\frac{M_T }{M_{\odot}}\right)^{1/3}
\left(\frac{100 {\rm \,d}}{P_b}\right)^{1/3}\,.
\end{equation}
As discussed in Sec.~\ref{sec:DMfield}, 
we expect $|\vec{V}|\sim 10^{-3}$, 
and hence the terms multiplying 
$\Delta \alpha$ in \eqref{LPE2} can be as large as $O(10)$ for binaries with long
orbital periods.

Equations (\ref{LPE2}) describe the new effects due to the ULDM
coupling. To obtain the full evolution of the orbital parameters,
one has to add the standard contributions due to the deviations
from Keplerian dynamics, notably, the general relativistic
corrections. Assuming that all corrections are small, they can be
combined linearly.

In what follows we focus on two representative cases: 

\vspace{.1 cm}
%\begin{itemize}
%%%%%%%%%%%%%%%%%%%%%%%%%%%%%%%
%\item{
{\bf Case 1:} $\Delta \alpha=0$ or negligible ULDM gradients 
$\nabla\Phi=0$, eccentric binaries.
%}\label{sec:boundsCase1}
 %%%%%%%%%%%%%%%%%%%%%%%%%%%%%%%
The ULDM interaction in this case does not introduce any preferred
direction, so the orbital elements $\Omega$ and $\iota$ describing the
orientation of the orbital plane remain
constant.
The effect on the orbital period $P_b\propto a^{3/2}$ in this case was 
already discussed in \cite{Blas:2016ddr}. Similarly to what
happens for pure
gravitational interaction, it vanishes for circular orbits, $e=0$.  
The use of the osculating method allows us to derive how {\it all} the
orbital parameters are modified in the presence of $\Phi$.
We postpone the numerical estimates of the new effects and the
discussion of resulting bounds on the ULDM couplings till 
Sec.~\ref{sec:UCbounds}. 

\vspace{.1 cm}
 %%%%%%%%%%%%%%%%%%%%%%%%%%%%%%%
% \item{
{\bf Case 2:} $\Delta \alpha\neq 0$ and $\nabla\Phi\neq0$, 
nearly circular orbits.
%}\label{subsec:circular}
 %%%%%%%%%%%%%%%%%%%%%%%%%%%%%%%
This case is important because many binary pulsars are in systems with low
eccentricities \cite{Manchester:2004bp,pscat}. By inspection of
Eqs.~(\ref{LPE2}) we see that for $e\ll 1$ non-trivial effects survive
only for the resonances on the principal ($N=1$) and the second
($N=2$) harmonics. We consider them in turn.

For $N=1$, $e\to 0$, the secular equations simplify to,
\begin{subequations}
\label{eqadeltaN1}
\begin{align}
\left\langle\frac{\dot a}{a}\right\rangle
=&\frac{\sqrt{2\rho_{\rm DM}}}{a\omega_b} \Delta\alpha \left[\vec{V}
  \cdot\hat{b}+\frac{\vec{S}\cdot (\hat{z}\times\hat{b})}{m_\Phi}
\right]\,, \label{eqadeltanPbN1}\\ 
 \langle \dot{e}\rangle=& \sqrt{2\rho_{\rm
     DM}}\,\frac{\alpha_T+2\alpha_\mu}{2}\,\sin\gamma_1\,,\label{esmalleN1}\\ 
\langle  \dot{\Omega}\rangle=&
-\frac{\sqrt{2\rho_{\rm DM}}}{2a\omega_b\sin\iota}\Delta\alpha
\left[{\vec{V}}\cdot \hat{z}\cos(\gamma_1-\omega)\right.\nonumber\\
&\left.\qquad\qquad\qquad\quad\;
+\frac{\vec{S}\cdot \hat{z}}{m_\Phi}\sin(\gamma_1-\omega)
\right]\,,\\ 
\langle\dot{\iota}\rangle=&\frac{\sqrt{2\rho_{\rm DM}}}{2a\omega_b}\Delta\alpha
\left[-{\vec{V}}\cdot \hat{z}\sin(\gamma_1-\omega)
\right.\nonumber\\
&\left.\qquad\qquad\qquad\quad\;
+\frac{\vec{S}\cdot \hat{z}}{m_\Phi}\cos(\gamma_1-\omega)
\right]\,,
\\
\langle \dot{\varpi}\rangle=& \sqrt{2\rho_{\rm
     DM}}\,\frac{\alpha_T+2\alpha_\mu}{2e}\,\cos\gamma_1+\ldots\,,
\label{wsmalleN1}\\
 \langle \dot{\epsilon}_1\rangle=&\frac{\sqrt{2\rho_{\rm
       DM}}}{a\omega_b} \Delta\alpha 
\left[
  \vec{V}\cdot(\hat{z}\times\hat{b})-\frac{\vec{S}\cdot\hat{b}}{m_\Phi}
\right]\nonumber\\
&+2\sin^2\left(\frac{\iota}{2}\right)\langle
\dot{\Omega}\rangle\,, 
\end{align}
\end{subequations}
where $\hat{b}(t)$ is the unit vector introduced in (\ref{bdef}) and
dots in (\ref{wsmalleN1}) stand for terms of order $O(1)$ at $e\to
0$. Note that, unlike the case of universal coupling, we now have a
non-trivial effect on the semimajor axis and hence on the orbital
period $P_b$. Besides, the orbital plane will be rotating 
whenever the ULDM velocity and/or density
gradient have non-vanishing projections on its normal.  

A comment is in order. As pointed out in footnote~\ref{foot:circ}, the
quantity $\gamma_1$ is ill-defined for circular orbits and should
appear in the equations for physical observables only in the
combination $\gamma_1-N\omega$. Recalling the expression (\ref{bdef1})
for $\hat{b}$ we see that this requirement is satisfied for all
Eqs.~(\ref{eqadeltaN1}), except those for the eccentricity and the
argument of the pericenter, Eqs.~(\ref{esmalleN1}),
(\ref{wsmalleN1}). The resolution of this apparent inconsistency lies
in the observation that the perturbation theory developed in terms of
$e$ and $\varpi$ breaks down when $e\to 0$. Instead, a pair of good
variables is provided by~\cite{Edwards:2006zg}
\be
\label{kappaeta}
\kappa=e\cos\omega~,~~~~~\eta=e\sin\omega\;,
\ee  
which are the components of the Runge--Lenz vector along the
directions $\hat n$ and $(\hat{z}\times\hat{n})$ respectively (see
Appendix~\ref{osculating}). To obtain the secular evolution of the new
observables we can use 
Eqs.~(\ref{esmalleN1}),
(\ref{wsmalleN1}) at small, but finite $e$ and then take the limit
$e\to 0$. This yields the equations,
\bseq
\label{kappaetaN1}
\begin{align}
\langle\dot\kappa\rangle=&\sqrt{2\rho_{\rm DM}}
\,\frac{\alpha_T+2\alpha_\mu}{2}\,\sin(\gamma_1-\omega)\;,\\ 
\langle\dot\eta\rangle=&\sqrt{2\rho_{\rm DM}}
\,\frac{\alpha_T+2\alpha_\mu}{2}\,\cos(\gamma_1-\omega)\;,
\end{align}
\eseq
which depend on $\gamma_1-\omega$, as they should.
  
For $N=2$ we have 
$\langle \dot a\rangle=\langle \dot \Omega\rangle
=\langle \dot \iota\rangle=\langle \dot{\epsilon}_1\rangle=0$ and
\bseq
\label{kappaetaN2}
\begin{align}
\langle \dot{\kappa}\rangle=&\frac{\sqrt{2\rho_{\rm DM}}}{4a\omega_b}
\Delta\alpha
\left[\vec{V}\cdot\hat{\tilde b}
+\frac{\vec{S}\cdot(\hat{z}\times\hat{\tilde b})}{m_\Phi}
\right]\,,\label{kappaN2}\\ 
\langle \dot{\eta}\rangle=& 
\frac{\sqrt{2\rho_{\rm DM}}}{4a\omega_b}
\Delta\alpha\left[-\vec{V}\cdot(\hat{z}\times\hat{\tilde b})
+\frac{\vec{S}\cdot\hat{\tilde b}}{m_\Phi}\right]
\,,\label{etaN2}
\end{align}
\eseq
where we have introduced the unit vector
\be
\label{btilde}
\hat{\tilde b}(t)=\hat{n}\cos\big(\gamma_1(t)-2\omega\big)
-(\hat{z}\times\hat{n})\sin\big(\gamma_1(t)-2\omega\big)\,,
\ee
which replaces $\hat b$ for $N=2$. We observe that if $\vec{V}$ and/or
$\vec{S}$ have non-vanishing projections on the orbital plane, the
orbit gets polarized. This is reminiscent of the Nordtvedt effect
\cite{Nordtvedt:1968qs} in SEP violating theories 
that was studied in the context of binary pulsars by Damour
and Sch\"afer \cite{Damour:1991rq}. In the latter case one considers
an approximately constant extra force
$\vec{F}^{\rm DS}=\Delta\vec{g}$ generated by the differential
acceleration of the two binary members in the galactic gravitational
field. For weakly circular orbits the resulting secular equations are, 
\be
\label{StandardDS}
\langle \dot{\kappa}\rangle= \frac{3}{2a\omega_b} \vec{F}^{\rm
    DS}\cdot(\hat z\times \hat n)\,,~~~~
\langle \dot{\eta}\rangle= -\frac{3}{2 a\omega_b}  \vec{F}^{\rm
  DS}\cdot \hat n\,,
\ee
with no change in the other orbital elements. One may wonder why the
effects of the constant (in the case of Damour--Sch\"afer) and
oscillating (in our case) forces turn out to be similar. To understand
this one recalls that what matters for the osculating orbit equations
(\ref{Lagrange}) are the components of the force in the frame $(\hat
r,\hat \theta)$ attached to the orbiting bodies. A constant force will
rotate in this frame {\it clockwise} (from $\hat\theta$ to $\hat r$)
with frequency $\omega_b$. This resonates with $\cos\theta$ and
$\sin\theta$ factors in Eqs.~(\ref{Lagrange2}), (\ref{Lagrange5}) to
produce the secular drift of the eccentricity parameters
(\ref{StandardDS}). Consider now the $N=2$ ULDM resonance. The
perturbing force oscillates in the fixed reference frame with nearly
twice the orbital frequency $m_\Phi\approx 2\omega_b$. Its components
in the rotating frame $(\hat r,\hat\theta)$ contain harmonics with
frequencies $m_\Phi+\omega_b\approx 3\omega_b$ and  
 $m_\Phi-\omega_b\approx \omega_b$. The latter harmonic corresponds
 to a contribution that rotates in the plane $(\hat r,\hat\theta)$
 {\it counter-clockwise} and resonates with the $\theta$-dependent
 factors in Eqs.~(\ref{Lagrange2}), (\ref{Lagrange5}). Apart from the
 flip of the angular velocity of the force in the $(\hat
 r,\hat\theta)$ plane, which only affects the overall coefficient in
 the secular equations, the similarity with the Damour--Sch\"afer case
 is clear.

\subsubsection{Evolution of the orbital elements: Quadratic coupling}
\label{sec:quadra}

In the case of quadratic coupling we encounter two types of effects:
resonant and non-resonant ones. The former are due to the oscillating
contribution in $\Phi^2/2$, see Eq.~(\ref{phi2}), and appear
if $m_\Phi$ is close to $N\omega_b/2$. They are the same as for linear
coupling. The corresponding secular equations are obtained from
(\ref{LPE2}) by the substitution (\ref{lintoquad}) supplemented with
\begin{align}
\label{lintoquadrnew}
&\alpha_T\mapsto\beta_T\frac{\sqrt{2\rho_{\rm
      DM}}}{N\omega_b}\;,
&\alpha_\mu\mapsto\beta_\mu\frac{\sqrt{2\rho_{\rm
      DM}}}{N\omega_b}\;.
\end{align}
Then, the rest of the discussion of the previous subsection applies
without change. 

The non-resonant effects come from the first term in
(\ref{phi2}). From Eq.~(\ref{eq:r12beta}) we see that it leads to the
perturbing force in the orbital equation,
\be
\label{Fnr}
\vec{F}^{\rm nr}=-\frac{\rho_{\rm DM}}{2m_\Phi^2}\beta_T
\frac{GM_T\vec{r}}{r^3}+
\frac{\rho_{\rm DM}}{m_\Phi^2}\Delta\beta \vec{S}\;,
\ee
where the superscript ``nr'' stands for non-resonant. The first
contribution 
has the same form as the Newtonian force and can be absorbed into the
redefinition of the total mass. We will drop it in what follows. The
second term is present only for non-universal coupling and represents
an approximately constant\footnote{It varies on time scales of the
  ULDM coherence time which we assume to be much longer than the
  orbital period.} force pointing along the gradient of ULDM
density. This is exactly the situation considered by Damour and
Sch\"afer \cite{Damour:1991rq}, with the only difference that, unlike
their case, the differential acceleration need not point towards
the galactic center. The resulting phenomenology is well-known. Being
conservative, the new force does not affect the orbital
period. However, it induces an evolution of the eccentricity and of
the orientation of the orbit. The full set of secular equations reads, 
\begin{subequations}
\label{eqnr}
\begin{align}
\left\langle\dot a\right\rangle^{\rm nr}
=&0 \label{anr}\\ 
 \langle \dot{e}\rangle^{\rm nr}=& \frac{\rho_{\rm DM}}{m_\Phi^2}\,\Delta\beta\,
\frac{3\sqrt{1-e^2}}{2a\omega_b}\,S_y\,,\label{enr}\\ 
\langle  \dot{\Omega}\rangle^{\rm nr}=&
-\frac{\rho_{\rm DM}}{m_\Phi^2}\,\Delta\beta\,
\frac{3e\sin\omega}{2a\omega_b\sqrt{1-e^2}\sin\iota}\,S_z\,,\\ 
\langle  \dot{\iota}\rangle^{\rm nr}=&
-\frac{\rho_{\rm DM}}{m_\Phi^2}\,\Delta\beta\,
\frac{3e\cos\omega}{2a\omega_b\sqrt{1-e^2}}\,S_z\,,\\ 
\langle \dot{\varpi}\rangle^{\rm nr}=& -\frac{\rho_{\rm
    DM}}{m_\Phi^2}\,\Delta\beta\,
\frac{3\sqrt{1-e^2}}{2ea\omega_b}\,S_x+
2\sin^2\frac{\iota}{2}\langle  \dot{\Omega}\rangle^{\rm nr}\;,
\label{wenr}\\
\langle \dot{\epsilon}_1\rangle^{\rm nr}=&\frac{\rho_{\rm
    DM}}{m_\Phi^2}\,\Delta\beta\,\frac{3e}{a\omega_b}\,S_x
+[1-\sqrt{1-e^2}]\langle \dot{\varpi}\rangle^{\rm nr} \nonumber\\
&+2\sqrt{1-e^2}\sin^2\left(\frac{\iota}{2}\right)\langle
\dot{\Omega}\rangle^{\rm nr}\,.
\end{align}
\end{subequations}
In the limit of circular orbits only $\langle \dot{e}\rangle^{\rm nr}$ and
$\langle \dot{\varpi}\rangle^{\rm nr}$ survive and combine into
equations of the form (\ref{StandardDS}) for the parameters $\kappa$
and $\eta$ with the replacement 
$\vec{F}^{\rm DS}\mapsto(\rho_{\rm DM}/m_\Phi^2)\Delta\beta\vec{S}$.

Due to non-resonant effects, timing even of a single binary pulsar 
allows us to probe a wide range of ULDM masses $m_\Phi\ll T^{-1}_{\rm
  obs} V_0^{-2}$, where $T_{\rm obs}$
is the observation time and $V_0$ is the DM virial velocity. The
upper limit comes from the requirement that $T_{\rm obs}$ is less than
the ULDM coherence time $t_{\rm coh}$ (see Eq.~(\ref{eq:tcoh})), so
that the ULDM density gradient does not change during the
observational campaign. 
This requirement is implied by the standard pulsar timing procedure
which assumes that the
  time derivatives of the orbital elements can be treated as
  constant. In principle, it can be relaxed if the analysis is
  appropriately modified to allow for time variation of $\vec S$ (see
  more on this below).

%%%%%%%%%%%%%%%%%%%%%%%%%%%%%%%
\section{Numerical estimates}\label{sec:bounds}
%%%%%%%%%%%%%%%%%%%%%%%%%%%%%%%

In this section we estimate the size of the effects
derived above and
compare it with the precision achieved in   pulsar timing 
measurements. We do not aim at a systematic exploration of the bounds
set by observations on the ULDM properties which is beyond the scope
of this paper. Our goal is to illustrate the potential reach of pulsar
timing constraints and identify the most promising directions for
further analysis.  

The standard timing models for binary pulsars parameterize the time of
arrival of the $i$th pulse as a function of $i$, the mass of the
pulsar $M_1$, its companion $M_2$, and the set of parameters $\{X\}$
which include the pulsar spin frequency, orbital elements, BB velocity
and acceleration with respect to the Solar System, etc. 
\cite{Teukolsky1976,1986AIHS...44..263D,Edwards:2006zg}. Secular
variations are introduced phenomenologically by allowing a given
parameter $X$ to be a linear function of time, $X(t)=X_0+\dot{X} t$.  
In our context, this approximation is valid if the system is
sufficiently close to resonance (for resonant effects) or if the
observation time does not exceed the ULDM coherence time (for
non-resonant effects). Violation of these conditions leads to a loss
of sensitivity that we will qualitatively discuss later in this
section. A systematic study should incorporate the modifications we
derived, including modulations due to deviations from
resonance or loss of coherence of the ULDM field, in the timing
package \cite{Edwards:2006zg}. We leave this study for the future. 

Within the assumption of a constant secular drift of the orbital
parameters  
Ref.~\cite{Teukolsky1976} estimated the {\it statistical}
uncertainty of the measured drift in a given
campaign\footnote{Ref.~\cite{Teukolsky1976} considers the binary
  pulsar PSR1913+16 and makes some simplifications specific to this
  system in the derivation of the covariance matrix for the orbital
  parameters. Still, the final results can be applied to other systems as
  order-of-magnitude estimates.}. 
We quote the
results for the time derivatives of the orbital period and
eccentricity which are typically the two quantities measured with the
highest precision\footnote{Another well-measured quantity is the
  precession of the pericenter $\dot\omega$. This, however, receives
  large GR contributions \cite{Damour:1991rd} and is used to extract
  the total mass of the binary $M_T$. Thus, it does not provide
  independent constraints on deviations from GR.}, 
\bseq
\label{errors}
\begin{align}
\label{errorPdot}
 \delta\dot{P}_b=C_P& \times 10^{-12}\,\frac{M_T}{M_2}
 \left(\frac{M_T}{M_\odot} \right)^{-1/3} 
\left(\frac{P_b}{100\, \rm{d}}\right)^{4/3}
\nonumber\\
& \times \left( \frac{\delta t}{1\mu s}\right)
\left(\frac{\dot n_{\rm dat}}{{\rm{d}}^{-1}}\right)^{-1/2}   
\left( \frac{T_{\rm obs}}{10 \,\rm{y}}\right)^{-5/2},\\
\label{erroredot} 
   \delta\dot{e}=C_e&\times 10^{-18}\, {\rm s}^{-1}\,\frac{M_T}{M_2}
\left( \frac{M_T}{M_\odot} \right)^{-1/3} 
\left(\frac{P_b}{100\, \rm{d}}\right)^{-2/3}
\nonumber\\
 &\times \left( \frac{\delta t}{1\mu s}\right)
\left(\frac{\dot n_{\rm dat}}{{\rm{d}}^{-1}}\right)^{-1/2}   
   \left( \frac{T_{\rm obs}}{10 \,\rm{y}}\right)^{-3/2},
\end{align} 
\eseq
where the coefficients $C_P$, $C_e$ depend on the
eccentricity and the orientation of the orbit and are typically 
of order one, $\dot n_{\rm dat}$ is
the rate at which the timing data are acquired with the precision 
$\delta t$ and $T_{\rm obs}$ is the total time of observation. Here
``d'' stands for days and ``y'' for years. Eqs.~(\ref{errors}) set the
benchmark for the magnitude of the effects that can be potentially
detected. 
It is worth noting that the numbers we are using are rather
conservative. The future Square Kilometer Array (SKA) is expected to
reach the timing precision of ${\delta t }\sim 80-230$\,ns 
with 10 min of integration of SKA \cite{Kramer11}, which suggests that
the data acquisition rate $\dot n_{\rm dat}$ can exceed $1/\rm d$.  

We observe from Eqs.~(\ref{errors}) that the statistical uncertainty
in determination of $\dot P_b$ decreases with observation time faster
than that of $\dot e$. However, the measured value $\dot P_b^{\rm
  meas}$ differs from the intrinsic one $\dot P_b^{\rm int}$ due to a
number of {\it systematic} effects.
The most important among them are related to the motion of the
BB with respect to the Solar System Barycenter: an apparent change of
the orbital period due to the galactic acceleration
\cite{1991ApJ...366..501D}
and the Shklovskii
effect \cite{1970SvA....13..562S}. The corresponding contributions
have to be subtracted 
to find $\dot P_b^{\rm int}$. This introduces additional uncertainty
in the comparison of the theoretical predictions made in terms of
$\dot P_b^{\rm int}$ with the data. On the other hand,  
the measurements of $\dot{e}$ are not contaminated by any known
external systematic effect \cite{Freire:2012nb}. This gives them an
advantage over measurements of $\dot P_b$.

 %%%%%%%%%%%%%%%%%%%%%%%%%%%%%%%
\subsection{Resonant effects}\label{sec:resonant}
%%%%%%%%%%%%%%%%%%%%%%%%%%%%%%%
We start with the resonant effects and assume that the deviations from
the resonance is small enough, so that the period of secular
modulations 
\begin{align}
\label{Tmod}
T_{\rm mod}\equiv 2\pi/{\delta\omega_{1,2}}
\end{align}
is significantly longer than the duration of the
observational campaign.
More precisely, in the numerical estimates we will assume
$|\delta\omega_{1,2}|\lesssim 1/T_{\rm obs}$. 
Then the modulation phase $\gamma_{1,2}$ in
the secular equations can
be treated as constant which corresponds to a constant drift of the
orbital parameters.

\subsubsection{Pure gravitational coupling} \label{sec:gravbounds}

This case was considered in
Sec.~\ref{pureGrav}. 
Replacing the semi-major axis $a$ in favour of the orbital period $P_b$ using  expression (\ref{Pb}) and 
substituting realistic values into Eqs.~\eqref{LPE1} we obtain the
estimates,
\begin{widetext}
\begin{subequations}\label{dotwgrav}
\begin{align}
\langle \dot{P_b} \rangle&\simeq
-1.6\times10^{-17}\left(\frac{P_b}{100\, {\rm d}}\right)^2
\left(\frac{ \rho_{\rm DM} }{0.3\, \frac{{\rm GeV}}{\rm
      cm^3}}\right)\frac{J_N(Ne)}{N}\sin\gamma_{2}\,,\label{dotPbgrav2}\\ 
\langle \dot{e} \rangle&\simeq
-6.2\times10^{-25}\, {\rm s}^{-1}\,\left(\frac{P_b}{100\, {\rm d}}\right)
\left(\frac{ \rho_{\rm DM} }{0.3\, \frac{{\rm GeV}}{\rm
      cm^3}}\right)(1-e^2)\frac{J_N(Ne)}{ e
  N}\sin\gamma_{2}\,,\label{dotegrav2}\\ 
 \langle\dot{\varpi}\rangle= \langle\dot{\omega}\rangle&\simeq 
-1.1\times10^{-15}\,\frac{\rm deg}{\rm
   yr}\left(\frac{P_b}{100\, {\rm d}}\right) \left(\frac{ \rho_{\rm
       DM}}{0.3\, \frac{{\rm GeV}}{\rm cm^3}}\right)
 \sqrt{1-e^2}\frac{J'_N(Ne)}{N
   e}\cos\gamma_{2}\,,\label{dotwgrav2}\\ 
 \langle\dot{\epsilon}_1\rangle&\simeq
   2.5\times10^{-24}\,{\rm s}^{-1}\,\left(\frac{P_b}{100\, {\rm d}}\right)
 \left(\frac{ \rho_{\rm DM}(\vec{x})}{0.3 \frac{{\rm GeV}}{\rm
       cm^3}}\right)\left[\frac{J_N(Ne)}{N^2}+ (1-e^2-
   \sqrt{1-e^2})\frac{J'_N(Ne)}{4 e N}\right]\cos\gamma_{2}\,. 
\end{align}
\end{subequations} 
\end{widetext}
As already pointed out in \cite{Blas:2016ddr}, detection of these
effects presents a strong challenge. It would require a determination
of  $\langle\dot{P}_b\rangle$ with the accuracy of $10^{-17}$ for
systems with orbital periods of $\sim 100$ days. This accuracy may be
achieved in the future for the double pulsar \cite{Kehl:2016mgp}. The
latter, however, has $P_b \sim 0.1$\,d, which makes the effects
described by Eqs.~(\ref{dotwgrav}) completely negligible. 
Eq.~\eqref{errorPdot} shows how the
precision deteriorates for the non-relativistic systems where the
effect of ULDM is more sizable. 

The consideration of the rest of orbital  parameters shown in
\eqref{dotwgrav} does not change the previous conclusion. Hence,
unless a binary system is found in a region of large DM density,
the prospects to detect the ULDM-induced oscillations of the galactic
gravitational field with binary pulsars are slim.

%%%%%%%%%%%%%%%%%%%%%%%%%%%%%%%
\subsubsection{Universal direct coupling}\label{sec:UCbounds}
%%%%%%%%%%%%%%%%%%%%%%%%%%%%%%%
\begin{table*}
\begin{ruledtabular}
\begin{tabular}{c|c|c||c|c}
$N$&$m_\Phi\,,~10^{-22}\,{\rm eV}$ & $|\alpha\sin\gamma_1|\,,~{\rm
  GeV}^{-1}$&$m_\Phi\,,~10^{-22}\,{\rm eV}$&$|\beta\sin\gamma_2|\,,~{\rm
  GeV}^{-2}$\\
\hline
\hline
1& [4.97 ; 5.11] & $1.7\times 10^{-21}$& [2.48 ; 2.55] & $3.7\times 10^{-31}$\\
\hline
2& [10.01 ; 10.15] & $4.0\times 10^{-21}$&[5.00 ; 5.07] & $1.8\times 10^{-30}$\\
\hline
3& [15.05 ; 15.19] & $8.4\times 10^{-21}$ & [7.52 ; 7.59] & $5.6\times 10^{-30}$\\
\hline
4& [20.09 ; 20.23] & $1.7\times 10^{-20}$ & [10.04 ; 10.11] & $1.5\times 10^{-29}$
\end{tabular}
\end{ruledtabular}
\caption{Upper limits on the absolute value of a universal linear
  (left) and quadratic (right) couplings of ULDM to matter following
  from the $\dot P_b$ measurements 
for the binary pulsar J1903+0327. The limits apply to
the indicated intervals of ULDM masses. We show the results
corresponding to resonances on the four lowest harmonics. The ULDM
density at the location of the binary is assumed to be 
$\rho_{\rm DM}=0.3\,{\rm GeV}/{\rm cm}^{3}$.}
\label{tab:univ1}
\end{table*}

We now show that timing of binary pulsars can give relevant bounds on
direct ULDM -- SM couplings. Let us first neglect the difference in
the effective couplings of the two binary members. Setting
$\Delta\alpha=0$, $\alpha_T=\alpha_\mu=\alpha$ in Eqs.~(\ref{LPE2}) we obtain,
\begin{widetext}
\begin{subequations}
\label{eq:dotalpha}
\begin{align}
\langle\dot{P_b}\rangle& \simeq 2.5\times 10^{-12} 
\left(\frac{\alpha}{10^{-23} {\rm{GeV}}^{-1}}\right)
 \left(\frac{P_b}{100\,
    {\rm d}}\right)\sqrt{\frac{\rho_{\rm DM}}{{0.3\, \frac{{\rm
          GeV}}{\rm cm^3}}}} \,  J_N(N e)\sin\gamma_1\,,\label{periodderDC}\\ 
\langle\dot{e}\rangle&\simeq 10^{-19}\,{\rm s}^{-1}
\left(\frac{\alpha}{10^{-23} {\rm{GeV}}^{-1}}\right)
\sqrt{\frac{\rho_{\rm DM}}{{0.3\, \frac{{\rm GeV}}{\rm cm^3}}}}\,
\frac{(1-e^2)}{e} J_N(N e)\sin\gamma_1\,,\label{ederDC}\\ 
\langle\dot{\varpi}\rangle=\langle\dot{\omega}\rangle&\simeq
1.8\times10^{-10}\,\frac{ \rm deg}{\rm yr} 
\left(\frac{\alpha}{10^{-23} {\rm{GeV}}^{-1}}\right)
\sqrt{\frac{\rho_{\rm DM}}{{0.3\, \frac{{\rm GeV}}{\rm
        cm^3}}}}\,\frac{  \sqrt{1-e^2} }{e}J_N'(N e)\cos\gamma_1\,. 
\end{align}
\end{subequations} 
For the case of quadratic coupling the substitution
(\ref{lintoquadrnew}) yields,
\begin{subequations}
\label{eq:dotbeta}
\begin{align}
\langle\dot{P_b}\rangle& \simeq 11.4\times 10^{-12} 
\left(\frac{\beta}{10^{-32} {\rm{GeV}}^{-2}}\right)
 \left(\frac{P_b}{100\,
    {\rm d}}\right)^2\left(\frac{\rho_{\rm DM}}{{0.3\, \frac{{\rm
          GeV}}{\rm cm^3}}}\right) 
\,  \frac{J_N(N e)}{N}\sin\gamma_2\,,\label{periodderbeta}\\ 
\langle\dot{e}\rangle&\simeq 4.4\times10^{-19}\,{\rm s}^{-1}
\left(\frac{\beta}{10^{-32} {\rm{GeV}}^{-2}}\right)
\left(\frac{P_b}{100\,{\rm d}}\right)
\left(\frac{\rho_{\rm DM}}{{0.3\, \frac{{\rm GeV}}{\rm cm^3}}}\right)\,
(1-e^2)\, \frac{J_N(N e)}{Ne}\sin\gamma_2\,,\label{ederbeta}\\ 
\langle\dot{\varpi}\rangle=\langle\dot{\omega}\rangle&\simeq
7.9\times 10^{-10}\,\frac{ \rm deg}{\rm yr} 
\left(\frac{\beta}{10^{-32} {\rm{GeV}}^{-2}}\right)
\left(\frac{P_b}{100\,{\rm d}}\right)
\left(\frac{\rho_{\rm DM}}{{0.3\, \frac{{\rm GeV}}{\rm
        cm^3}}}\right)\,\sqrt{1-e^2}\,\frac{J_N'(N e)}{Ne}\cos\gamma_2\,. 
\end{align}
\end{subequations}
\end{widetext}
In both cases $\langle\dot\Omega\rangle=\langle\dot\iota\rangle=0$ and
$\langle\dot{\epsilon_1}\rangle =
[1-\sqrt{1-e^2}]\langle\dot{\omega}\rangle$.
Comparison of the above expressions with the estimated precision
(\ref{errors}) suggests that  pulsar timing can probe the values of
the linear (quadratic) coupling down to\footnote{This under the assumption
  that the phase $\gamma_{1,2}$ takes a random value between $0$ and
  $2\pi$, so that $\sin\gamma_{1,2}$ is a number of order one.} 
$\alpha\sim10^{-23}\,{\rm GeV}^{-1}$ ($\beta\sim10^{-32}\,{\rm
  GeV}^{-2}$). 
The range of ULDM masses accessible at the maximal sensitivity is
restricted by the resonant condition, which reads 
$|m_\Phi-N\omega_b|\lesssim T_{\rm obs}^{-1}$ 
( $|2 m_\Phi-N\omega_b|\lesssim T_{\rm obs}^{-1}$ ) for the $N$th
resonance. In systems with large eccentricity several first
resonances have comparable constraining power.

As an example, let us consider the binary pulsar 
J1903+0327. The system consists of a millisecond pulsar with mass
$M_1\simeq 1.67 M_{\odot}$ and a main-sequence companion star 
with $M_2\sim M_{\odot}$. The orbital period and eccentricity are 
$P_b= 95  \, {\rm d}$, $e= 0.44$. For this system both 
$\dot{P}_{b}$ and $\dot{e}$ have been measured using the data
collected over $T_{\rm obs}\sim 3\,{\rm years}$ \cite{Freire:2010tf}: 
$\dot{P}_{b}=(-53\pm33)\times 10^{-12}$ and $\dot{e}=(14\pm 6)\times
10^{-17} {\rm s}^{-1}$. Taking conservatively the measured value of
$\dot P_b$ as the upper limit on the possible effects of ULDM we
obtain the constraints shown in Table~\ref{tab:univ1}.  
The constraints from the drift of the eccentricity are a factor
of $20$ weaker.   

The numbers in the third column of the table can be compared with the
bounds on linearly coupled fields following from the measurements
of light deflection by the Sun \cite{Bertotti:2003rm} 
and planetary dynamics \cite{Iorio:2005fk}. These are at the level 
$\alpha\lesssim 10^{-21}\,{\rm GeV}^{-1}$ in the considered mass
range. We see that the pulsar bounds are competitive. Moreover, as
emphasized in Sec.~\ref{sec:Lag}, they probe effective scalar couplings
in the regime of strong gravity, which are in general different from
the weak-field coupling tested by the Solar System observations. For
the quadratically coupled ULDM the existing bounds in this mass range
are very mild (see the discussion in \cite{Blas:2016ddr}). This makes
binary pulsar timing particularly promising for testing this type of
models. As highlighted in \cite{Freire:2012nb}, the determination of
the orbital parameters of J1903+0327 is expected to improve in
future. Apart from the sheer increase of statistics due to a longer
campaign, the improvement will come from infrared interferometric
observations of the main sequence star companion, allowing to better
determine the location of the system in the Galaxy and the
orientation of its orbit.  

For systems with nearly circular orbits a universal interaction with
ULDM does not lead to any change in the orbital period. Then the  
strongest effect is the
drift of eccentricity occurring at the $N=1$ resonance. To illustrate
the ensuing constraints, we consider the system J1713+0747 that has
been timed for more than two decades, $T_{\rm obs}\sim 20\,{\rm y}$
\cite{1993ApJ...410L..91F,Zhu:2015mdo,Zhu:2018etc}.    
It consists
of a millisecond pulsar with mass $M_1\simeq 1.3 \,  M_{\odot}$ in a
  wide nearly circular orbit ($P_b\simeq 67.8 \,{\rm d}$, 
$e\simeq 7\times 10^{-5} $) with a white dwarf companion of mass
  $M_2\simeq  0.29 \, M_{\odot}$. 
The residual drifts of its orbital period and eccentricity parameters
(after subtracting the known standard contributions) have been
constrained at the level \cite{Zhu:2018etc}:
$\dot{P}_b= (0.03\pm 0.15) \times 10^{-12}$,
$\dot\kappa=(0.4\pm 0.4)\times 10^{-17}\,{\rm s}^{-1}$,
$\dot\eta=(0.7\pm 0.4)\times 10^{-17}\,{\rm s}^{-1}$. Comparing the two
latter bounds with the predicted effect of ULDM,
Eqs.~(\ref{kappaetaN1}) and analogous equations for the quadratic
coupling, we obtain the constraints presented in
Table~\ref{tab:univ2}.
\begin{table}
\begin{ruledtabular}
\begin{tabular}{c|c|c|c}
$N$&$m_\Phi\,,~10^{-22}\,{\rm eV}$ & $|\alpha|\,,~{\rm
  GeV}^{-1}$&$ |\beta|\,,~{\rm
  GeV}^{-2}$\\
\hline
\hline
1& [3.52 ; 3.53] & --- & $9.1\times 10^{-31}$\\
\hline
1& [7.05 ; 7.07] & $2.8\times 10^{-21}$& ---
\end{tabular}
\end{ruledtabular}
\caption{Upper limits on ULDM -- matter couplings
  from the measurements of the drift of the eccentricity parameters
for the binary pulsar J1713+0747. The constrained mass intervals
  correspond to the $N=1$ resonance. The first row corresponds to the quadratic coupling and the second to the linear coupling.  We assume
$\rho_{\rm DM}=0.3\,{\rm GeV}/{\rm cm}^{3}$.}
\label{tab:univ2}
\end{table}  
These bounds are comparable to those in Table~\ref{tab:univ1}, but
apply to different resonant ULDM masses, as a consequence of different
orbital periods of J1713+0747 and J1903+0327. We stress that they come entirely
from the measurement of the eccentricity parameters.
Note that, unlike the bounds in Table~\ref{tab:univ1}, these
constraints are independent of the relative phase $\gamma_{1,2}$
between the orbital motion and the ULDM oscillations. Indeed, the
expressions (\ref{kappaetaN1}) for the drift of the eccentricity
parameters $\kappa$ and $\eta$ involve both $\sin\gamma_{1,2}$ and
$\cos\gamma_{1,2}$. Hence, by taking the combination 
$\sqrt{\langle \dot\kappa\rangle^2+\langle \dot\eta\rangle^2}$ the
dependence on $\gamma_{1,2}$ can be eliminated.

%%%%%%%%%%%%%%%%%%%%%%%%%%%%%%%
\subsubsection{Non-universal coupling}\label{sec:genericbounds}
%%%%%%%%%%%%%%%%%%%%%%%%%%%%%%%

All orbital elements of a binary system will in general drift under
the influence of ULDM, once the non-universality of the coupling is
taken into account. In our preliminary analysis we focus on the
orbital period and eccentricity. We also restrict ourselves to systems with
nearly circular orbits, which constitute the majority of observed
pulsar binaries \cite{pscat,Manchester:2004bp}. As derived in
Sec.~\ref{sec:bina}, only resonances on the first two harmonics
survive in this case. The precise magnitude of the effects depends on
the directions of the ULDM velocity and density gradient with respect
to the orbital plane. It is convenient to introduce the following
notations, 
\bseq
\label{qdefs}
\begin{align}
\label{q1def}
q_1&=A\cos(\gamma_1-\omega)+B\sin(\gamma_1-\omega)\,,\\
\label{q1kappadef}
\tilde
q_{1,\kappa}&=A\cos(\gamma_1-2\omega)+B\sin(\gamma_1-2\omega)\,,\\
\label{q1etadef}
\tilde
q_{1,\eta}&=-A\sin(\gamma_1-2\omega)+B\cos(\gamma_1-2\omega)\,,\\
\label{qsimpledef}
q&=\sqrt{A^2+B^2}\;,
\end{align}
\eseq
where
\begin{align}
A=\frac{\vec V\cdot\hat n}{V}+\frac{\vec S\cdot (\hat z\times\hat
  n)}{m_\Phi V}\;,~~~
B=\frac{\vec S\cdot\hat n}{m_\Phi V}-\frac{\vec V\cdot (\hat z\times\hat
  n)}{V}\;,
\end{align}
and $V$ is the absolute value of the relative velocity between the
ULDM and the binary. Generically, one expects the quantities $q_1$,
$\tilde q_{1,\kappa}$, $\tilde q_{1,\eta}$, $q$ to be $O(1)$ numbers. 
Note that $q$ depends only on the projections of
$\vec V$ and $\vec S$ on the orbital plane and is independent of the
phase difference between the ULDM oscillations and the orbital
motion. 
Substituting the numerical values into Eqs.~(\ref{eqadeltanPbN1}),
(\ref{kappaetaN2}) and using the definitions of the vectors $\hat b$,
$\hat{\tilde b}$ from (\ref{bdef1}), (\ref{btilde}) we obtain:
\begin{widetext}
\bseq
\begin{align}
\label{PbdotDal}
\langle\dot P_b\rangle&\simeq 2.7\times 10^{-12}
\left(\frac{\Delta\alpha}{10^{-23}\,{\rm GeV}^{-1}}\right)
\left(\frac{M_T}{M_\odot}\right)^{-1/3}
\left(\frac{P_b}{100\,{\rm d}}\right)^{4/3}
\sqrt{\frac{\rho_{\rm DM}}{0.3\frac{\rm GeV}{{\rm cm}^3}}}
\left(\frac{V}{10^{-3}}\right) q_1 &\text{for}~N=1\;,\\
\label{kappadotDal}
\langle\dot \kappa\rangle&\simeq 5.3\times 10^{-20}\,{\rm s}^{-1}
\left(\frac{\Delta\alpha}{10^{-23}\,{\rm GeV}^{-1}}\right)
\left(\frac{M_T}{M_\odot}\right)^{-1/3}
\left(\frac{P_b}{100\,{\rm d}}\right)^{1/3}
\sqrt{\frac{\rho_{\rm DM}}{0.3\frac{\rm GeV}{{\rm cm}^3}}}
\left(\frac{V}{10^{-3}}\right) \tilde q_{1,\kappa}
&\text{for}~N=2\;.
\end{align}
\eseq
The expression for $\langle \dot\eta\rangle$ is the same as
(\ref{kappadotDal}) 
with the
replacement $\tilde q_{1,\kappa}\mapsto\tilde q_{1,\eta}$. 

Using the substitution (\ref{lintoquad}) one obtains the
equations for the case of quadratic coupling, 
\bseq
\begin{align}
\label{PbdotDbe}
\langle\dot P_b\rangle& \simeq 12.1\times 10^{-12}
\left(\frac{\Delta\beta}{10^{-32}\,{\rm GeV}^{-2}}\right)
\left(\frac{M_T}{M_\odot}\right)^{-1/3}
\left(\frac{P_b}{100\,{\rm d}}\right)^{7/3}
\left(\frac{\rho_{\rm DM}}{0.3\frac{\rm GeV}{{\rm cm}^3}}\right)
\left(\frac{V}{10^{-3}}\right) q_2
&\text{for}~N=1\;,\\
\label{kappadotDbe}
\langle\dot \kappa\rangle&\simeq 11.9\times 10^{-20}\,{\rm s}^{-1}
\left(\frac{\Delta\beta}{10^{-32}\,{\rm GeV}^{-2}}\right)
\left(\frac{M_T}{M_\odot}\right)^{-1/3}
\left(\frac{P_b}{100\,{\rm d}}\right)^{4/3}
\left(\frac{\rho_{\rm DM}}{0.3\frac{\rm GeV}{{\rm cm}^3}}\right)
\left(\frac{V}{10^{-3}}\right) \tilde q_{2,\kappa}
&\text{for}~N=2\;,
\end{align}
\eseq
and similarly for $\langle\dot\eta\rangle$ with $\tilde q_{2,\kappa}$
changed to $\tilde q_{2,\eta}$. The coefficients $q_2$, $\tilde
q_{2,\kappa}$, $\tilde q_{2,\eta}$ have the same form as
Eqs.~(\ref{qdefs}) with $\gamma_1$ replaced by $\gamma_2$. 
\end{widetext}

To illustrate the attainable sensitivity, let us again consider the
system J1713+0747. The constraints on the drift of its orbital period
and eccentricity yield the limits on the ULDM couplings shown in
Table~\ref{tab:nonuniv}. The limits set by $\dot P_b$ are the
strongest current constraints in the relevant mass ranges, both for
the quadratic and linear couplings. In particular, they are by a few
orders of magnitude stronger than the bounds in Table~\ref{tab:univ2}
obtained for the same system assuming universal couplings. On the
other hand, the constraints in Table~\ref{tab:nonuniv} do not restrict
the values of the ULDM couplings directly, but only their difference
for the two members of the binary. As discussed in Sec.~\ref{DCcase},
in general one expects $\Delta\alpha\sim 0.1 \alpha$ and similarly for
$\beta$, but the precise relation may depend on the model. Besides,
the bounds in Table~\ref{tab:nonuniv} depend on unknown order-one
coefficients $q_{1,2}$, so inferring constraints on ULDM couplings from
them requires an additional assumption that these coefficients do not
vanish accidentally. In this respect, the bounds in Table~\ref{tab:univ2}
are more robust.
\begin{table}
\begin{ruledtabular}
\begin{tabular}{c|c|c}
$N$&$m_\Phi\,,~10^{-22}\,{\rm eV}$ & --- \\
\hline
\hline
1& [3.52 ; 3.53] & $|\Delta\beta\,q_2|<4.3\times 10^{-34}\,{\rm GeV}^{-2}$\\
\hline
2& [7.05 ; 7.06] & $|\Delta\beta\,q|<2.2\times 10^{-30}\,{\rm GeV}^{-2}$\\
\hline
1& [7.05 ; 7.07] & $|\Delta\alpha\,q_1|<1.3\times 10^{-24}\,{\rm GeV}^{-1}$ \\
\hline
2& [14.11 ; 14.13] & $|\Delta\alpha\,q|<3.4\times 10^{-21}\,{\rm GeV}^{-1}$
\end{tabular}
\end{ruledtabular}
\caption{Limits on the difference of the ULDM couplings to the members
  of the binary pulsar system J1713+0747. For masses 
  corresponding to the $N=1$ resonance the limits come from the
  measurement of $\dot P_b$, whereas the $N=2$ resonances are
  constrained by $\dot \kappa$, $\dot \eta$. 
We assume the ULDM density
  at the system location  
$\rho_{\rm DM}=0.3\,{\rm GeV}/{\rm cm}^{3}$ and the relative
velocity between the binary and ULDM $V=10^{-3}$.}
\label{tab:nonuniv}
\end{table}

The bounds from $\dot\kappa$,
$\dot \eta$ listed in Table~\ref{tab:nonuniv} are
significantly weaker compared to those from $\dot P_b$, 
but they probe different masses.
In general, constraining ULDM couplings from eccentricity measurements
can have certain advantages. First, as already mentioned, such
measurements are less subject to systematic uncertainties than those
of $\dot P_b$. Second, the ULDM contribution into 
$\dot\kappa$, $\dot\eta$ has one less power of 
$\left(\frac{P_b}{100\,{\rm d}}\right)$ which makes it more important than
$\dot P_b$ 
for not so slow systems. Yet more powerful constraints can
potentially be derived from a dedicated analysis of timing data using
the complete set of equations for the orbital parameters. 

The analogy with the Nordtvedt effect mentioned at the end of
Sec.~\ref{sec:bina} may suggest to consider 
the measurements of the absolute value of eccentricity,
rather that its time derivative, as a possible probe of ULDM along the
lines of the Damour--Sch{\"a}fer test of SEP violation
\cite{Damour:1991rq}.   
Notice, however, that the standard  Damour--Sch{\"a}fer test assumes
the rate of periastron advance $\dot{\omega}_{PN}$, given by the 
post-Newtonian GR
corrections, to be sufficiently large so
that the projection of the galactic acceleration 
$\vec{g}$ onto the orbit can be considered
constant over time scales $2\pi/\dot{\omega}_{PN}$. The
corresponding assumption in our case would be that the 
r.h.s. of Eqs.~(\ref{kappaetaN2}) can be regarded as
approximately constant over the same time scales implying 
the condition $\dot{\gamma}_{1,2}(t)=\delta\omega_{1,2} \ll
\dot{\omega}_{PN}$. In other words, the system must be extremely close
to resonance, which is a highly improbable situation. Thus, the 
Damour--Sch\"afer-type analysis does not appear promising for the
study of resonant ULDM effects. Its relevance for probing non-resonant
effects will be discussed in Sec.~\ref{sec:nonresonant}.

Finally, let us discuss the resonant BB acceleration which appears in
the case of non-universal coupling and which we have disregarded so
far. Using Eq.~(\ref{linearBBcirc}) for circular orbits we obtain for
the $N=1$ resonance,
\be
\begin{split}
\langle\ddot{\vec R}\rangle\simeq
-2.7&\times 10^{-27}\,{\rm
  s}^{-1}
\left(\frac{\Delta\alpha}{10^{-23}\,{\rm GeV}^{-1}}\right)
\frac{4\mu}{M_T}
\left(\frac{M_T}{M_\odot}\right)^{1/3}\\
&\times
\left(\frac{P_b}{100\,{\rm d}}\right)^{2/3}
\sqrt{\frac{\rho_{\rm DM}}{0.3\frac{\rm GeV}{{\rm cm}^3}}}
\left(\frac{\delta\omega_1}{0.1\,{\rm y}^{-1}}\right)
\hat b\;.
\end{split}
\ee
Notice that we have normalized the frequency of secular modulations
$\delta\omega_1$ to the value $0.1\,{\rm y}^{-1}$ which ensures that
the acceleration stays constant over an observational campaign of
$T_{\rm obs}\sim 10\,{\rm y}$. To estimate the experimental
precision, with which $\ddot{R}$ can be measured, we use the
following reasoning. An acceleration along the line of sight
contributes through the time dependent Doppler effect into the
apparent change of the pulsar spin frequency $\dot\nu^{\rm
  meas}/\nu\sim \ddot R_{\parallel}$. The latter can me measured in a
given campaign with the uncertainty \cite{Teukolsky1976},
\be
\label{nuprecision}
\frac{\delta\dot\nu}{\nu}\simeq
1.8\times 10^{-23}\,{\rm s}^{-1}
\left(\frac{\delta t}{\mu{\rm s}}\right)
\left(\frac{\dot n_{\rm dat}}{{\rm d}^{-1}}\right)^{-1/2}
\left(\frac{T_{\rm obs}}{10\,{\rm y}}\right)^{-5/2}.
\ee
This provides us with a proxy for $\ddot R$
statistical uncertainty. We see that for the interesting values of 
$\Delta\alpha\sim 10^{-23}\,{\rm GeV}^{-1}$ it is
almost 4 orders of magnitude bigger than the expected effect. Similar
result holds for the case of quadratic coupling.

Moreover, using (\ref{nuprecision}) as an estimate of $\delta\ddot R$
is certainly overly optimistic. The ULDM contribution must be
disentangled from other sources of $\dot\nu^{\rm meas}$, such as the
intrinsic spin-down, the galactic acceleration and the Shklovskii
effect. In principle, this might be possible by observing the secular
modulations. However, for a fixed campaign duration $T_{\rm obs}$,
observing modulations leads to a further increase of the statistical
error (see below). We conclude that the BB acceleration has
less constraining power than the study of the orbital elements and can
be safely neglected.  

\subsubsection{Estimate of back-reaction on ULDM distribution}
In all previous calculations we have assumed that the ULDM
configuration is not modified by the interaction with the binary
system. The conditions for this approximation are formulated in
(\ref{BackRe}). Let is verify them. The average rate of the momentum
transfer from ULDM to the binary is given by the acceleration of its
barycenter, $\dot p_{\rm b}=M_T \langle \ddot R\rangle$, and the first
condition in (\ref{BackRe}) becomes,
\begin{align}
3 \times 10^{-14}\;&\bigg(\frac{m_\Phi}{10^{-21}\,{\rm eV}}\bigg)^2
\bigg(\frac{\langle\ddot R\rangle}{10^{-27}\,{\rm s}^{-1}}\bigg)\notag\\
&\times\bigg(\frac{M_T}{M_\odot}\bigg)
\bigg(\frac{\rho_{\rm DM}}{0.3\frac{\rm GeV}{{\rm cm}^3}}\bigg)^{-1}\ll
1\;.
\label{BackReP}
\end{align}  
On the other hand, the energy transfer affects primarily the energy of
the Keplerian motion. The latter is related to the binary semimajor axis by
the standard expression,
\be
{\cal E}_{\rm b}=-\frac{G M_T\mu}{2a}\;.
\ee  
Connecting the latter to the orbital period, we
obtain that the second condition in (\ref{BackRe}) amounts to
\begin{align}
&3 \times 10^{-15}\;\bigg(\frac{m_\Phi}{10^{-21}\,{\rm eV}}\bigg)^2
\bigg(\frac{\langle\dot P_{b}\rangle}{10^{-10}}\bigg)
\bigg(\frac{P_{b}}{100\,{\rm d}}\bigg)^{-5/3}
\notag\\
&\times\bigg(\frac{M_T}{M_\odot}\bigg)^{2/3}
\bigg(\frac{\mu}{M_\odot}\bigg)
\bigg(\frac{\rho_{\rm DM}}{0.3\frac{\rm GeV}{{\rm cm}^3}}\bigg)^{-1}
\bigg(\frac{V_0}{10^{-3}}\bigg)
\ll
1\;.
\label{BackReE}
\end{align}    
We see that both conditions are safely fulfilled in the
examples considered above.

%%%%%%%%%%%%%%%%%%%%%%%%%%%%%%%
\subsection{Detuning from resonance}\label{sec:detuning}
%%%%%%%%%%%%%%%%%%%%%%%%%%%%%%%

\begin{figure} [tbp]
        \center{\includegraphics[width=0.45\textwidth]{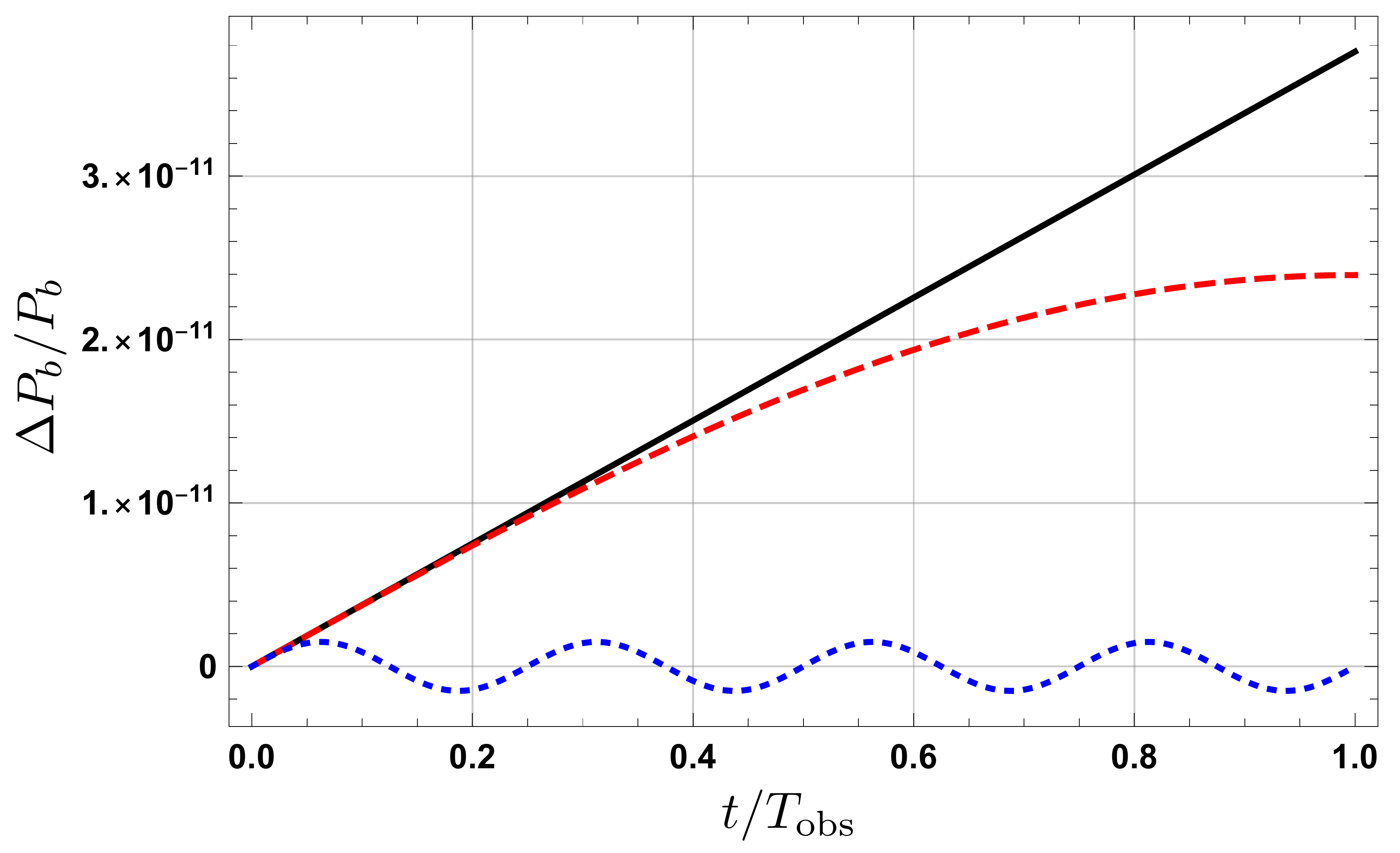}}
\caption{The relative change of the orbital period due to the ULDM
  coupling as a function of time. The three lines correspond to 
cases differing by the
  detuning from resonance: exact resonance, $\delta\omega_1=0$ (black
  solid); mild detuning, $\delta\omega_1=\pi/2T_{\rm obs}$ (red
  dashed); strong detuning, $\delta\omega_1=8\pi/T_{\rm obs}$ (blue
  dotted). We consider the $N=1$ resonance and take the 
initial resonant phase $\gamma_1(0)=\pi/2$. Other parameters are:
  $e=0.44$,
   $\alpha=10^{-23}\,{\rm GeV}^{-1}$, $\rho_{\rm DM}=0.3\,{\rm
   GeV/cm}^3$, $T_{\rm obs}=20\,{\rm y}$.} 
\label{dpboverpbfig}
\end{figure}

We now consider the case when the modulation period  
$T_{\rm mod}$ defined in Eq.~(\ref{Tmod}) 
is comparable or shorter than the observation time $T_{\rm obs}$,  
but still longer than the
binary period. 
More precisely, we are interested in the behavior of the system at
time scales $t$, such that 
\be
\label{thierarchy}
P_b\ll t\sim T_{\rm mod} < T_{\rm obs}\;.
\ee
Therefore, secular effects can still be isolated by averaging over
a time interval $\Delta t$ encompassing several orbital
periods around $t$, but shorter than $T_{\rm mod}$,
\be
\label{Dthierarchy}
P_b\ll \Delta t\ll T_{\rm mod}\;.
\ee 
The new feature is that averaged quantities obtained in this way will
now depend on time due to the detuning from resonance.
We want to understand how the timing sensitivity changes in this
situation. For simplicity, we focus on the variation of the orbital period
in the case of universal linearly coupling. 
Keeping in the expression for 
$\langle\dot P_b\rangle$ only the contribution of the resonant mode
$N$ and integrating this expression from $0$ to time $t$,  
we
obtain the orbital period as a function of time,   
\begin{align}
\frac{\Delta P_b(t)}{P_b} \simeq &\, 9.1\times 10^{-11}  
\left(\frac{\alpha}{10^{-23}\,{\rm GeV}^{-1}}\right)
\sqrt{\frac{\rho_{\rm DM}}{{0.3\, \frac{{\rm GeV}}{\rm cm^3}}}}\notag\\
&\times  
\left(\frac{0.1\,{\rm y}^{-1}}{\delta\omega_1}\right)
J_N(N e)\big[\cos\gamma_{1}(0)-\cos\gamma_{1}(t)\big] \,,
\label{periodderDCMod}
\end{align} 
where $\Delta P_b(t)\equiv P_b(t)-P_b(0)$. The behavior of $\Delta P_b$
during an observational campaign with $T_{\rm obs}\sim
20\,{\rm y}$ is illustrated in Fig.~\ref{dpboverpbfig} for three
choices of $\delta\omega_1$. We observe that the amplitude of the
modulations decreases inversely proportional to $\delta\omega_1$. 
Thus, although the modulations provide a
characteristic signature of ULDM, they become harder to detect for a
given precision of the $P_b$ measurements.

The extraction of the sensitivity to this signal from observations
requires the reanalysis of the data with a modification of  the data
analysis package (e.g. TEMPO2 \cite{Edwards:2006zg}). We leave this
for future work. Here we provide a rough estimate using the following
strategy. We chop the oscillating signal into intervals of almost resonant
behavior,    
bound the secular drift of the orbital parameters using data from each
interval, and combine the measurements from different intervals
assuming that their errors are uncorrelated. 
In other words, we consider the signal as a {\it
  triangle wave}, starting at an initial time $T_0$ and with $\dot
P_b$ constant over typical
intervals of duration $T_I=T_{\rm mod}/4$.  For simplicity, we assume
that $T_0$ coincides with the first point of the measurement.  
This can be achieved  by scanning the data for different $T_{0}$. 
The error on $\dot P_b$ obtained by fitting the data  
in an interval $T_I$ is read off from (\ref{errorPdot}),
  \begin{equation}
  (\delta{\dot{P}_b})_I\simeq (\delta\dot P_b)_{T_{\rm obs}} \left(\frac{T_{\rm obs}}{T_I}\right)^{5/2}=(\delta\dot P_b)_{T_{\rm obs}} \left(\frac{4T_{\rm obs}}{T_{\rm mod}}\right)^{5/2},
  \end{equation}
where $(\delta\dot P_b)_{T_{\rm obs}}$ is the would-be uncertainty for
the case when $\dot P_b$ stays constant for the whole observational
campaign. The total 
sensitivity is then estimated by dividing $(\delta{\dot{P}_b})_I$ by
the square-root of the number of intervals,   
    \begin{equation} \label{Amplit}
    \begin{split}
  (\delta{\dot{P}_b})_{\sum_I}  &\simeq  (\delta\dot P_b)_{T_{\rm obs}}
  \left(\frac{4T_{\rm obs}}{T_{\rm mod}}\right)^{5/2}
  \left(\frac{T_{\rm mod}}{4T_{\rm obs}}\right)^{1/2}\\
&= (\delta\dot P_b)_{T_{\rm obs}}\left(\frac{2\delta\omega_1 T_{\rm
      obs}}{\pi}\right)^{2}, 
 \end{split}
    \end{equation}  
where in the final step we used \eqref{Tmod}. This formula applies
when $\delta\omega_1\geq \pi/2 T_{\rm obs}$; otherwise, one should
just use $(\delta\dot P_b)_{T_{\rm obs}}$. We see that the
uncertainty in the determination of $\dot P_b$ grows as the system
deviates more from resonance, which degrades the
sensitivity to ULDM couplings. 

\begin{figure} [tbp]
 % \center{\includegraphics[width=0.45\textwidth]{fig2_new.pdf}}
      \center{\includegraphics[width=0.5\textwidth]{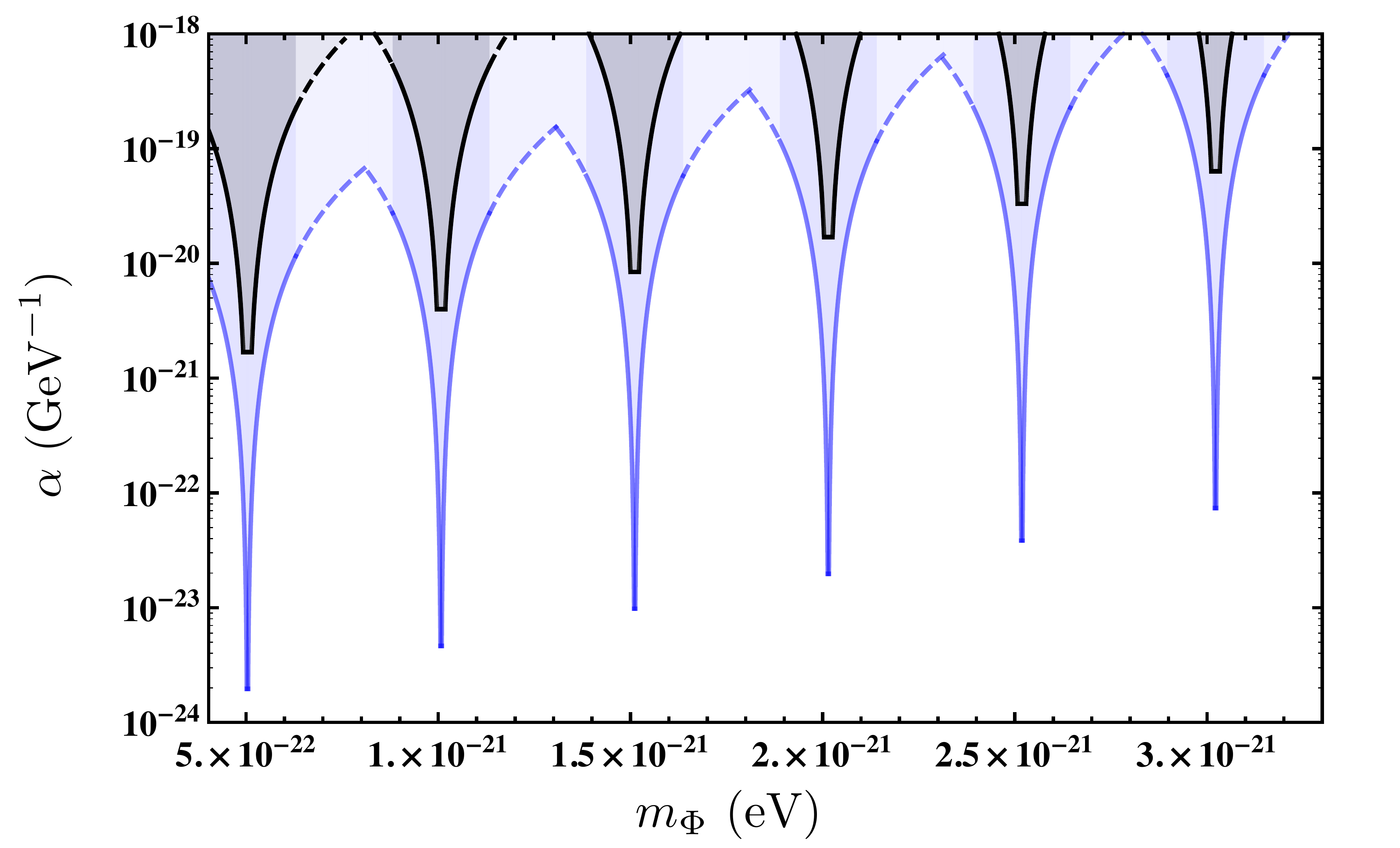}}
\caption{Constraint on the linear coupling 
$\alpha$ as a function of the ULDM mass from observations of a binary
system with parameters similar to J1903+0327 
  ($P_b\simeq 95\,{\rm d}$, $e\simeq 0.44$). Black lines show bounds
  obtained using Eq.~(\ref{Amplit}) with 
$(\delta\dot P_b)_{T_{\rm obs}}\lesssim  10^{-10}$ which is the
current precision from
3 years of timing data. The shaded regions above the curves
are excluded.
Blue lines show 
prospective constraints assuming future precision 
$(\delta\dot P_b)_{T_{\rm obs}}\lesssim  10^{-13}$ after 20 years of
observation time.  
Notice the logarithmic scale on
  the vertical axis. 
Dark shading above the solid lines shows the limits on $\alpha$ in
resonant bands 
where $|\delta\omega_1|<0.2\,\omega_b$.
The
leftmost  resonance corresponds to $N=1$ with $N$ increasing towards
the right up to $N=6$. 
Light shading above dashed lines represents the extrapolation of the
bounds outside the 
resonant regime and should be taken with a grain of salt. We assume
the dark matter density 
$\rho_{\rm DM}\simeq 0.3\,{\rm GeV/cm}^3$.} 
\label{Figresw}
\end{figure}

To illustrate this point, we consider the binary pulsar J1903+0327
which we already discussed in Sec.~\ref{sec:UCbounds}. Current
measurements over $T_{\rm obs}\sim 3$ years constrain the drift of its
orbital period at the level $(\delta \dot P)_{T_{\rm obs}}\lesssim
10^{-10}$. Using this value we obtained the estimates of the
constraints on the ULDM coupling at resonances listed in
Table~\ref{tab:univ1}. In Fig.~\ref{Figresw} we show with black lines how the
constraints degrade when the ULDM mass $m_\Phi$ is away from an exact
resonance. The curves are obtained by taking Eq.~(\ref{Amplit})
for the upper limit on the amplitude of $\langle\dot P_b\rangle$,
which is admittedly a rather crude estimate. Thus, the bounds in
Fig.~\ref{Figresw} should be interpreted as indicative. 
A robust exclusion requires a dedicated analysis of J1903+0327 timing
data, which is outside the scope of this paper.
We have restricted our analysis to the 
vicinity of resonances where a single harmonic dominates. Thus, the
bounds apply within resonant bands of width 
$\delta \omega_1 \ll \omega_b$; they are shown by solid lines. In
principle, one can extend the calculation beyond the resonant case by
summing up the contributions from multiple harmonics or solving the orbital motion numerically~\cite{Rozner:2019gba}. We do not
perform such calculation in this paper. Instead, to get a rough idea
of the expected result, we extrapolate the resonant expressions to  
$\delta \omega_1 \sim \omega_b$; this extrapolation is shown with dashed
lines.

For comparison, we also show how the constraints can be improved assuming
J1903+0327 or a system with similar parameters is timed for  
$T_{\rm obs}\sim
20\,{\rm y}$ and the drift of its orbital period is constrained at the
level 
$(\delta {\dot{P}_b})_{T_{\rm obs}}  \lesssim   10^{-13} $
(blue curves). The sensitivity to the parameter $\alpha$ precisely at
resonances (at the dips of the V-shaped curves in Fig.~\ref{Figresw})
scales with the observation time as $T_{\rm
  obs}^{-5/2}$. In contrast,  it scales as 
 $T_{\rm obs}^{-5/2}\cdot T_{\rm obs}^{2}=T_{\rm
   obs}^{-1/2}$, outside of the resonance, cf. Eq.~(\ref{Amplit}). Thus, the V-shaped
 exclusion regions become deeper and have sharper tips as the
 observation time increases.
  The plot suggests that 
future data even for a single system will be able to cover a 
substantial part of the ULDM parameter space.
This, together with the possibility of detecting many systems with
a variety of orbital parameters, motivates a more dedicated study.

\subsection{Non-resonant effects for quadratic coupling}
\label{sec:nonresonant}

As we discussed previously, in the case of the quadratic coupling the
ULDM force acting on the bodies has a non-oscillating
contribution proportional to the DM density gradients. The 
latter are set by the de Broglie
wavelength of ULDM in the halo. Thus, the vector
\be
\label{smalls}
\vec s\equiv \vec{S}/(m_\Phi V_0)\;,
\ee 
where $V_0$ is the DM virial velocity,
is expected to have order-one norm.
The non-resonant force
affects both the motion of the BB, Eq.~(\ref{quadraticBB}), and the
orbital elements, Eq.~(\ref{eqnr}). Let us start with the first
effect. Substituting the numerical values we obtain,
\begin{align}
\langle\ddot{\vec R}\rangle \simeq &\,3.5\times 10^{-23}\,{\rm s}^{-1}
\left(\frac{\beta_T}{10^{-32}\,{\rm GeV}^{-2}}\right)
\left(\frac{\rho_{\rm DM}}{0.3\,\frac{\rm GeV}{{\rm cm}^3}}\right)\notag\\
&\times\left(\frac{V_0}{10^{-3}}\right)
\left(\frac{m_\Phi}{10^{-21}\,{\rm eV}}\right)^{-1}
\vec s\;.
\label{ddotRnr}
\end{align}
According to the discussion at the end of
Sec.~\ref{sec:genericbounds}, detecting this small extra acceleration
on top of the standard contributions will be extremely
challenging. Still, it is worth stressing that this effect exists
not only for binaries, 
independently of their orbital frequency, but even for
solitary pulsars. Hence, one could think of increasing the sensitivity
by a joint analysis of many systems. We do not pursue this possibility
here.

The study of the
orbital parameters appears more promising. 
Here the leading effect is the drift of the
eccentricity given by Eq.~(\ref{enr}), which upon inserting the
numbers takes the form,
\begin{widetext}
\begin{align}
\label{dotenr}
\langle \dot e\rangle \simeq 3.4\times 10^{-19}\,{\rm s}^{-1}\,
\sqrt{1-e^2}
\left(\frac{\Delta\beta}{10^{-32}\,{\rm GeV}^{-2}}\right)
\left(\frac{M_T}{M_\odot}\right)^{-1/3}
\left(\frac{P_b}{100\,{\rm d}}\right)^{1/3}
\left(\frac{\rho_{\rm DM}}{0.3\,\frac{\rm GeV}{{\rm cm}^3}}\right)
\left(\frac{V_0}{10^{-3}}\right)
\left(\frac{m_\Phi}{10^{-21}\,{\rm eV}}\right)^{-1}
s_y\;.
\end{align}
\end{widetext}
For nearly circular orbits this expression should be substituted by
the equations for $\dot \kappa$ and $\dot \eta$, which differ only by
the replacement $s_y\mapsto \vec{s}\cdot(\hat z\times \hat n)$
and $s_y\mapsto - \vec{s}\cdot\hat n$, respectively.
Consider, for example, the system J1713+0747. From the constraints on
the drift of its eccentricity parameters we obtain the bound,
\be
\label{Dbetanr}
|\Delta \beta\, \vec{s}|<5.3\times 10^{-31}\,{\rm GeV}^{-2}
\left(\frac{m_\Phi}{10^{-21}\,{\rm eV}}\right)\;,
\ee
where we have assumed the typical values for the DM density and virial
velocity, $\rho_{\rm DM}=0.3\,{\rm GeV}/{\rm cm}^3$,
$V_0=10^{-3}$. Importantly, this bound applies in a wide range of ULDM
masses,
\be
\label{mrange}
m_{\Phi}\lesssim 10^{-18}\,{\rm eV}\,,
\ee  
where it is comparable or stronger than the other existing bounds 
(discussed e.g. in Ref.~\cite{Blas:2016ddr}).
The limitation comes from the standard assumption, made in inferring
the bounds on the eccentricity drift, that the time derivatives of the
orbital elements are constant over the observation time, which for
J1713+0747 is $T_{\rm obs}\simeq 20\,{\rm y}$. This requires the
constancy of the ULDM gradients on the same time scales. In other
words, the ULDM coherence time $t_{\rm coh}$ should be longer than
$T_{\rm obs}$. Recalling the expression (\ref{eq:tcoh}) one obtains
the condition~(\ref{mrange}).

In principle, one could extend the reach of pulsar constraints to
higher masses by generalizing the timing analysis to non-constant
parameter drift along the lines of Sec.~\ref{sec:detuning}. However,
there is one more restriction that prevents from significantly
extending the accessible parameter space, at least for the case of
J1713+0747.   
This is non-negligible back-reaction of the binary on the ULDM
configuration at high $m_\Phi$. Indeed, substituting the expression
for the barycenter acceleration (\ref{ddotRnr}) into the condition
(\ref{BackReP}) we find that the back-reaction is small 
provided that,
\be
\label{nobackreact}
\left(\frac{\beta_T}{10^{-23}\,{\rm GeV}^{-2}}\right)
\left(\frac{m_\Phi}{10^{-21}\,{\rm eV}}\right)
|\vec s|\ll
\left(\frac{M_\odot}{M_T}\right)
\left(\frac{10^{-3}}{V_0}\right).
\ee 
Assuming
$\Delta\beta\sim 0.1\beta_T$, we see that for the case of J1713+0747
neglecting the back-reaction in the derivation of the bound
(\ref{Dbetanr}) is indeed justified for DM masses satisfying the
condition (\ref{mrange}). However, for higher masses one should
include the back-reaction, which is beyond the scope of this
paper. 

It is worth comparing the constraint (\ref{Dbetanr}) to the bounds on
$\Delta \beta$ derived from the consideration of resonant effects 
in J1713+0747 listed in Table~\ref{tab:nonuniv}. We observe
that the resonant drift of the orbital period leads to a stronger bound
within the $N=1$ resonance band. However, for all other masses the
broadband constraint (\ref{Dbetanr}) dominates. In particular, the
$N=2$ resonant contribution to the eccentricity drift turns out to be
smaller (by about a factor 7) than the non-resonant one, as it is
apparent from comparison of Eqs.~(\ref{kappadotDbe}) and (\ref{dotenr}).

Observation of binary pulsars with nearly circular orbits yields yet
another type of constraints. The interaction with ULDM leads to the
growth of eccentricity, even if originally it was exactly zero. The
growth continues until the ULDM force can no longer be considered as
constant after the coherence time $t_{\rm coh}$, or until the orbit
significantly changes the orientation due to the post-Newtonian
periastron precession $\dot \omega_{PN}$ \cite{Damour:1991rq}. Thus,
the typical increment of the eccentricity induced by ULDM is,
\be
\label{Deltae}
\Delta e \simeq \langle\dot e \rangle \min\{t_{\rm coh},
{\dot\omega_{PN}}^{-1}\}\;,
\ee 
where $\langle\dot e \rangle$ is given by Eq.~(\ref{dotenr}), $t_{\rm
  coh}$ is given by Eq.~(\ref{eq:tcoh}), and the precession time 
reads in GR,
\be
\label{omegaPNdot}
\begin{split}
{\dot\omega_{PN}}^{-1}=&\,\frac{1-e^2}{3\omega_b}(GM_T\omega_b)^{-2/3}\\
=&\,6.2\times 10^5\,{\rm y}\,(1-e^2)
\left(\frac{M_T}{M_\odot}\right)^{-2/3}
\left(\frac{P_b}{100\,{\rm d}}\right)^{5/3}.
\end{split}
\ee
Imposing that $\Delta e$ be smaller than the measured value of
the eccentricity for a given system, yields constraints on the ULDM
coupling. 

As an example we consider the system J0024-7201X \cite{freire,pscat}
which has the orbital period $P_b\simeq 10.9\,{\rm d}$ and very low
measured eccentricity $e=(4.8\pm 1.5)\times 10^{-7}$
\cite{Ridolfi:2016fet}. It consists of a pulsar with mass $M_1\simeq
1.4\,M_\odot$ and a companion with $M_2\simeq 0.4\,M_\odot$; the
timing data were collected over a period $T_{\rm obs}\simeq 15\,{\rm
  y}$. We find that for this system ${\dot\omega_{PN}}^{-1}$ is
shorter than $t_{\rm coh}$ for $m_\Phi\lesssim 6.2\times
10^{-21}\,{\rm eV}$. Therefore, we
impose $\langle \dot e\rangle\,{\dot\omega_{PN}}^{-1}<e$ for ULDM masses
below this value and
$\langle \dot e\rangle\, t_{\rm coh}<e$ for higher masses. This gives
the bounds,
\be
\label{DSbounds}
|\Delta\beta\,\vec s|<\begin{cases}
1.4\times 10^{-31}\,{\rm GeV}^{-2}
\left(\frac{m_\Phi}{10^{-21}\,{\rm eV}}\right)\\
\quad\text{for}~m_\Phi\lesssim 6.2\times
10^{-21}\,{\rm eV}\;;\\
\\
2.3\times 10^{-30}\,{\rm GeV}^{-2}
\left(\frac{m_\Phi}{10^{-20}\,{\rm eV}}\right)^2\\
\quad\text{for}~6.2\times
10^{-21}\,{\rm eV}\lesssim m_\Phi\lesssim 5\times 10^{-19}\,{\rm eV}\,,
\end{cases}
\ee
where we have assumed our usual reference values for $\rho_{\rm DM}$
and $V_0$. The accessible ULDM mass range is limited from above by the
requirement of negligible back-reaction. One observes that 
for $m_\Phi\lesssim 2\times 10^{-20}\,{\rm eV}$, the
bound
(\ref{DSbounds}) is somewhat stronger than the one obtained from a
direct timing measurement, Eq.~(\ref{Dbetanr}). It is, however, less
robust (see e.g. \cite{Freire:2012nb}). Indeed, one cannot exclude a
 cancellation between the change in the eccentricity induced by
ULDM and its initial value, so that the net eccentricity happens to be
small during the observational epoch. This caveat can be removed by
considering a population of low-eccentricity binary pulsars and
performing a statistical analysis along the lines of 
\cite{wex_2000,Stairs:2005hu,Gonzalez:2011kt}. We leave the
exploration of this promising direction for future.

 %%%%%%%%%%%%%%%%%%%%%%%%%%%%%%%
\section{Conclusions and outlook}\label{sec:concl}
%%%%%%%%%%%%%%%%%%%%%%%%%%%%%%%

In this paper we have made a detailed study of the influence of
ultralight dark matter (ULDM) candidates on the evolution of binary
systems and have identified potentially observable signatures in
high-precision timing measurements of binary pulsars.
We have focused on secular effects that accumulate over many orbital
periods. We have shown that such effects appear due to the time
evolution of the DM field $\Phi$ when the frequency of the $\Phi$
oscillations is in resonance with the orbital motion. Another class on
(non-resonant) secular effects arises when $\Phi$ couples
quadratically to the masses of the binary system members. 

The resonant effects appear both due to the oscillations of
the curvature of the spacetime generated by the DM stress energy
tensor, and from the possible direct coupling of
the oscillating DM field to the SM fields. The first effect is 
generic for any ULDM candidate as it is based purely on
the gravitational interaction of DM with the stars.
However, from our estimate of the resulting drift
of the orbital elements of binary pulsars (Eqs.~\eqref{dotwgrav}) 
we have concluded  that it is too
small to be constrained by current data. 
The prospects for reaching the required sensitivity 
in the near future are not
very optimistic. Still, Nature has been quite indulgent so far in
providing convenient binary pulsar systems to test fundamental
laws. Hopefully, the new generation of radio surveys, 
with SKA as flagship
\cite{Kramer:2015bea}, will also benefit from  this generosity
regarding DM. 

We have performed a comprehensive analysis of the scenario where ULDM
couples directly to SM through the dilatonic portal
(linear or quadratic). We have argued that even if the interaction is
universal (WEP preserving) at the fundamental level, the effective
couplings to the binary system members will be non-universal due to violation
of SEP. We have derived the complete set of equations for the orbital
parameters in this scenario, both for the resonant (Eqs.~\eqref{LPE2}) and
for non-resonant (Eqs.~\ref{eqnr}) cases. We have identified the two
most prominent effects. The first one is the time evolution of the
orbital period $P_b$ in the resonant case. In contrast to the previous studies
that considered only universally coupled ULDM, we have found that
this effect does not vanish for circular orbit due to the
non-universal coupling. The second effect is the change of the orbital
eccentricity $e$ that appears in the resonant, as well as non resonant
cases. It is similar to the polarization of orbits obtained in 
Refs.~\cite{LopezNacir:2018epg,Armaleo:2019gil} for the scenarios with
vector and tensor DM. We have discussed its relationship to the   
Damour--Sch{\"a}fer effect in SEP violating theories
\cite{Damour:1991rq}.

We have estimated the sensitivity of the timing measurements to the
above effects. For this purpose, we used     
 the phenomenological  pulsar timing model  developed in
 \cite{Teukolsky1976,1986AIHS...44..263D,Edwards:2006zg}. In this
 model, the orbital parameters are allowed to change during the
 campaign at a constant rate; the time derivatives of 
different parameters are treated as independent variables.
 We have found that inside the resonant
 bands the strongest constraints come from the drift of the orbital period,
 provided the systematic error in the determination of $\dot P_b$ is
 kept under control. One of the main foregrounds here is an apparent change of
 $P_b$ due to the acceleration of the binary with respect to the Solar
 System Barycenter~\cite{1991ApJ...366..501D}. A better determination
 of the galactic acceleration in the vicinity of the binary (e.g. by
 tracking of near-by stars) would help to reduce this systematics. 
 We have estimated the dependence of the
 constraints on the deviation of the ULDM mass $m_\Phi$ from the
 resonance value (see Fig.~\ref{Figresw}). Despite a significant loss
 of sensitivity outside the resonant bands, we proposed that the
 combination of timing data from multiple binaries with different
 orbital periods may efficiently constrain wide regions in the
 parameter space of the ULDM candidates. 

Eccentricity measurements are more robust against foreground
contamination~\cite{Freire:2012nb}. They
provide complementary information that
potentially can help to disentangle the ULDM effects from possible
systematics. Moreover, they strongly constrain the non-resonant
eccentricity drift present in the case of quadratically coupled $\Phi$.
These non-resonant constraints have an important advantage:
the timing data even for a single binary pulsar allow one to put
bounds on the quadratic coupling in a broad range of ULDM masses,
$m_\Phi\lesssim 10^{-18}\,{\rm eV}$. The accessible mass interval is
restricted only by the requirement that the back-reaction of the
binary system on the field $\Phi$ should be negligible. All in all, we
have concluded that pulsar timing can probe an interesting portion of
the ULDM parameter space, yet unconstrained by other observations.   

We have also pointed out that the direct coupling leads to an additional
secular acceleration of the binary barycenter which, however, appears
to be unobservable.

There are several future directions to explore. An immediate next step
would be the incorporation of the effects derived in this work into the
timing package TEMPO2~\cite{Edwards:2006zg}. This will allow us 
to explore how the sensitivity can be improved by taking into account
the correlated drift of all orbital elements and put rigorous
constraints outside the resonant bands. Another promising direction is
the statistical analysis of a population of low-eccentricity binaries
in order to constrain the non-resonant effects of quadratically
coupled ULDM along the lines of the Damour--Sch\"afer test 
\cite{Damour:1991rq,wex_2000,Stairs:2005hu,Gonzalez:2011kt}.
 It would be also interesting to explore if ULDM may help to explain the gaps in the $P_b-e$ plane 
in the population of  binary pulsars \cite{Barr:2016vxv,Hui:2018mkc}.

We have focused on binary pulsars with non-relativistic orbits.  The
reason is not only simplicity, but the fact that 
most of the effects we discussed become weaker 
as the orbital period decreases.  
However, given that many fast binaries present excellent timing
opportunities, it will be worth extending  
our calculations to include
post-Newtonian corrections.    
Furthermore, the interaction of a binary system with $\Phi$ 
will
lead to the emission of dipolar
radiation which is tightly constrained \cite{Wex:2014nva}. 
It is important to understand the implications of this process in the
present context.
Finally, our analysis has been formulated in terms of effective ULDM
couplings to the masses of the binary system members. It will be interesting
to relate these effective couplings to the fundamental parameters in
specific models, in particular when non-perturbative scalarization
effects may show up.
  
Before closing, let us notice that all our estimates assume the value
$\rho_{\rm DM}=0.3\,{\rm GeV}/{\rm cm}^3$ corresponding to the
expectations in the neighborhood of the Solar System.
An enhancement of this number would generate stronger
constraints. In particular, the discovery of binary pulsars in the
center of the galaxy would boost all our bounds and estimates.

\paragraph*{Acknowledgments}  We thank  Vitor Cardoso,  Paulo Freire,
Gian Giudice, 
David Pirtskhalava, Valery Rubakov and Federico Urban for
discussions. D.L.N. thanks CERN Theory Department and King's College
London for hospitality during the work on this paper. The work of DNL has been supported by CONICET, ANPCyT y UBA.
The work of S.S. is 
supported by the Swiss National Science Foundation, the Tomalla
Foundation and the RFBR grant
17-02-00651.  

\appendix

%%%%%%%%%%%%%%%%%%%%%%%%%%%%%%%
\section{Keplerian and osculating orbits}\label{osculating}
%%%%%%%%%%%%%%%%%%%%%%%%%%%%%%%

In this appendix we summarize some properties of the Keplerian motion
and its perturbations. A detailed discussion can be found, e.g., in
Ref.~\cite{poisson2014gravity}.  

We consider an elliptic orbit oriented as in Fig.~\ref{orbits}. The
unperturbed solution to the Kepler problem is described in terms of
the relative distance vector  
$\vec{r}(t)=r(t)\hat{r}(t)$, where
\bseq
\begin{align}
\label{rE}
&r(t)=a (1-e\cos E(t))\,,\\
&\hat{r}(t)= [\cos{\Omega} \cos(\omega\!+\!\theta(t)) 
-\cos{\iota} \sin{\Omega} \sin(\omega\!+\!\theta(t))]\hat{X}\nonumber\\
&\qquad+[\sin{\Omega} \cos(\omega\!+\!\theta(t)) -\cos{\iota} \cos{\Omega} 
\sin(\omega\!+\!\theta(t))]\hat{Y}\nonumber\\
&\qquad+\sin{\iota} \sin(\omega+\theta(t))\hat{Z}\,,\\
\label{thetaE}
&\tan\frac{\theta(t)}{2}=\sqrt{\frac{1+e}{1-e}}\tan\frac{E(t)}{2}\,.
\end{align} 
\eseq
The {\it eccentric anomaly} $E$ is implicitly defined as a function of
time by the equation,
\be
\label{eccan}
E(t)-e\sin E(t)=\omega_b (t-t_0)\;,
\ee
where $\omega_b$ is the orbital frequency given by
Eq.~(\ref{orbfreq}). The eccentric anomaly changes by $2\pi$ over one
orbital period. The motion is completely characterized by six constant
parameters ({\it orbital elements}): semimajor axis $a$, eccentricity
$e$, the longitude of the ascending node $\Omega$,
the inclination angle $\iota$, the argument of the
pericenter $\omega$ (not to be confused with the orbital frequency!)
and the time of the pericenter passage 
$t_0$.    

The system possesses two vector integrals of motion: the angular
momentum,
\be
\vec{L}=\vec{r}\times\dot{\vec{r}}\equiv L\hat z\;,
\ee 
and the Runge--Lenz vector,
\be
\vec{A}=\frac{\dot{\vec{r}}\times\vec{L}}{GM_T}-\hat r\equiv A\hat x\;.
\ee
The orbital elements are related to them in the following way,
\bseq
\label{orbelLA*}
\begin{align}
&a(1-e^2)=\frac{L^2}{GM_T}\,,\label{orbelLA1}\\
&e=A\,,\\
&\cos\iota=\hat z\cdot\hat Z=\frac{\vec L\cdot\hat Z}{L}\,,\\
&\sin\iota\cos\Omega=\hat z\cdot\hat X=\frac{\vec L\cdot\hat X}{L}\,,\\
&\sin\iota\cos\omega=\hat x\cdot\hat Z=\frac{\vec A\cdot\hat Z}{A}\,.
\end{align}
\eseq

In the main text we need the Fourier decompositions of the Keplerian
functions $r(t)$, $\theta(t)$ and several related quantities. These
are conveniently 
derived by rewriting the relevant Fourier integrals as integrals over
the eccentric anomaly $E$ and using Eqs.~(\ref{rE}), (\ref{thetaE}),
(\ref{eccan}). In this way one obtains \cite{watson1995treatise},
\begin{subequations} 
\label{KeplFour}
\begin{align}
& \frac{r}{a}= 1+\frac{e^2}{2}-2e\sum_{n=1}^{\infty}
\frac{J_n'(ne)}{n}  \,\cos n\omega_b \tilde t\,,\\ 
& \frac{r^2}{a^2}= 1+{ \frac{3}{2}}e^2-4\sum_{n=1}^{\infty}
\frac{J_n(ne)}{n^2}  \,\cos n\omega_b \tilde t\,,\\ 
&\cos \theta=-e+\frac{2(1-e^2)}{e} \sum_{n=1}^{\infty} J_n(ne) \,
\cos n\omega_b \tilde t\,,\\
&\sin \theta= 2\sqrt{1-e^2}\sum_{n=1}^{\infty}  J_n'(ne) \,\sin
n\omega_b \tilde t\,,\\
&\frac{r}{a}\cos \theta=-\frac{3e}{2}+2
\sum_{n=1}^{\infty} \frac{J_n'(ne)}{n} \,\cos n\omega_b \tilde t\,,\\ 
&\frac{r}{a}\sin \theta=\frac{2\sqrt{1-e^2}}{e}
\sum_{n=1}^{\infty} \frac{J_n(ne)}{n} \,\sin n\omega_b \tilde t\,,\\ 
&\frac{a^2}{r^2}\cos \theta= \sum_{n=1}^{\infty}2n J_n'(ne) \,\cos
n\omega_b \tilde t\,,\\ 
&\frac{a^2}{r^2}\sin \theta=
\frac{\sqrt{1-e^2}}{e}\sum_{n=1}^{\infty}2n  J_n(ne) \,\sin n\omega_b
\tilde t\,, 
\end{align}
\end{subequations} where $\tilde t=t-t_0$, $J_n(z)$ is the Bessel
function and $J_n'(z)$ is its derivative with respect to $z$.  

We now turn to the perturbed Keplerian equations,
\begin{equation}\label{PE}
\ddot{\vec{r}}=-\frac{GM_T}{r^3}\vec{r}+\vec{F}\,.
\end{equation}  
The method of osculating orbits represents an application of the
method of variation of constants to these equations. One starts from
the solution of the unperturbed problem,
\be
\vec{r}_{(0)}(t;\mu_i)~,~~~~
\vec{v}_{(0)}(t;\mu_i)\equiv \dot{\vec{r}}_{(0)}\,,
\ee
understood as a function of time and orbital elements 
$\{\mu_1,\ldots,\mu_6\}=\{a,e,\Omega,\iota,\omega,t_0\}$.
To find the solution of (\ref{PE}) one allows the orbital elements to
depend on time. Namely, one uses the following Ansatz,
\be
\label{oscul}
\vec{r}(t)=\vec{r}_{(0)}(t;\mu_i(t))~,~~~~
\vec{v}(t)=\vec{v}_{(0)}(t;\mu_i(t))\;.
\ee
The compatibility of these two expressions requires
\bseq
\label{osculeq}
\be
\label{osculeq1}
\sum_i \frac{\partial\vec{r}_{(0)}}{\partial\mu_i}\dot\mu_i=0\,,
\ee
which gives 3 equations for the time derivatives of the 6 variables
$\mu_i$. The remaining equations are obtained by substituting
(\ref{oscul}) into (\ref{PE}), which yields 
\be
\label{osculeq2}
\sum_i \frac{\partial\vec{v}_{(0)}}{\partial\mu_i}\dot\mu_i=\vec{F}\,.
\ee
\eseq
Resolving the linear equations (\ref{osculeq}) with respect to
$\dot\mu_i$ one obtains a system of first-order differential equations
for the evolution of the orbital elements.

In practice, a straightforward application of the above procedure
leads to very lengthy calculations. It is possible to bypass this
difficulty by studying the effect of the perturbing force on the
angular momentum and Runge--Lenz vector. An elementary calculation
gives, 
\bseq
\label{LAdot}
\begin{align}
&\dot{\vec{L}}=rF_\theta \hat z-rF_z\hat\theta\;,\\
&\dot{\vec{A}}=\frac{1}{GM_T}[2LF_\theta \hat{r}
-(LF_r+r\dot rF_\theta)\hat\theta
-r\dot{r} F_z\hat{z}]\;,
\end{align}
\eseq
where we have decomposed the perturbing force as in
Eq.~(\ref{Fdecomp}). Substituting these expressions into the time
derivatives of Eqs.~(\ref{orbelLA*}) we obtain
Eqs.~(\ref{Lagrange1})--(\ref{Lagrange5}) from the main text. The last
Eq.~(\ref{Lagrange6}) can then be obtained, for example, by requiring
that $\dot r$ is given by the Keplerian expression for the radial
velocity.

%%%%%%%%%%%%%%%%%%%%%%%%%%%%%%%
\section{Osculating orbit equations for binary coupled to ULDM}
\label{AppendixLPE2}
%%%%%%%%%%%%%%%%%%%%%%%%%%%%%%%
Substituting the expressions (\ref{forceDC}) for the force into
(\ref{Lagrange}) and using the equations of the unperturbed orbit to
simplify the expressions, we find in the case of the linear coupling,  
\begin{widetext}
\begin{subequations}
\begin{align}\label{eqa}
\frac{\dot{a}}{a}=& -\frac{2\alpha_T\Phi e\omega_b}{\sqrt{1-e^2}}
\frac{a^2}{r^2}\sin\theta
-\frac{2\alpha_{\mu}\dot{\Phi}(1+e^2+2e\cos\theta)}{(1-e^2)}
-\frac{2\Delta\alpha}{a\omega_b\sqrt{1-e^2}}[(\dot{\Phi} 
V_{x}+\Phi S_{x})\sin\theta- (\dot{\Phi} V_{y}+\Phi
S_{y})(e+\cos\theta)]\,,\\ 
\dot{e}=&-\alpha_{T}\Phi\sqrt{1-e^2}\omega_b\frac{a^2}{r^2}\sin\theta
-2\alpha_{\mu}\dot{\Phi}(\cos\theta+e) 
+\frac{\Delta\alpha\sqrt{1-e^2}}{a e\omega_b}
\left[(\dot{\Phi} V_{x}+\Phi S_{x})
\left(\frac{r}{a}\sin\theta-\sin\theta\right) \right.\nonumber\\
 &\qquad\qquad\qquad\qquad\qquad\qquad\qquad\qquad\qquad\qquad
\qquad\qquad\qquad
+\left. (\dot{\Phi} V_{y}+\Phi S_{y})
\left(e+\cos\theta-\frac{r}{a}\cos\theta\right)\right]\,,\label{eqe}\\
 \dot{\Omega}=&\Delta\alpha (\dot{\Phi} V_{z}+\Phi S_{z})
\frac{1}{a\omega_b\sqrt{1-e^2}\sin\iota} 
\Big(\frac{r}{a}\cos\theta\sin\omega+\frac{r}{a}\sin\theta\cos\omega\Big)\,,\\
\dot{\iota}=& \Delta\alpha(\dot{\Phi} V_{z}+\Phi S_{z})
\frac{1}{a\omega_b\sqrt{1-e^2}}
\Big(\frac{r}{a}\cos\theta\cos\omega-\frac{r}{a}\sin\theta\sin\omega\Big)\,,\\
\dot{\varpi}=&\frac{\alpha_{T}\Phi\omega_b\sqrt{1-e^2}}{e} 
\frac{a^2}{r^2}\cos\theta
-\frac{2\alpha_{\mu}\dot{\Phi}}{e} \sin\theta
 +\frac{\Delta\alpha\sqrt{1-e^2}}{ae\omega_b}\left[
- (\dot{\Phi} V_{x}+\Phi S_{x})
  \left(1+\frac{r}{a(1-e^2)}+\frac{r\cos\theta}{a
      e(1-e^2)}-\frac{\cos\theta}{e}\right)\right.
\nonumber\\
&\qquad\qquad\qquad\qquad\qquad\qquad\qquad\qquad\qquad\qquad
\left.+(\dot{\Phi}
   V_{y}+\Phi S_{y})  \left(\frac{\sin\theta}{e}-\frac{r\sin\theta}{a
       e(1-e^2)}\right)\right]
+2\sin^2\left(\frac{\iota}{2}\right)\dot{\Omega}\,,\\
\dot{\epsilon_1}=&\frac{2\alpha_{T}\Phi\omega_b}{(1-e^2)}(1+e\cos\theta)
+\frac{2\alpha_{\mu} \dot{\Phi}e}{\sqrt{1-e^2}}\frac{r}{a}\sin\theta
-\frac{2\Delta\alpha}{a^2\omega_b} 
\left[(\dot{\Phi}V_{x}+\Phi S_x)\frac{r}{a}\cos\theta 
+(\dot\Phi V_{y}+\Phi S_{y})\frac{r}{a}\sin\theta\right]\nonumber\\
 &+[1-\sqrt{1-e^2}]\dot{\varpi}+2\sqrt{1-e^2}  \sin^2\left(\frac{\iota}{2}\right)\dot{\Omega}\,. 
\end{align}
\end{subequations}
The equations for quadratic coupling are obtained by replacing 
$\{\alpha_T,\alpha_\mu,\Delta\alpha\}\mapsto\{\beta_T,\beta_\mu,\Delta\beta\}$, 
$\Phi\mapsto\Phi^2/2$, $\vec{S}\mapsto 2\vec{S}$.

\end{widetext}

\bibliography{biblio}    

\begin{thebibliography}{92}%
\makeatletter
\providecommand \@ifxundefined [1]{%
 \@ifx{#1\undefined}
}%
\providecommand \@ifnum [1]{%
 \ifnum #1\expandafter \@firstoftwo
 \else \expandafter \@secondoftwo
 \fi
}%
\providecommand \@ifx [1]{%
 \ifx #1\expandafter \@firstoftwo
 \else \expandafter \@secondoftwo
 \fi
}%
\providecommand \natexlab [1]{#1}%
\providecommand \enquote  [1]{``#1''}%
\providecommand \bibnamefont  [1]{#1}%
\providecommand \bibfnamefont [1]{#1}%
\providecommand \citenamefont [1]{#1}%
\providecommand \href@noop [0]{\@secondoftwo}%
\providecommand \href [0]{\begingroup \@sanitize@url \@href}%
\providecommand \@href[1]{\@@startlink{#1}\@@href}%
\providecommand \@@href[1]{\endgroup#1\@@endlink}%
\providecommand \@sanitize@url [0]{\catcode `\\12\catcode `\$12\catcode
  `\&12\catcode `\#12\catcode `\^12\catcode `\_12\catcode `\%12\relax}%
\providecommand \@@startlink[1]{}%
\providecommand \@@endlink[0]{}%
\providecommand \url  [0]{\begingroup\@sanitize@url \@url }%
\providecommand \@url [1]{\endgroup\@href {#1}{\urlprefix }}%
\providecommand \urlprefix  [0]{URL }%
\providecommand \Eprint [0]{\href }%
\providecommand \doibase [0]{http://dx.doi.org/}%
\providecommand \selectlanguage [0]{\@gobble}%
\providecommand \bibinfo  [0]{\@secondoftwo}%
\providecommand \bibfield  [0]{\@secondoftwo}%
\providecommand \translation [1]{[#1]}%
\providecommand \BibitemOpen [0]{}%
\providecommand \bibitemStop [0]{}%
\providecommand \bibitemNoStop [0]{.\EOS\space}%
\providecommand \EOS [0]{\spacefactor3000\relax}%
\providecommand \BibitemShut  [1]{\csname bibitem#1\endcsname}%
\let\auto@bib@innerbib\@empty
%</preamble>
\bibitem [{\citenamefont {Marsh}(2016)}]{Marsh:2015xka}%
  \BibitemOpen
  \bibfield  {author} {\bibinfo {author} {\bibfnamefont {David. J.~E.}\
  \bibnamefont {Marsh}},\ }\bibfield  {title} {\enquote {\bibinfo {title}
  {{Axion Cosmology}},}\ }\href {\doibase 10.1016/j.physrep.2016.06.005}
  {\bibfield  {journal} {\bibinfo  {journal} {Phys. Rept.}\ }\textbf {\bibinfo
  {volume} {643}},\ \bibinfo {pages} {1--79} (\bibinfo {year} {2016})},\
  \Eprint {http://arxiv.org/abs/1510.07633} {arXiv:1510.07633 [astro-ph.CO]}
  \BibitemShut {NoStop}%
%%CITATION = ARXIV:1510.07633;%%
\bibitem [{\citenamefont {Damour}\ and\ \citenamefont
  {Donoghue}(2010)}]{Damour:2010rp}%
  \BibitemOpen
  \bibfield  {author} {\bibinfo {author} {\bibfnamefont {Thibault}\
  \bibnamefont {Damour}}\ and\ \bibinfo {author} {\bibfnamefont {John~F.}\
  \bibnamefont {Donoghue}},\ }\bibfield  {title} {\enquote {\bibinfo {title}
  {{Equivalence Principle Violations and Couplings of a Light Dilaton}},}\
  }\href {\doibase 10.1103/PhysRevD.82.084033} {\bibfield  {journal} {\bibinfo
  {journal} {Phys. Rev.}\ }\textbf {\bibinfo {volume} {D82}},\ \bibinfo {pages}
  {084033} (\bibinfo {year} {2010})},\ \Eprint {http://arxiv.org/abs/1007.2792}
  {arXiv:1007.2792 [gr-qc]} \BibitemShut {NoStop}%
%%CITATION = ARXIV:1007.2792;%%
\bibitem [{\citenamefont {Arvanitaki}\ \emph
  {et~al.}(2015{\natexlab{a}})\citenamefont {Arvanitaki}, \citenamefont
  {Huang},\ and\ \citenamefont {Van~Tilburg}}]{Arvanitaki:2014faa}%
  \BibitemOpen
  \bibfield  {author} {\bibinfo {author} {\bibfnamefont {Asimina}\ \bibnamefont
  {Arvanitaki}}, \bibinfo {author} {\bibfnamefont {Junwu}\ \bibnamefont
  {Huang}}, \ and\ \bibinfo {author} {\bibfnamefont {Ken}\ \bibnamefont
  {Van~Tilburg}},\ }\bibfield  {title} {\enquote {\bibinfo {title} {{Searching
  for dilaton dark matter with atomic clocks}},}\ }\href {\doibase
  10.1103/PhysRevD.91.015015} {\bibfield  {journal} {\bibinfo  {journal} {Phys.
  Rev.}\ }\textbf {\bibinfo {volume} {D91}},\ \bibinfo {pages} {015015}
  (\bibinfo {year} {2015}{\natexlab{a}})},\ \Eprint
  {http://arxiv.org/abs/1405.2925} {arXiv:1405.2925 [hep-ph]} \BibitemShut
  {NoStop}%
%%CITATION = ARXIV:1405.2925;%%
\bibitem [{\citenamefont {Hu}\ \emph {et~al.}(2000)\citenamefont {Hu},
  \citenamefont {Barkana},\ and\ \citenamefont {Gruzinov}}]{Hu:2000ke}%
  \BibitemOpen
  \bibfield  {author} {\bibinfo {author} {\bibfnamefont {Wayne}\ \bibnamefont
  {Hu}}, \bibinfo {author} {\bibfnamefont {Rennan}\ \bibnamefont {Barkana}}, \
  and\ \bibinfo {author} {\bibfnamefont {Andrei}\ \bibnamefont {Gruzinov}},\
  }\bibfield  {title} {\enquote {\bibinfo {title} {{Cold and fuzzy dark
  matter}},}\ }\href {\doibase 10.1103/PhysRevLett.85.1158} {\bibfield
  {journal} {\bibinfo  {journal} {Phys.Rev.Lett.}\ }\textbf {\bibinfo {volume}
  {85}},\ \bibinfo {pages} {1158--1161} (\bibinfo {year} {2000})},\ \Eprint
  {http://arxiv.org/abs/astro-ph/0003365} {arXiv:astro-ph/0003365 [astro-ph]}
  \BibitemShut {NoStop}%
%%CITATION = ASTRO-PH/0003365;%%
\bibitem [{\citenamefont {Amendola}\ and\ \citenamefont
  {Barbieri}(2006)}]{Amendola:2005ad}%
  \BibitemOpen
  \bibfield  {author} {\bibinfo {author} {\bibfnamefont {Luca}\ \bibnamefont
  {Amendola}}\ and\ \bibinfo {author} {\bibfnamefont {Riccardo}\ \bibnamefont
  {Barbieri}},\ }\bibfield  {title} {\enquote {\bibinfo {title} {{Dark matter
  from an ultra-light pseudo-Goldsone-boson}},}\ }\href {\doibase
  10.1016/j.physletb.2006.08.069} {\bibfield  {journal} {\bibinfo  {journal}
  {Phys. Lett.}\ }\textbf {\bibinfo {volume} {B642}},\ \bibinfo {pages}
  {192--196} (\bibinfo {year} {2006})},\ \Eprint
  {http://arxiv.org/abs/hep-ph/0509257} {arXiv:hep-ph/0509257 [hep-ph]}
  \BibitemShut {NoStop}%
%%CITATION = HEP-PH/0509257;%%
\bibitem [{\citenamefont {Hui}\ \emph {et~al.}(2017)\citenamefont {Hui},
  \citenamefont {Ostriker}, \citenamefont {Tremaine},\ and\ \citenamefont
  {Witten}}]{Hui:2016ltb}%
  \BibitemOpen
  \bibfield  {author} {\bibinfo {author} {\bibfnamefont {Lam}\ \bibnamefont
  {Hui}}, \bibinfo {author} {\bibfnamefont {Jeremiah~P.}\ \bibnamefont
  {Ostriker}}, \bibinfo {author} {\bibfnamefont {Scott}\ \bibnamefont
  {Tremaine}}, \ and\ \bibinfo {author} {\bibfnamefont {Edward}\ \bibnamefont
  {Witten}},\ }\bibfield  {title} {\enquote {\bibinfo {title} {{Ultralight
  scalars as cosmological dark matter}},}\ }\href {\doibase
  10.1103/PhysRevD.95.043541} {\bibfield  {journal} {\bibinfo  {journal} {Phys.
  Rev.}\ }\textbf {\bibinfo {volume} {D95}},\ \bibinfo {pages} {043541}
  (\bibinfo {year} {2017})},\ \Eprint {http://arxiv.org/abs/1610.08297}
  {arXiv:1610.08297 [astro-ph.CO]} \BibitemShut {NoStop}%
%%CITATION = ARXIV:1610.08297;%%
\bibitem [{\citenamefont {Hlo\v{z}ek}\ \emph {et~al.}(2015)\citenamefont
  {Hlo\v{z}ek}, \citenamefont {Grin}, \citenamefont {Marsh},\ and\
  \citenamefont {Ferreira}}]{Hlozek:2014lca}%
  \BibitemOpen
  \bibfield  {author} {\bibinfo {author} {\bibfnamefont {Ren\'e}\ \bibnamefont
  {Hlo\v{z}ek}}, \bibinfo {author} {\bibfnamefont {Daniel}\ \bibnamefont
  {Grin}}, \bibinfo {author} {\bibfnamefont {David J.~E.}\ \bibnamefont
  {Marsh}}, \ and\ \bibinfo {author} {\bibfnamefont {Pedro~G.}\ \bibnamefont
  {Ferreira}},\ }\bibfield  {title} {\enquote {\bibinfo {title} {{A search for
  ultralight axions using precision cosmological data}},}\ }\href {\doibase
  10.1103/PhysRevD.91.103512} {\bibfield  {journal} {\bibinfo  {journal} {Phys.
  Rev.}\ }\textbf {\bibinfo {volume} {D91}},\ \bibinfo {pages} {103512}
  (\bibinfo {year} {2015})},\ \Eprint {http://arxiv.org/abs/1410.2896}
  {arXiv:1410.2896 [astro-ph.CO]} \BibitemShut {NoStop}%
%%CITATION = ARXIV:1410.2896;%%
\bibitem [{\citenamefont {Hlo\v{z}ek}\ \emph {et~al.}(2017)\citenamefont
  {Hlo\v{z}ek}, \citenamefont {Marsh}, \citenamefont {Grin}, \citenamefont
  {Allison}, \citenamefont {Dunkley},\ and\ \citenamefont
  {Calabrese}}]{Hlozek:2016lzm}%
  \BibitemOpen
  \bibfield  {author} {\bibinfo {author} {\bibfnamefont {Ren\'ee}\ \bibnamefont
  {Hlo\v{z}ek}}, \bibinfo {author} {\bibfnamefont {David J.~E.}\ \bibnamefont
  {Marsh}}, \bibinfo {author} {\bibfnamefont {Daniel}\ \bibnamefont {Grin}},
  \bibinfo {author} {\bibfnamefont {Rupert}\ \bibnamefont {Allison}}, \bibinfo
  {author} {\bibfnamefont {Jo}~\bibnamefont {Dunkley}}, \ and\ \bibinfo
  {author} {\bibfnamefont {Erminia}\ \bibnamefont {Calabrese}},\ }\bibfield
  {title} {\enquote {\bibinfo {title} {{Future CMB tests of dark matter:
  Ultralight axions and massive neutrinos}},}\ }\href {\doibase
  10.1103/PhysRevD.95.123511} {\bibfield  {journal} {\bibinfo  {journal} {Phys.
  Rev.}\ }\textbf {\bibinfo {volume} {D95}},\ \bibinfo {pages} {123511}
  (\bibinfo {year} {2017})},\ \Eprint {http://arxiv.org/abs/1607.08208}
  {arXiv:1607.08208 [astro-ph.CO]} \BibitemShut {NoStop}%
%%CITATION = ARXIV:1607.08208;%%
\bibitem [{\citenamefont {Ir\v{s}i\v{c}}\ \emph {et~al.}(2017)\citenamefont
  {Ir\v{s}i\v{c}}, \citenamefont {Viel}, \citenamefont {Haehnelt},
  \citenamefont {Bolton},\ and\ \citenamefont {Becker}}]{Irsic:2017yje}%
  \BibitemOpen
  \bibfield  {author} {\bibinfo {author} {\bibfnamefont {Vid}\ \bibnamefont
  {Ir\v{s}i\v{c}}}, \bibinfo {author} {\bibfnamefont {Matteo}\ \bibnamefont
  {Viel}}, \bibinfo {author} {\bibfnamefont {Martin~G.}\ \bibnamefont
  {Haehnelt}}, \bibinfo {author} {\bibfnamefont {James~S.}\ \bibnamefont
  {Bolton}}, \ and\ \bibinfo {author} {\bibfnamefont {George~D.}\ \bibnamefont
  {Becker}},\ }\bibfield  {title} {\enquote {\bibinfo {title} {{First
  constraints on fuzzy dark matter from Lyman-$\alpha$ forest data and
  hydrodynamical simulations}},}\ }\href {\doibase
  10.1103/PhysRevLett.119.031302} {\bibfield  {journal} {\bibinfo  {journal}
  {Phys. Rev. Lett.}\ }\textbf {\bibinfo {volume} {119}},\ \bibinfo {pages}
  {031302} (\bibinfo {year} {2017})},\ \Eprint
  {http://arxiv.org/abs/1703.04683} {arXiv:1703.04683 [astro-ph.CO]}
  \BibitemShut {NoStop}%
%%CITATION = ARXIV:1703.04683;%%
\bibitem [{\citenamefont {Armengaud}\ \emph {et~al.}(2017)\citenamefont
  {Armengaud}, \citenamefont {Palanque-Delabrouille}, \citenamefont {Yeche},
  \citenamefont {Marsh},\ and\ \citenamefont {Baur}}]{Armengaud:2017nkf}%
  \BibitemOpen
  \bibfield  {author} {\bibinfo {author} {\bibfnamefont {Eric}\ \bibnamefont
  {Armengaud}}, \bibinfo {author} {\bibfnamefont {Nathalie}\ \bibnamefont
  {Palanque-Delabrouille}}, \bibinfo {author} {\bibfnamefont {Christophe}\
  \bibnamefont {Yeche}}, \bibinfo {author} {\bibfnamefont {David J.~E.}\
  \bibnamefont {Marsh}}, \ and\ \bibinfo {author} {\bibfnamefont {Julien}\
  \bibnamefont {Baur}},\ }\bibfield  {title} {\enquote {\bibinfo {title}
  {{Constraining the mass of light bosonic dark matter using SDSS
  Lyman-$\alpha$ forest}},}\ }\href {\doibase 10.1093/mnras/stx1870} {\bibfield
   {journal} {\bibinfo  {journal} {Mon. Not. Roy. Astron. Soc.}\ }\textbf
  {\bibinfo {volume} {471}},\ \bibinfo {pages} {4606--4614} (\bibinfo {year}
  {2017})},\ \Eprint {http://arxiv.org/abs/1703.09126} {arXiv:1703.09126
  [astro-ph.CO]} \BibitemShut {NoStop}%
%%CITATION = ARXIV:1703.09126;%%
\bibitem [{\citenamefont {Kobayashi}\ \emph {et~al.}(2017)\citenamefont
  {Kobayashi}, \citenamefont {Murgia}, \citenamefont {De~Simone}, \citenamefont
  {Ir\v{s}i\v{c}},\ and\ \citenamefont {Viel}}]{Kobayashi:2017jcf}%
  \BibitemOpen
  \bibfield  {author} {\bibinfo {author} {\bibfnamefont {Takeshi}\ \bibnamefont
  {Kobayashi}}, \bibinfo {author} {\bibfnamefont {Riccardo}\ \bibnamefont
  {Murgia}}, \bibinfo {author} {\bibfnamefont {Andrea}\ \bibnamefont
  {De~Simone}}, \bibinfo {author} {\bibfnamefont {Vid}\ \bibnamefont
  {Ir\v{s}i\v{c}}}, \ and\ \bibinfo {author} {\bibfnamefont {Matteo}\
  \bibnamefont {Viel}},\ }\bibfield  {title} {\enquote {\bibinfo {title}
  {{Lyman-$\alpha$ constraints on ultralight scalar dark matter: Implications
  for the early and late universe}},}\ }\href {\doibase
  10.1103/PhysRevD.96.123514} {\bibfield  {journal} {\bibinfo  {journal} {Phys.
  Rev.}\ }\textbf {\bibinfo {volume} {D96}},\ \bibinfo {pages} {123514}
  (\bibinfo {year} {2017})},\ \Eprint {http://arxiv.org/abs/1708.00015}
  {arXiv:1708.00015 [astro-ph.CO]} \BibitemShut {NoStop}%
%%CITATION = ARXIV:1708.00015;%%
\bibitem [{\citenamefont {Bar}\ \emph {et~al.}(2018)\citenamefont {Bar},
  \citenamefont {Blas}, \citenamefont {Blum},\ and\ \citenamefont
  {Sibiryakov}}]{Bar:2018acw}%
  \BibitemOpen
  \bibfield  {author} {\bibinfo {author} {\bibfnamefont {Nitsan}\ \bibnamefont
  {Bar}}, \bibinfo {author} {\bibfnamefont {Diego}\ \bibnamefont {Blas}},
  \bibinfo {author} {\bibfnamefont {Kfir}\ \bibnamefont {Blum}}, \ and\
  \bibinfo {author} {\bibfnamefont {Sergey}\ \bibnamefont {Sibiryakov}},\
  }\bibfield  {title} {\enquote {\bibinfo {title} {{Galactic rotation curves
  versus ultralight dark matter: Implications of the soliton-host halo
  relation}},}\ }\href {\doibase 10.1103/PhysRevD.98.083027} {\bibfield
  {journal} {\bibinfo  {journal} {Phys. Rev.}\ }\textbf {\bibinfo {volume}
  {D98}},\ \bibinfo {pages} {083027} (\bibinfo {year} {2018})},\ \Eprint
  {http://arxiv.org/abs/1805.00122} {arXiv:1805.00122 [astro-ph.CO]}
  \BibitemShut {NoStop}%
%%CITATION = ARXIV:1805.00122;%%
\bibitem [{\citenamefont {Bar}\ \emph {et~al.}(2019)\citenamefont {Bar},
  \citenamefont {Blum}, \citenamefont {Eby},\ and\ \citenamefont
  {Sato}}]{Bar:2019bqz}%
  \BibitemOpen
  \bibfield  {author} {\bibinfo {author} {\bibfnamefont {Nitsan}\ \bibnamefont
  {Bar}}, \bibinfo {author} {\bibfnamefont {Kfir}\ \bibnamefont {Blum}},
  \bibinfo {author} {\bibfnamefont {Joshua}\ \bibnamefont {Eby}}, \ and\
  \bibinfo {author} {\bibfnamefont {Ryosuke}\ \bibnamefont {Sato}},\ }\bibfield
   {title} {\enquote {\bibinfo {title} {{Ultralight dark matter in disk
  galaxies}},}\ }\href {\doibase 10.1103/PhysRevD.99.103020} {\bibfield
  {journal} {\bibinfo  {journal} {Phys. Rev.}\ }\textbf {\bibinfo {volume}
  {D99}},\ \bibinfo {pages} {103020} (\bibinfo {year} {2019})},\ \Eprint
  {http://arxiv.org/abs/1903.03402} {arXiv:1903.03402 [astro-ph.CO]}
  \BibitemShut {NoStop}%
%%CITATION = ARXIV:1903.03402;%%
\bibitem [{\citenamefont {Safarzadeh}\ and\ \citenamefont
  {Spergel}(2019)}]{Safarzadeh:2019sre}%
  \BibitemOpen
  \bibfield  {author} {\bibinfo {author} {\bibfnamefont {Mohammadtaher}\
  \bibnamefont {Safarzadeh}}\ and\ \bibinfo {author} {\bibfnamefont {David~N.}\
  \bibnamefont {Spergel}},\ }\bibfield  {title} {\enquote {\bibinfo {title}
  {{Ultra-light Dark Matter is Incompatible with the Milky Way's Dwarf
  Satellites}},}\ }\href@noop {} {\  (\bibinfo {year} {2019})},\ \Eprint
  {http://arxiv.org/abs/1906.11848} {arXiv:1906.11848 [astro-ph.CO]}
  \BibitemShut {NoStop}%
%%CITATION = ARXIV:1906.11848;%%
\bibitem [{\citenamefont {Khmelnitsky}\ and\ \citenamefont
  {Rubakov}(2014)}]{Khmelnitsky:2013lxt}%
  \BibitemOpen
  \bibfield  {author} {\bibinfo {author} {\bibfnamefont {Andrei}\ \bibnamefont
  {Khmelnitsky}}\ and\ \bibinfo {author} {\bibfnamefont {Valery}\ \bibnamefont
  {Rubakov}},\ }\bibfield  {title} {\enquote {\bibinfo {title} {{Pulsar timing
  signal from ultralight scalar dark matter}},}\ }\href {\doibase
  10.1088/1475-7516/2014/02/019} {\bibfield  {journal} {\bibinfo  {journal}
  {JCAP}\ }\textbf {\bibinfo {volume} {1402}},\ \bibinfo {pages} {019}
  (\bibinfo {year} {2014})},\ \Eprint {http://arxiv.org/abs/1309.5888}
  {arXiv:1309.5888 [astro-ph.CO]} \BibitemShut {NoStop}%
%%CITATION = ARXIV:1309.5888;%%
\bibitem [{\citenamefont {Porayko}\ and\ \citenamefont
  {Postnov}(2014)}]{Porayko:2014rfa}%
  \BibitemOpen
  \bibfield  {author} {\bibinfo {author} {\bibfnamefont {N.~K.}\ \bibnamefont
  {Porayko}}\ and\ \bibinfo {author} {\bibfnamefont {K.~A.}\ \bibnamefont
  {Postnov}},\ }\bibfield  {title} {\enquote {\bibinfo {title} {{Constraints on
  ultralight scalar dark matter from pulsar timing}},}\ }\href {\doibase
  10.1103/PhysRevD.90.062008} {\bibfield  {journal} {\bibinfo  {journal} {Phys.
  Rev.}\ }\textbf {\bibinfo {volume} {D90}},\ \bibinfo {pages} {062008}
  (\bibinfo {year} {2014})},\ \Eprint {http://arxiv.org/abs/1408.4670}
  {arXiv:1408.4670 [astro-ph.CO]} \BibitemShut {NoStop}%
%%CITATION = ARXIV:1408.4670;%%
\bibitem [{\citenamefont {Marsh}\ and\ \citenamefont
  {Niemeyer}(2018)}]{Marsh:2018zyw}%
  \BibitemOpen
  \bibfield  {author} {\bibinfo {author} {\bibfnamefont {David J.~E.}\
  \bibnamefont {Marsh}}\ and\ \bibinfo {author} {\bibfnamefont {Jens~C.}\
  \bibnamefont {Niemeyer}},\ }\bibfield  {title} {\enquote {\bibinfo {title}
  {{Strong Constraints on Fuzzy Dark Matter from Ultrafaint Dwarf Galaxy
  Eridanus II}},}\ }\href@noop {} {\  (\bibinfo {year} {2018})},\ \Eprint
  {http://arxiv.org/abs/1810.08543} {arXiv:1810.08543 [astro-ph.CO]}
  \BibitemShut {NoStop}%
%%CITATION = ARXIV:1810.08543;%%
\bibitem [{\citenamefont {Schneider}(2018)}]{Schneider:2018xba}%
  \BibitemOpen
  \bibfield  {author} {\bibinfo {author} {\bibfnamefont {Aurel}\ \bibnamefont
  {Schneider}},\ }\bibfield  {title} {\enquote {\bibinfo {title} {{Constraining
  noncold dark matter models with the global 21-cm signal}},}\ }\href {\doibase
  10.1103/PhysRevD.98.063021} {\bibfield  {journal} {\bibinfo  {journal} {Phys.
  Rev.}\ }\textbf {\bibinfo {volume} {D98}},\ \bibinfo {pages} {063021}
  (\bibinfo {year} {2018})},\ \Eprint {http://arxiv.org/abs/1805.00021}
  {arXiv:1805.00021 [astro-ph.CO]} \BibitemShut {NoStop}%
%%CITATION = ARXIV:1805.00021;%%
\bibitem [{\citenamefont {Lidz}\ and\ \citenamefont
  {Hui}(2018)}]{Lidz:2018fqo}%
  \BibitemOpen
  \bibfield  {author} {\bibinfo {author} {\bibfnamefont {Adam}\ \bibnamefont
  {Lidz}}\ and\ \bibinfo {author} {\bibfnamefont {Lam}\ \bibnamefont {Hui}},\
  }\bibfield  {title} {\enquote {\bibinfo {title} {{Implications of a
  prereionization 21-cm absorption signal for fuzzy dark matter}},}\ }\href
  {\doibase 10.1103/PhysRevD.98.023011} {\bibfield  {journal} {\bibinfo
  {journal} {Phys. Rev.}\ }\textbf {\bibinfo {volume} {D98}},\ \bibinfo {pages}
  {023011} (\bibinfo {year} {2018})},\ \Eprint
  {http://arxiv.org/abs/1805.01253} {arXiv:1805.01253 [astro-ph.CO]}
  \BibitemShut {NoStop}%
%%CITATION = ARXIV:1805.01253;%%
\bibitem [{\citenamefont {Boyarsky}\ \emph {et~al.}(2019)\citenamefont
  {Boyarsky}, \citenamefont {Iakubovskyi}, \citenamefont {Ruchayskiy},
  \citenamefont {Rudakovskyi},\ and\ \citenamefont
  {Valkenburg}}]{Boyarsky:2019fgp}%
  \BibitemOpen
  \bibfield  {author} {\bibinfo {author} {\bibfnamefont {Alexey}\ \bibnamefont
  {Boyarsky}}, \bibinfo {author} {\bibfnamefont {Dmytro}\ \bibnamefont
  {Iakubovskyi}}, \bibinfo {author} {\bibfnamefont {Oleg}\ \bibnamefont
  {Ruchayskiy}}, \bibinfo {author} {\bibfnamefont {Anton}\ \bibnamefont
  {Rudakovskyi}}, \ and\ \bibinfo {author} {\bibfnamefont {Wessel}\
  \bibnamefont {Valkenburg}},\ }\bibfield  {title} {\enquote {\bibinfo {title}
  {{21-cm observations and warm dark matter models}},}\ }\href@noop {} {\
  (\bibinfo {year} {2019})},\ \Eprint {http://arxiv.org/abs/1904.03097}
  {arXiv:1904.03097 [astro-ph.CO]} \BibitemShut {NoStop}%
%%CITATION = ARXIV:1904.03097;%%
\bibitem [{\citenamefont {Marsh}(2015)}]{Marsh:2015daa}%
  \BibitemOpen
  \bibfield  {author} {\bibinfo {author} {\bibfnamefont {David J.~E.}\
  \bibnamefont {Marsh}},\ }\bibfield  {title} {\enquote {\bibinfo {title}
  {{Nonlinear hydrodynamics of axion dark matter: Relative velocity effects and
  quantum forces}},}\ }\href {\doibase 10.1103/PhysRevD.91.123520} {\bibfield
  {journal} {\bibinfo  {journal} {Phys. Rev.}\ }\textbf {\bibinfo {volume}
  {D91}},\ \bibinfo {pages} {123520} (\bibinfo {year} {2015})},\ \Eprint
  {http://arxiv.org/abs/1504.00308} {arXiv:1504.00308 [astro-ph.CO]}
  \BibitemShut {NoStop}%
%%CITATION = ARXIV:1504.00308;%%
\bibitem [{\citenamefont {Arvanitaki}\ and\ \citenamefont
  {Dubovsky}(2011)}]{Arvanitaki:2010sy}%
  \BibitemOpen
  \bibfield  {author} {\bibinfo {author} {\bibfnamefont {Asimina}\ \bibnamefont
  {Arvanitaki}}\ and\ \bibinfo {author} {\bibfnamefont {Sergei}\ \bibnamefont
  {Dubovsky}},\ }\bibfield  {title} {\enquote {\bibinfo {title} {{Exploring the
  String Axiverse with Precision Black Hole Physics}},}\ }\href {\doibase
  10.1103/PhysRevD.83.044026} {\bibfield  {journal} {\bibinfo  {journal}
  {Phys.Rev.}\ }\textbf {\bibinfo {volume} {D83}},\ \bibinfo {pages} {044026}
  (\bibinfo {year} {2011})},\ \Eprint {http://arxiv.org/abs/1004.3558}
  {arXiv:1004.3558 [hep-th]} \BibitemShut {NoStop}%
%%CITATION = ARXIV:1004.3558;%%
\bibitem [{\citenamefont {Arvanitaki}\ \emph
  {et~al.}(2015{\natexlab{b}})\citenamefont {Arvanitaki}, \citenamefont
  {Baryakhtar},\ and\ \citenamefont {Huang}}]{Arvanitaki:2014wva}%
  \BibitemOpen
  \bibfield  {author} {\bibinfo {author} {\bibfnamefont {Asimina}\ \bibnamefont
  {Arvanitaki}}, \bibinfo {author} {\bibfnamefont {Masha}\ \bibnamefont
  {Baryakhtar}}, \ and\ \bibinfo {author} {\bibfnamefont {Xinlu}\ \bibnamefont
  {Huang}},\ }\bibfield  {title} {\enquote {\bibinfo {title} {{Discovering the
  QCD Axion with Black Holes and Gravitational Waves}},}\ }\href {\doibase
  10.1103/PhysRevD.91.084011} {\bibfield  {journal} {\bibinfo  {journal} {Phys.
  Rev.}\ }\textbf {\bibinfo {volume} {D91}},\ \bibinfo {pages} {084011}
  (\bibinfo {year} {2015}{\natexlab{b}})},\ \Eprint
  {http://arxiv.org/abs/1411.2263} {arXiv:1411.2263 [hep-ph]} \BibitemShut
  {NoStop}%
%%CITATION = ARXIV:1411.2263;%%
\bibitem [{\citenamefont {Arvanitaki}\ \emph {et~al.}(2017)\citenamefont
  {Arvanitaki}, \citenamefont {Baryakhtar}, \citenamefont {Dimopoulos},
  \citenamefont {Dubovsky},\ and\ \citenamefont
  {Lasenby}}]{Arvanitaki:2016qwi}%
  \BibitemOpen
  \bibfield  {author} {\bibinfo {author} {\bibfnamefont {Asimina}\ \bibnamefont
  {Arvanitaki}}, \bibinfo {author} {\bibfnamefont {Masha}\ \bibnamefont
  {Baryakhtar}}, \bibinfo {author} {\bibfnamefont {Savas}\ \bibnamefont
  {Dimopoulos}}, \bibinfo {author} {\bibfnamefont {Sergei}\ \bibnamefont
  {Dubovsky}}, \ and\ \bibinfo {author} {\bibfnamefont {Robert}\ \bibnamefont
  {Lasenby}},\ }\bibfield  {title} {\enquote {\bibinfo {title} {{Black Hole
  Mergers and the QCD Axion at Advanced LIGO}},}\ }\href {\doibase
  10.1103/PhysRevD.95.043001} {\bibfield  {journal} {\bibinfo  {journal} {Phys.
  Rev.}\ }\textbf {\bibinfo {volume} {D95}},\ \bibinfo {pages} {043001}
  (\bibinfo {year} {2017})},\ \Eprint {http://arxiv.org/abs/1604.03958}
  {arXiv:1604.03958 [hep-ph]} \BibitemShut {NoStop}%
%%CITATION = ARXIV:1604.03958;%%
\bibitem [{\citenamefont {Brito}\ \emph {et~al.}(2017)\citenamefont {Brito},
  \citenamefont {Ghosh}, \citenamefont {Barausse}, \citenamefont {Berti},
  \citenamefont {Cardoso}, \citenamefont {Dvorkin}, \citenamefont {Klein},\
  and\ \citenamefont {Pani}}]{Brito:2017zvb}%
  \BibitemOpen
  \bibfield  {author} {\bibinfo {author} {\bibfnamefont {Richard}\ \bibnamefont
  {Brito}}, \bibinfo {author} {\bibfnamefont {Shrobana}\ \bibnamefont {Ghosh}},
  \bibinfo {author} {\bibfnamefont {Enrico}\ \bibnamefont {Barausse}}, \bibinfo
  {author} {\bibfnamefont {Emanuele}\ \bibnamefont {Berti}}, \bibinfo {author}
  {\bibfnamefont {Vitor}\ \bibnamefont {Cardoso}}, \bibinfo {author}
  {\bibfnamefont {Irina}\ \bibnamefont {Dvorkin}}, \bibinfo {author}
  {\bibfnamefont {Antoine}\ \bibnamefont {Klein}}, \ and\ \bibinfo {author}
  {\bibfnamefont {Paolo}\ \bibnamefont {Pani}},\ }\bibfield  {title} {\enquote
  {\bibinfo {title} {{Gravitational wave searches for ultralight bosons with
  LIGO and LISA}},}\ }\href {\doibase 10.1103/PhysRevD.96.064050} {\bibfield
  {journal} {\bibinfo  {journal} {Phys. Rev.}\ }\textbf {\bibinfo {volume}
  {D96}},\ \bibinfo {pages} {064050} (\bibinfo {year} {2017})},\ \Eprint
  {http://arxiv.org/abs/1706.06311} {arXiv:1706.06311 [gr-qc]} \BibitemShut
  {NoStop}%
%%CITATION = ARXIV:1706.06311;%%
\bibitem [{\citenamefont {Davoudiasl}\ and\ \citenamefont
  {Denton}(2019)}]{Davoudiasl:2019nlo}%
  \BibitemOpen
  \bibfield  {author} {\bibinfo {author} {\bibfnamefont {Hooman}\ \bibnamefont
  {Davoudiasl}}\ and\ \bibinfo {author} {\bibfnamefont {Peter~B}\ \bibnamefont
  {Denton}},\ }\bibfield  {title} {\enquote {\bibinfo {title} {{Ultralight
  Boson Dark Matter and Event Horizon Telescope Observations of M87*}},}\
  }\href {\doibase 10.1103/PhysRevLett.123.021102} {\bibfield  {journal}
  {\bibinfo  {journal} {Phys. Rev. Lett.}\ }\textbf {\bibinfo {volume} {123}},\
  \bibinfo {pages} {021102} (\bibinfo {year} {2019})},\ \Eprint
  {http://arxiv.org/abs/1904.09242} {arXiv:1904.09242 [astro-ph.CO]}
  \BibitemShut {NoStop}%
%%CITATION = ARXIV:1904.09242;%%
\bibitem [{\citenamefont {Damour}\ and\ \citenamefont
  {Esposito-Far{\'e}se}(1992)}]{Damour:1992we}%
  \BibitemOpen
  \bibfield  {author} {\bibinfo {author} {\bibfnamefont {T.}~\bibnamefont
  {Damour}}\ and\ \bibinfo {author} {\bibfnamefont {G.}~\bibnamefont
  {Esposito-Far{\'e}se}},\ }\bibfield  {title} {\enquote {\bibinfo {title}
  {{Tensor multiscalar theories of gravitation}},}\ }\href {\doibase
  10.1088/0264-9381/9/9/015} {\bibfield  {journal} {\bibinfo  {journal} {Class.
  Quant. Grav.}\ }\textbf {\bibinfo {volume} {9}},\ \bibinfo {pages}
  {2093--2176} (\bibinfo {year} {1992})}\BibitemShut {NoStop}%
\bibitem [{\citenamefont {Williams}\ \emph {et~al.}(2012)\citenamefont
  {Williams}, \citenamefont {Turyshev},\ and\ \citenamefont
  {Boggs}}]{Williams:2012nc}%
  \BibitemOpen
  \bibfield  {author} {\bibinfo {author} {\bibfnamefont {James~G.}\
  \bibnamefont {Williams}}, \bibinfo {author} {\bibfnamefont {Slava~G.}\
  \bibnamefont {Turyshev}}, \ and\ \bibinfo {author} {\bibfnamefont {Dale}\
  \bibnamefont {Boggs}},\ }\bibfield  {title} {\enquote {\bibinfo {title}
  {{Lunar Laser Ranging Tests of the Equivalence Principle}},}\ }\href
  {\doibase 10.1088/0264-9381/29/18/184004} {\bibfield  {journal} {\bibinfo
  {journal} {Class. Quant. Grav.}\ }\textbf {\bibinfo {volume} {29}},\ \bibinfo
  {pages} {184004} (\bibinfo {year} {2012})},\ \Eprint
  {http://arxiv.org/abs/1203.2150} {arXiv:1203.2150 [gr-qc]} \BibitemShut
  {NoStop}%
%%CITATION = ARXIV:1203.2150;%%
\bibitem [{\citenamefont {{Genova}}\ \emph {et~al.}(2018)\citenamefont
  {{Genova}}, \citenamefont {{Mazarico}}, \citenamefont {{Goossens}},
  \citenamefont {{Lemoine}}, \citenamefont {{Neumann}}, \citenamefont
  {{Smith}},\ and\ \citenamefont {{Zuber}}}]{2018NatCo...9..289G}%
  \BibitemOpen
  \bibfield  {author} {\bibinfo {author} {\bibfnamefont {A.}~\bibnamefont
  {{Genova}}}, \bibinfo {author} {\bibfnamefont {E.}~\bibnamefont
  {{Mazarico}}}, \bibinfo {author} {\bibfnamefont {S.}~\bibnamefont
  {{Goossens}}}, \bibinfo {author} {\bibfnamefont {F.~G.}\ \bibnamefont
  {{Lemoine}}}, \bibinfo {author} {\bibfnamefont {G.~A.}\ \bibnamefont
  {{Neumann}}}, \bibinfo {author} {\bibfnamefont {D.~E.}\ \bibnamefont
  {{Smith}}}, \ and\ \bibinfo {author} {\bibfnamefont {M.~T.}\ \bibnamefont
  {{Zuber}}},\ }\bibfield  {title} {\enquote {\bibinfo {title} {{Solar system
  expansion and strong equivalence principle as seen by the NASA MESSENGER
  mission}},}\ }\href {\doibase 10.1038/s41467-017-02558-1} {\bibfield
  {journal} {\bibinfo  {journal} {Nature Communications}\ }\textbf {\bibinfo
  {volume} {9}},\ \bibinfo {eid} {289} (\bibinfo {year} {2018})}\BibitemShut
  {NoStop}%
\bibitem [{\citenamefont {Talmadge}\ \emph {et~al.}(1988)\citenamefont
  {Talmadge}, \citenamefont {Berthias}, \citenamefont {Hellings},\ and\
  \citenamefont {Standish}}]{Talmadge:1988qz}%
  \BibitemOpen
  \bibfield  {author} {\bibinfo {author} {\bibfnamefont {C.}~\bibnamefont
  {Talmadge}}, \bibinfo {author} {\bibfnamefont {J.~P.}\ \bibnamefont
  {Berthias}}, \bibinfo {author} {\bibfnamefont {R.~W.}\ \bibnamefont
  {Hellings}}, \ and\ \bibinfo {author} {\bibfnamefont {E.~M.}\ \bibnamefont
  {Standish}},\ }\bibfield  {title} {\enquote {\bibinfo {title} {{Model
  Independent Constraints on Possible Modifications of Newtonian Gravity}},}\
  }\href {\doibase 10.1103/PhysRevLett.61.1159} {\bibfield  {journal} {\bibinfo
   {journal} {Phys. Rev. Lett.}\ }\textbf {\bibinfo {volume} {61}},\ \bibinfo
  {pages} {1159--1162} (\bibinfo {year} {1988})}\BibitemShut {NoStop}%
%%CITATION = PRLTA,61,1159;%%
\bibitem [{\citenamefont {Adelberger}\ \emph {et~al.}(2003)\citenamefont
  {Adelberger}, \citenamefont {Heckel},\ and\ \citenamefont
  {Nelson}}]{Adelberger:2003zx}%
  \BibitemOpen
  \bibfield  {author} {\bibinfo {author} {\bibfnamefont {E.~G.}\ \bibnamefont
  {Adelberger}}, \bibinfo {author} {\bibfnamefont {Blayne~R.}\ \bibnamefont
  {Heckel}}, \ and\ \bibinfo {author} {\bibfnamefont {A.~E.}\ \bibnamefont
  {Nelson}},\ }\bibfield  {title} {\enquote {\bibinfo {title} {{Tests of the
  gravitational inverse square law}},}\ }\href {\doibase
  10.1146/annurev.nucl.53.041002.110503} {\bibfield  {journal} {\bibinfo
  {journal} {Ann. Rev. Nucl. Part. Sci.}\ }\textbf {\bibinfo {volume} {53}},\
  \bibinfo {pages} {77--121} (\bibinfo {year} {2003})},\ \Eprint
  {http://arxiv.org/abs/hep-ph/0307284} {arXiv:hep-ph/0307284 [hep-ph]}
  \BibitemShut {NoStop}%
%%CITATION = HEP-PH/0307284;%%
\bibitem [{\citenamefont {Iorio}(2007)}]{Iorio:2005fk}%
  \BibitemOpen
  \bibfield  {author} {\bibinfo {author} {\bibfnamefont {Lorenzo}\ \bibnamefont
  {Iorio}},\ }\bibfield  {title} {\enquote {\bibinfo {title} {{First evidence
  of the general relativistic gravitomagnetic field of the Sun and new
  constraints on a Yukawa-like fifth force}},}\ }\href {\doibase
  10.1016/j.pss.2007.04.001} {\bibfield  {journal} {\bibinfo  {journal}
  {Planet. Space Sci.}\ }\textbf {\bibinfo {volume} {55}},\ \bibinfo {pages}
  {1290--1298} (\bibinfo {year} {2007})},\ \Eprint
  {http://arxiv.org/abs/gr-qc/0507041} {arXiv:gr-qc/0507041 [gr-qc]}
  \BibitemShut {NoStop}%
%%CITATION = GR-QC/0507041;%%
\bibitem [{\citenamefont {Bertotti}\ \emph {et~al.}(2003)\citenamefont
  {Bertotti}, \citenamefont {Iess},\ and\ \citenamefont
  {Tortora}}]{Bertotti:2003rm}%
  \BibitemOpen
  \bibfield  {author} {\bibinfo {author} {\bibfnamefont {B.}~\bibnamefont
  {Bertotti}}, \bibinfo {author} {\bibfnamefont {L.}~\bibnamefont {Iess}}, \
  and\ \bibinfo {author} {\bibfnamefont {P.}~\bibnamefont {Tortora}},\
  }\bibfield  {title} {\enquote {\bibinfo {title} {{A test of general
  relativity using radio links with the Cassini spacecraft}},}\ }\href
  {\doibase 10.1038/nature01997} {\bibfield  {journal} {\bibinfo  {journal}
  {Nature}\ }\textbf {\bibinfo {volume} {425}},\ \bibinfo {pages} {374}
  (\bibinfo {year} {2003})}\BibitemShut {NoStop}%
%%CITATION = NATUA,425,374;%%
\bibitem [{\citenamefont {{Armstrong}}\ \emph {et~al.}(2003)\citenamefont
  {{Armstrong}}, \citenamefont {{Iess}}, \citenamefont {{Tortora}},\ and\
  \citenamefont {{Bertotti}}}]{2003ApJ...599..806A}%
  \BibitemOpen
  \bibfield  {author} {\bibinfo {author} {\bibfnamefont {J.~W.}\ \bibnamefont
  {{Armstrong}}}, \bibinfo {author} {\bibfnamefont {L.}~\bibnamefont {{Iess}}},
  \bibinfo {author} {\bibfnamefont {P.}~\bibnamefont {{Tortora}}}, \ and\
  \bibinfo {author} {\bibfnamefont {B.}~\bibnamefont {{Bertotti}}},\ }\bibfield
   {title} {\enquote {\bibinfo {title} {{Stochastic Gravitational Wave
  Background: Upper Limits in the 10$^{-6}$ to 10$^{-3}$ Hz Band}},}\ }\href
  {\doibase 10.1086/379505} {\bibfield  {journal} {\bibinfo  {journal} {\apj}\
  }\textbf {\bibinfo {volume} {599}},\ \bibinfo {pages} {806--813} (\bibinfo
  {year} {2003})}\BibitemShut {NoStop}%
\bibitem [{\citenamefont {Blas}\ \emph {et~al.}(2017)\citenamefont {Blas},
  \citenamefont {L\'{o}pez~Nacir},\ and\ \citenamefont
  {Sibiryakov}}]{Blas:2016ddr}%
  \BibitemOpen
  \bibfield  {author} {\bibinfo {author} {\bibfnamefont {Diego}\ \bibnamefont
  {Blas}}, \bibinfo {author} {\bibfnamefont {Diana}\ \bibnamefont
  {L\'{o}pez~Nacir}}, \ and\ \bibinfo {author} {\bibfnamefont {Sergey}\
  \bibnamefont {Sibiryakov}},\ }\bibfield  {title} {\enquote {\bibinfo {title}
  {{Ultralight Dark Matter Resonates with Binary Pulsars}},}\ }\href {\doibase
  10.1103/PhysRevLett.118.261102} {\bibfield  {journal} {\bibinfo  {journal}
  {Phys. Rev. Lett.}\ }\textbf {\bibinfo {volume} {118}},\ \bibinfo {pages}
  {261102} (\bibinfo {year} {2017})},\ \Eprint
  {http://arxiv.org/abs/1612.06789} {arXiv:1612.06789 [hep-ph]} \BibitemShut
  {NoStop}%
%%CITATION = ARXIV:1612.06789;%%
\bibitem [{\citenamefont {L\'opez~Nacir}\ and\ \citenamefont
  {Urban}(2018)}]{LopezNacir:2018epg}%
  \BibitemOpen
  \bibfield  {author} {\bibinfo {author} {\bibfnamefont {Diana}\ \bibnamefont
  {L\'opez~Nacir}}\ and\ \bibinfo {author} {\bibfnamefont {Federico~R.}\
  \bibnamefont {Urban}},\ }\bibfield  {title} {\enquote {\bibinfo {title}
  {{Vector Fuzzy Dark Matter, Fifth Forces, and Binary Pulsars}},}\ }\href
  {\doibase 10.1088/1475-7516/2018/10/044} {\bibfield  {journal} {\bibinfo
  {journal} {JCAP}\ }\textbf {\bibinfo {volume} {1810}},\ \bibinfo {pages}
  {044} (\bibinfo {year} {2018})},\ \Eprint {http://arxiv.org/abs/1807.10491}
  {arXiv:1807.10491 [astro-ph.CO]} \BibitemShut {NoStop}%
%%CITATION = ARXIV:1807.10491;%%
\bibitem [{\citenamefont {Armaleo}\ \emph {et~al.}(2020)\citenamefont
  {Armaleo}, \citenamefont {L\'opez~Nacir},\ and\ \citenamefont
  {Urban}}]{Armaleo:2019gil}%
  \BibitemOpen
  \bibfield  {author} {\bibinfo {author} {\bibfnamefont {Juan~Manuel}\
  \bibnamefont {Armaleo}}, \bibinfo {author} {\bibfnamefont {Diana}\
  \bibnamefont {L\'opez~Nacir}}, \ and\ \bibinfo {author} {\bibfnamefont
  {Federico~R.}\ \bibnamefont {Urban}},\ }\bibfield  {title} {\enquote
  {\bibinfo {title} {{Binary Pulsars as probes for Spin-2 Ultralight Dark
  Matter}},}\ }\href {\doibase 10.1088/1475-7516/2020/01/053} {\bibfield
  {journal} {\bibinfo  {journal} {JCAP}\ }\textbf {\bibinfo {volume} {2001}},\
  \bibinfo {pages} {053} (\bibinfo {year} {2020})},\ \Eprint
  {http://arxiv.org/abs/1909.13814} {arXiv:1909.13814 [astro-ph.HE]}
  \BibitemShut {NoStop}%
%%CITATION = ARXIV:1909.13814;%%
\bibitem [{\citenamefont {Rozner}\ \emph {et~al.}(2019)\citenamefont {Rozner},
  \citenamefont {Grishin}, \citenamefont {Ginat}, \citenamefont {Igoshev},\
  and\ \citenamefont {Desjacques}}]{Rozner:2019gba}%
  \BibitemOpen
  \bibfield  {author} {\bibinfo {author} {\bibfnamefont {Mor}\ \bibnamefont
  {Rozner}}, \bibinfo {author} {\bibfnamefont {Evgeni}\ \bibnamefont
  {Grishin}}, \bibinfo {author} {\bibfnamefont {Yonadav~Barry}\ \bibnamefont
  {Ginat}}, \bibinfo {author} {\bibfnamefont {Andrei~P.}\ \bibnamefont
  {Igoshev}}, \ and\ \bibinfo {author} {\bibfnamefont {Vincent}\ \bibnamefont
  {Desjacques}},\ }\bibfield  {title} {\enquote {\bibinfo {title} {{Axion
  resonances in binary pulsar systems}},}\ }\href@noop {} {\  (\bibinfo {year}
  {2019})},\ \Eprint {http://arxiv.org/abs/1904.01958} {arXiv:1904.01958
  [astro-ph.CO]} \BibitemShut {NoStop}%
%%CITATION = ARXIV:1904.01958;%%
\bibitem [{\citenamefont {{Blandford}}\ and\ \citenamefont
  {{Teukolsky}}(1976)}]{Teukolsky1976}%
  \BibitemOpen
  \bibfield  {author} {\bibinfo {author} {\bibfnamefont {R.}~\bibnamefont
  {{Blandford}}}\ and\ \bibinfo {author} {\bibfnamefont {S.~A.}\ \bibnamefont
  {{Teukolsky}}},\ }\bibfield  {title} {\enquote {\bibinfo {title}
  {{Arrival-time analysis for a pulsar in a binary system.}}}\ }\href {\doibase
  10.1086/154315} {\bibfield  {journal} {\bibinfo  {journal} {\apj}\ }\textbf
  {\bibinfo {volume} {205}},\ \bibinfo {pages} {580--591} (\bibinfo {year}
  {1976})}\BibitemShut {NoStop}%
\bibitem [{\citenamefont {{Damour}}\ and\ \citenamefont
  {{Deruelle}}(1986)}]{1986AIHS...44..263D}%
  \BibitemOpen
  \bibfield  {author} {\bibinfo {author} {\bibfnamefont {T.}~\bibnamefont
  {{Damour}}}\ and\ \bibinfo {author} {\bibfnamefont {N.}~\bibnamefont
  {{Deruelle}}},\ }\bibfield  {title} {\enquote {\bibinfo {title} {{General
  relativistic celestial mechanics of binary systems. II. The post-Newtonian
  timing formula.}}}\ }\href@noop {} {\bibfield  {journal} {\bibinfo  {journal}
  {Ann.~Inst.~Henri Poincar{\'e} Phys.~Th{\'e}or., Vol.~44, No.~3, p.~263 -
  292}\ }\textbf {\bibinfo {volume} {44}},\ \bibinfo {pages} {263--292}
  (\bibinfo {year} {1986})}\BibitemShut {NoStop}%
\bibitem [{\citenamefont {Edwards}\ \emph {et~al.}(2006)\citenamefont
  {Edwards}, \citenamefont {Hobbs},\ and\ \citenamefont
  {Manchester}}]{Edwards:2006zg}%
  \BibitemOpen
  \bibfield  {author} {\bibinfo {author} {\bibfnamefont {Russell~T.}\
  \bibnamefont {Edwards}}, \bibinfo {author} {\bibfnamefont {G.~B.}\
  \bibnamefont {Hobbs}}, \ and\ \bibinfo {author} {\bibfnamefont {R.~N.}\
  \bibnamefont {Manchester}},\ }\bibfield  {title} {\enquote {\bibinfo {title}
  {{Tempo2, a new pulsar timing package. 2. The timing model and precision
  estimates}},}\ }\href {\doibase 10.1111/j.1365-2966.2006.10870.x} {\bibfield
  {journal} {\bibinfo  {journal} {Mon. Not. Roy. Astron. Soc.}\ }\textbf
  {\bibinfo {volume} {372}},\ \bibinfo {pages} {1549--1574} (\bibinfo {year}
  {2006})},\ \Eprint {http://arxiv.org/abs/astro-ph/0607664}
  {arXiv:astro-ph/0607664 [astro-ph]} \BibitemShut {NoStop}%
%%CITATION = ASTRO-PH/0607664;%%
\bibitem [{\citenamefont {Lorimer}(2008)}]{Lorimer:2008se}%
  \BibitemOpen
  \bibfield  {author} {\bibinfo {author} {\bibfnamefont {D.~R.}\ \bibnamefont
  {Lorimer}},\ }\bibfield  {title} {\enquote {\bibinfo {title} {{Binary and
  Millisecond Pulsars}},}\ }\href {\doibase 10.12942/lrr-2008-8} {\bibfield
  {journal} {\bibinfo  {journal} {Living Rev. Rel.}\ }\textbf {\bibinfo
  {volume} {11}},\ \bibinfo {pages} {8} (\bibinfo {year} {2008})},\ \Eprint
  {http://arxiv.org/abs/0811.0762} {arXiv:0811.0762 [astro-ph]} \BibitemShut
  {NoStop}%
%%CITATION = ARXIV:0811.0762;%%
\bibitem [{\citenamefont {Manchester}\ \emph {et~al.}(2005)\citenamefont
  {Manchester}, \citenamefont {Hobbs}, \citenamefont {Teoh},\ and\
  \citenamefont {Hobbs}}]{Manchester:2004bp}%
  \BibitemOpen
  \bibfield  {author} {\bibinfo {author} {\bibfnamefont {R~N}\ \bibnamefont
  {Manchester}}, \bibinfo {author} {\bibfnamefont {G~B}\ \bibnamefont {Hobbs}},
  \bibinfo {author} {\bibfnamefont {A}~\bibnamefont {Teoh}}, \ and\ \bibinfo
  {author} {\bibfnamefont {M}~\bibnamefont {Hobbs}},\ }\bibfield  {title}
  {\enquote {\bibinfo {title} {{The Australia Telescope National Facility
  pulsar catalogue}},}\ }\href {\doibase 10.1086/428488} {\bibfield  {journal}
  {\bibinfo  {journal} {Astron. J.}\ }\textbf {\bibinfo {volume} {129}},\
  \bibinfo {pages} {1993} (\bibinfo {year} {2005})},\ \Eprint
  {http://arxiv.org/abs/astro-ph/0412641} {arXiv:astro-ph/0412641 [astro-ph]}
  \BibitemShut {NoStop}%
%%CITATION = ASTRO-PH/0412641;%%
\bibitem [{psc()}]{pscat}%
  \BibitemOpen
  \href@noop {} {}\bibinfo {note} {{\it ATNF Pulsar Catalogue},
  \url{http://www.atnf.csiro.au/people/pulsar/psrcat/}}\BibitemShut {NoStop}%
\bibitem [{\citenamefont {Schive}\ \emph
  {et~al.}(2014{\natexlab{a}})\citenamefont {Schive}, \citenamefont {Chiueh},\
  and\ \citenamefont {Broadhurst}}]{Schive:2014dra}%
  \BibitemOpen
  \bibfield  {author} {\bibinfo {author} {\bibfnamefont {Hsi-Yu}\ \bibnamefont
  {Schive}}, \bibinfo {author} {\bibfnamefont {Tzihong}\ \bibnamefont
  {Chiueh}}, \ and\ \bibinfo {author} {\bibfnamefont {Tom}\ \bibnamefont
  {Broadhurst}},\ }\bibfield  {title} {\enquote {\bibinfo {title} {{Cosmic
  Structure as the Quantum Interference of a Coherent Dark Wave}},}\ }\href
  {\doibase 10.1038/nphys2996} {\bibfield  {journal} {\bibinfo  {journal}
  {Nature Phys.}\ }\textbf {\bibinfo {volume} {10}},\ \bibinfo {pages}
  {496--499} (\bibinfo {year} {2014}{\natexlab{a}})},\ \Eprint
  {http://arxiv.org/abs/1406.6586} {arXiv:1406.6586 [astro-ph.GA]} \BibitemShut
  {NoStop}%
%%CITATION = ARXIV:1406.6586;%%
\bibitem [{\citenamefont {Schive}\ \emph
  {et~al.}(2014{\natexlab{b}})\citenamefont {Schive}, \citenamefont {Liao},
  \citenamefont {Woo}, \citenamefont {Wong}, \citenamefont {Chiueh},
  \citenamefont {Broadhurst},\ and\ \citenamefont {Hwang}}]{Schive:2014hza}%
  \BibitemOpen
  \bibfield  {author} {\bibinfo {author} {\bibfnamefont {Hsi-Yu}\ \bibnamefont
  {Schive}}, \bibinfo {author} {\bibfnamefont {Ming-Hsuan}\ \bibnamefont
  {Liao}}, \bibinfo {author} {\bibfnamefont {Tak-Pong}\ \bibnamefont {Woo}},
  \bibinfo {author} {\bibfnamefont {Shing-Kwong}\ \bibnamefont {Wong}},
  \bibinfo {author} {\bibfnamefont {Tzihong}\ \bibnamefont {Chiueh}}, \bibinfo
  {author} {\bibfnamefont {Tom}\ \bibnamefont {Broadhurst}}, \ and\ \bibinfo
  {author} {\bibfnamefont {W.~Y.~Pauchy}\ \bibnamefont {Hwang}},\ }\bibfield
  {title} {\enquote {\bibinfo {title} {{Understanding the Core-Halo Relation of
  Quantum Wave Dark Matter from 3D Simulations}},}\ }\href {\doibase
  10.1103/PhysRevLett.113.261302} {\bibfield  {journal} {\bibinfo  {journal}
  {Phys. Rev. Lett.}\ }\textbf {\bibinfo {volume} {113}},\ \bibinfo {pages}
  {261302} (\bibinfo {year} {2014}{\natexlab{b}})},\ \Eprint
  {http://arxiv.org/abs/1407.7762} {arXiv:1407.7762 [astro-ph.GA]} \BibitemShut
  {NoStop}%
%%CITATION = ARXIV:1407.7762;%%
\bibitem [{\citenamefont {Schwabe}\ \emph {et~al.}(2016)\citenamefont
  {Schwabe}, \citenamefont {Niemeyer},\ and\ \citenamefont
  {Engels}}]{Schwabe:2016rze}%
  \BibitemOpen
  \bibfield  {author} {\bibinfo {author} {\bibfnamefont {Bodo}\ \bibnamefont
  {Schwabe}}, \bibinfo {author} {\bibfnamefont {Jens~C.}\ \bibnamefont
  {Niemeyer}}, \ and\ \bibinfo {author} {\bibfnamefont {Jan~F.}\ \bibnamefont
  {Engels}},\ }\bibfield  {title} {\enquote {\bibinfo {title} {{Simulations of
  solitonic core mergers in ultralight axion dark matter cosmologies}},}\
  }\href {\doibase 10.1103/PhysRevD.94.043513} {\bibfield  {journal} {\bibinfo
  {journal} {Phys. Rev.}\ }\textbf {\bibinfo {volume} {D94}},\ \bibinfo {pages}
  {043513} (\bibinfo {year} {2016})},\ \Eprint
  {http://arxiv.org/abs/1606.05151} {arXiv:1606.05151 [astro-ph.CO]}
  \BibitemShut {NoStop}%
%%CITATION = ARXIV:1606.05151;%%
\bibitem [{\citenamefont {Veltmaat}\ \emph {et~al.}(2018)\citenamefont
  {Veltmaat}, \citenamefont {Niemeyer},\ and\ \citenamefont
  {Schwabe}}]{Veltmaat:2018dfz}%
  \BibitemOpen
  \bibfield  {author} {\bibinfo {author} {\bibfnamefont {Jan}\ \bibnamefont
  {Veltmaat}}, \bibinfo {author} {\bibfnamefont {Jens~C.}\ \bibnamefont
  {Niemeyer}}, \ and\ \bibinfo {author} {\bibfnamefont {Bodo}\ \bibnamefont
  {Schwabe}},\ }\bibfield  {title} {\enquote {\bibinfo {title} {{Formation and
  structure of ultralight bosonic dark matter halos}},}\ }\href {\doibase
  10.1103/PhysRevD.98.043509} {\bibfield  {journal} {\bibinfo  {journal} {Phys.
  Rev.}\ }\textbf {\bibinfo {volume} {D98}},\ \bibinfo {pages} {043509}
  (\bibinfo {year} {2018})},\ \Eprint {http://arxiv.org/abs/1804.09647}
  {arXiv:1804.09647 [astro-ph.CO]} \BibitemShut {NoStop}%
%%CITATION = ARXIV:1804.09647;%%
\bibitem [{\citenamefont {Lisanti}(2017)}]{Lisanti:2016jxe}%
  \BibitemOpen
  \bibfield  {author} {\bibinfo {author} {\bibfnamefont {Mariangela}\
  \bibnamefont {Lisanti}},\ }\bibfield  {title} {\enquote {\bibinfo {title}
  {{Lectures on Dark Matter Physics}},}\ }\href {\doibase
  10.1142/9789813149441_0007} {\bibfield  {journal} {\bibinfo  {journal}
  {{Proceedings, Theoretical Advanced Study Institute in Elementary Particle
  Physics: New Frontiers in Fields and Strings (TASI 2015): Boulder, CO, USA,
  June 1-26, 2015}}\ ,\ \bibinfo {pages} {399--446}} (\bibinfo {year}
  {2017})},\ \Eprint {http://arxiv.org/abs/1603.03797} {arXiv:1603.03797
  [hep-ph]} \BibitemShut {NoStop}%
\bibitem [{\citenamefont {Piffl}\ \emph {et~al.}(2014)\citenamefont {Piffl}
  \emph {et~al.}}]{Piffl:2014mfa}%
  \BibitemOpen
  \bibfield  {author} {\bibinfo {author} {\bibfnamefont {T.}~\bibnamefont
  {Piffl}} \emph {et~al.},\ }\bibfield  {title} {\enquote {\bibinfo {title}
  {{Constraining the Galaxy's dark halo with RAVE stars}},}\ }\href {\doibase
  10.1093/mnras/stu1948} {\bibfield  {journal} {\bibinfo  {journal} {Mon. Not.
  Roy. Astron. Soc.}\ }\textbf {\bibinfo {volume} {445}},\ \bibinfo {pages}
  {3133--3151} (\bibinfo {year} {2014})},\ \Eprint
  {http://arxiv.org/abs/1406.4130} {arXiv:1406.4130 [astro-ph.GA]} \BibitemShut
  {NoStop}%
%%CITATION = ARXIV:1406.4130;%%
\bibitem [{\citenamefont {McKee}\ \emph {et~al.}(2015)\citenamefont {McKee},
  \citenamefont {Parravano},\ and\ \citenamefont {Hollenbach}}]{McKee:2015hwa}%
  \BibitemOpen
  \bibfield  {author} {\bibinfo {author} {\bibfnamefont {Christopher~F.}\
  \bibnamefont {McKee}}, \bibinfo {author} {\bibfnamefont {Antonio}\
  \bibnamefont {Parravano}}, \ and\ \bibinfo {author} {\bibfnamefont
  {David~J.}\ \bibnamefont {Hollenbach}},\ }\bibfield  {title} {\enquote
  {\bibinfo {title} {{Stars, Gas, and Dark Matter in the Solar
  Neighborhood}},}\ }\href {\doibase 10.1088/0004-637X,
  10.1088/0004-637X/814/1/13} {\bibfield  {journal} {\bibinfo  {journal}
  {Astrophys. J.}\ }\textbf {\bibinfo {volume} {814}},\ \bibinfo {pages} {13}
  (\bibinfo {year} {2015})},\ \Eprint {http://arxiv.org/abs/1509.05334}
  {arXiv:1509.05334 [astro-ph.GA]} \BibitemShut {NoStop}%
%%CITATION = ARXIV:1509.05334;%%
\bibitem [{\citenamefont {Evans}\ \emph {et~al.}(2019)\citenamefont {Evans},
  \citenamefont {O'Hare},\ and\ \citenamefont {McCabe}}]{Evans:2018bqy}%
  \BibitemOpen
  \bibfield  {author} {\bibinfo {author} {\bibfnamefont {N.~Wyn}\ \bibnamefont
  {Evans}}, \bibinfo {author} {\bibfnamefont {Ciaran A.~J.}\ \bibnamefont
  {O'Hare}}, \ and\ \bibinfo {author} {\bibfnamefont {Christopher}\
  \bibnamefont {McCabe}},\ }\bibfield  {title} {\enquote {\bibinfo {title}
  {{Refinement of the standard halo model for dark matter searches in light of
  the Gaia Sausage}},}\ }\href {\doibase 10.1103/PhysRevD.99.023012} {\bibfield
   {journal} {\bibinfo  {journal} {Phys. Rev.}\ }\textbf {\bibinfo {volume}
  {D99}},\ \bibinfo {pages} {023012} (\bibinfo {year} {2019})},\ \Eprint
  {http://arxiv.org/abs/1810.11468} {arXiv:1810.11468 [astro-ph.GA]}
  \BibitemShut {NoStop}%
%%CITATION = ARXIV:1810.11468;%%
\bibitem [{\citenamefont {Will}(1993)}]{will1993theory}%
  \BibitemOpen
  \bibfield  {author} {\bibinfo {author} {\bibfnamefont {C.M.}\ \bibnamefont
  {Will}},\ }\href@noop {} {\emph {\bibinfo {title} {Theory and Experiment in
  Gravitational Physics}}}\ (\bibinfo  {publisher} {Cambridge University
  Press},\ \bibinfo {year} {1993})\BibitemShut {NoStop}%
\bibitem [{\citenamefont {Damour}\ and\ \citenamefont
  {Taylor}(1992)}]{Damour:1991rd}%
  \BibitemOpen
  \bibfield  {author} {\bibinfo {author} {\bibfnamefont {Thibault}\
  \bibnamefont {Damour}}\ and\ \bibinfo {author} {\bibfnamefont {Joseph~H.}\
  \bibnamefont {Taylor}},\ }\bibfield  {title} {\enquote {\bibinfo {title}
  {{Strong field tests of relativistic gravity and binary pulsars}},}\ }\href
  {\doibase 10.1103/PhysRevD.45.1840} {\bibfield  {journal} {\bibinfo
  {journal} {Phys. Rev.}\ }\textbf {\bibinfo {volume} {D45}},\ \bibinfo {pages}
  {1840--1868} (\bibinfo {year} {1992})}\BibitemShut {NoStop}%
%%CITATION = PHRVA,D45,1840;%%
\bibitem [{\citenamefont {{Mashhoon}}(1978)}]{1978ApJ...223..285M}%
  \BibitemOpen
  \bibfield  {author} {\bibinfo {author} {\bibfnamefont {B.}~\bibnamefont
  {{Mashhoon}}},\ }\bibfield  {title} {\enquote {\bibinfo {title} {{On tidal
  resonance}},}\ }\href {\doibase 10.1086/156262} {\bibfield  {journal}
  {\bibinfo  {journal} {Astrophys. J.}\ }\textbf {\bibinfo {volume} {223}},\
  \bibinfo {pages} {285--298} (\bibinfo {year} {1978})}\BibitemShut {NoStop}%
\bibitem [{\citenamefont {Graham}\ \emph {et~al.}(2016)\citenamefont {Graham},
  \citenamefont {Kaplan}, \citenamefont {Mardon}, \citenamefont {Rajendran},\
  and\ \citenamefont {Terrano}}]{Graham:2015ifn}%
  \BibitemOpen
  \bibfield  {author} {\bibinfo {author} {\bibfnamefont {Peter~W.}\
  \bibnamefont {Graham}}, \bibinfo {author} {\bibfnamefont {David~E.}\
  \bibnamefont {Kaplan}}, \bibinfo {author} {\bibfnamefont {Jeremy}\
  \bibnamefont {Mardon}}, \bibinfo {author} {\bibfnamefont {Surjeet}\
  \bibnamefont {Rajendran}}, \ and\ \bibinfo {author} {\bibfnamefont
  {William~A.}\ \bibnamefont {Terrano}},\ }\bibfield  {title} {\enquote
  {\bibinfo {title} {{Dark Matter Direct Detection with Accelerometers}},}\
  }\href {\doibase 10.1103/PhysRevD.93.075029} {\bibfield  {journal} {\bibinfo
  {journal} {Phys. Rev.}\ }\textbf {\bibinfo {volume} {D93}},\ \bibinfo {pages}
  {075029} (\bibinfo {year} {2016})},\ \Eprint
  {http://arxiv.org/abs/1512.06165} {arXiv:1512.06165 [hep-ph]} \BibitemShut
  {NoStop}%
%%CITATION = ARXIV:1512.06165;%%
\bibitem [{\citenamefont {Damour}\ and\ \citenamefont
  {Esposito-Far{\`e}se}(1993)}]{Damour:1993hw}%
  \BibitemOpen
  \bibfield  {author} {\bibinfo {author} {\bibfnamefont {T.}~\bibnamefont
  {Damour}}\ and\ \bibinfo {author} {\bibfnamefont {G.}~\bibnamefont
  {Esposito-Far{\`e}se}},\ }\bibfield  {title} {\enquote {\bibinfo {title}
  {{Nonperturbative strong field effects in tensor - scalar theories of
  gravitation}},}\ }\href {\doibase 10.1103/PhysRevLett.70.2220} {\bibfield
  {journal} {\bibinfo  {journal} {Phys. Rev. Lett.}\ }\textbf {\bibinfo
  {volume} {70}},\ \bibinfo {pages} {2220--2223} (\bibinfo {year}
  {1993})}\BibitemShut {NoStop}%
\bibitem [{\citenamefont {Damour}\ and\ \citenamefont
  {Esposito-Farese}(1996)}]{Damour:1996ke}%
  \BibitemOpen
  \bibfield  {author} {\bibinfo {author} {\bibfnamefont {Thibault}\
  \bibnamefont {Damour}}\ and\ \bibinfo {author} {\bibfnamefont {Gilles}\
  \bibnamefont {Esposito-Farese}},\ }\bibfield  {title} {\enquote {\bibinfo
  {title} {{Tensor - scalar gravity and binary pulsar experiments}},}\ }\href
  {\doibase 10.1103/PhysRevD.54.1474} {\bibfield  {journal} {\bibinfo
  {journal} {Phys. Rev.}\ }\textbf {\bibinfo {volume} {D54}},\ \bibinfo {pages}
  {1474--1491} (\bibinfo {year} {1996})},\ \Eprint
  {http://arxiv.org/abs/gr-qc/9602056} {arXiv:gr-qc/9602056 [gr-qc]}
  \BibitemShut {NoStop}%
%%CITATION = GR-QC/9602056;%%
\bibitem [{\citenamefont {Barausse}\ \emph {et~al.}(2013)\citenamefont
  {Barausse}, \citenamefont {Palenzuela}, \citenamefont {Ponce},\ and\
  \citenamefont {Lehner}}]{Barausse:2012da}%
  \BibitemOpen
  \bibfield  {author} {\bibinfo {author} {\bibfnamefont {Enrico}\ \bibnamefont
  {Barausse}}, \bibinfo {author} {\bibfnamefont {Carlos}\ \bibnamefont
  {Palenzuela}}, \bibinfo {author} {\bibfnamefont {Marcelo}\ \bibnamefont
  {Ponce}}, \ and\ \bibinfo {author} {\bibfnamefont {Luis}\ \bibnamefont
  {Lehner}},\ }\bibfield  {title} {\enquote {\bibinfo {title} {{Neutron-star
  mergers in scalar-tensor theories of gravity}},}\ }\href {\doibase
  10.1103/PhysRevD.87.081506} {\bibfield  {journal} {\bibinfo  {journal}
  {Phys.Rev.}\ }\textbf {\bibinfo {volume} {D87}},\ \bibinfo {pages} {081506}
  (\bibinfo {year} {2013})},\ \Eprint {http://arxiv.org/abs/1212.5053}
  {arXiv:1212.5053 [gr-qc]} \BibitemShut {NoStop}%
%%CITATION = ARXIV:1212.5053;%%
\bibitem [{\citenamefont {Palenzuela}\ \emph {et~al.}(2014)\citenamefont
  {Palenzuela}, \citenamefont {Barausse}, \citenamefont {Ponce},\ and\
  \citenamefont {Lehner}}]{Palenzuela:2013hsa}%
  \BibitemOpen
  \bibfield  {author} {\bibinfo {author} {\bibfnamefont {Carlos}\ \bibnamefont
  {Palenzuela}}, \bibinfo {author} {\bibfnamefont {Enrico}\ \bibnamefont
  {Barausse}}, \bibinfo {author} {\bibfnamefont {Marcelo}\ \bibnamefont
  {Ponce}}, \ and\ \bibinfo {author} {\bibfnamefont {Luis}\ \bibnamefont
  {Lehner}},\ }\bibfield  {title} {\enquote {\bibinfo {title} {{Dynamical
  scalarization of neutron stars in scalar-tensor gravity theories}},}\ }\href
  {\doibase 10.1103/PhysRevD.89.044024} {\bibfield  {journal} {\bibinfo
  {journal} {Phys.Rev.}\ }\textbf {\bibinfo {volume} {D89}},\ \bibinfo {pages}
  {044024} (\bibinfo {year} {2014})},\ \Eprint {http://arxiv.org/abs/1310.4481}
  {arXiv:1310.4481 [gr-qc]} \BibitemShut {NoStop}%
%%CITATION = ARXIV:1310.4481;%%
\bibitem [{\citenamefont {Horbatsch}\ and\ \citenamefont
  {Burgess}(2011)}]{Horbatsch:2010hj}%
  \BibitemOpen
  \bibfield  {author} {\bibinfo {author} {\bibfnamefont {M.~W.}\ \bibnamefont
  {Horbatsch}}\ and\ \bibinfo {author} {\bibfnamefont {C.~P.}\ \bibnamefont
  {Burgess}},\ }\bibfield  {title} {\enquote {\bibinfo {title} {{Semi-Analytic
  Stellar Structure in Scalar-Tensor Gravity}},}\ }\href {\doibase
  10.1088/1475-7516/2011/08/027} {\bibfield  {journal} {\bibinfo  {journal}
  {JCAP}\ }\textbf {\bibinfo {volume} {1108}},\ \bibinfo {pages} {027}
  (\bibinfo {year} {2011})},\ \Eprint {http://arxiv.org/abs/1006.4411}
  {arXiv:1006.4411 [gr-qc]} \BibitemShut {NoStop}%
%%CITATION = ARXIV:1006.4411;%%
\bibitem [{\citenamefont {Mendes}(2015)}]{Mendes:2014ufa}%
  \BibitemOpen
  \bibfield  {author} {\bibinfo {author} {\bibfnamefont {Raissa F.~P.}\
  \bibnamefont {Mendes}},\ }\bibfield  {title} {\enquote {\bibinfo {title}
  {{Possibility of setting a new constraint to scalar-tensor theories}},}\
  }\href {\doibase 10.1103/PhysRevD.91.064024} {\bibfield  {journal} {\bibinfo
  {journal} {Phys. Rev.}\ }\textbf {\bibinfo {volume} {D91}},\ \bibinfo {pages}
  {064024} (\bibinfo {year} {2015})},\ \Eprint {http://arxiv.org/abs/1412.6789}
  {arXiv:1412.6789 [gr-qc]} \BibitemShut {NoStop}%
%%CITATION = ARXIV:1412.6789;%%
\bibitem [{\citenamefont {Palenzuela}\ and\ \citenamefont
  {Liebling}(2016)}]{Palenzuela:2015ima}%
  \BibitemOpen
  \bibfield  {author} {\bibinfo {author} {\bibfnamefont {Carlos}\ \bibnamefont
  {Palenzuela}}\ and\ \bibinfo {author} {\bibfnamefont {Steven~L.}\
  \bibnamefont {Liebling}},\ }\bibfield  {title} {\enquote {\bibinfo {title}
  {{Constraining scalar-tensor theories of gravity from the most massive
  neutron stars}},}\ }\href {\doibase 10.1103/PhysRevD.93.044009} {\bibfield
  {journal} {\bibinfo  {journal} {Phys. Rev.}\ }\textbf {\bibinfo {volume}
  {D93}},\ \bibinfo {pages} {044009} (\bibinfo {year} {2016})},\ \Eprint
  {http://arxiv.org/abs/1510.03471} {arXiv:1510.03471 [gr-qc]} \BibitemShut
  {NoStop}%
%%CITATION = ARXIV:1510.03471;%%
\bibitem [{\citenamefont {Mendes}\ and\ \citenamefont
  {Ortiz}(2016)}]{Mendes:2016fby}%
  \BibitemOpen
  \bibfield  {author} {\bibinfo {author} {\bibfnamefont {Raissa F.~P.}\
  \bibnamefont {Mendes}}\ and\ \bibinfo {author} {\bibfnamefont {Néstor}\
  \bibnamefont {Ortiz}},\ }\bibfield  {title} {\enquote {\bibinfo {title}
  {{Highly compact neutron stars in scalar-tensor theories of gravity:
  Spontaneous scalarization versus gravitational collapse}},}\ }\href {\doibase
  10.1103/PhysRevD.93.124035} {\bibfield  {journal} {\bibinfo  {journal} {Phys.
  Rev.}\ }\textbf {\bibinfo {volume} {D93}},\ \bibinfo {pages} {124035}
  (\bibinfo {year} {2016})},\ \Eprint {http://arxiv.org/abs/1604.04175}
  {arXiv:1604.04175 [gr-qc]} \BibitemShut {NoStop}%
%%CITATION = ARXIV:1604.04175;%%
\bibitem [{\citenamefont {Damour}\ and\ \citenamefont
  {Esposito-Farese}(1998)}]{Damour:1998jk}%
  \BibitemOpen
  \bibfield  {author} {\bibinfo {author} {\bibfnamefont {Thibault}\
  \bibnamefont {Damour}}\ and\ \bibinfo {author} {\bibfnamefont {Gilles}\
  \bibnamefont {Esposito-Farese}},\ }\bibfield  {title} {\enquote {\bibinfo
  {title} {{Gravitational wave versus binary - pulsar tests of strong field
  gravity}},}\ }\href {\doibase 10.1103/PhysRevD.58.042001} {\bibfield
  {journal} {\bibinfo  {journal} {Phys.Rev.}\ }\textbf {\bibinfo {volume}
  {D58}},\ \bibinfo {pages} {042001} (\bibinfo {year} {1998})},\ \Eprint
  {http://arxiv.org/abs/gr-qc/9803031} {arXiv:gr-qc/9803031 [gr-qc]}
  \BibitemShut {NoStop}%
%%CITATION = GR-QC/9803031;%%
\bibitem [{\citenamefont {Shao}\ \emph {et~al.}(2017)\citenamefont {Shao},
  \citenamefont {Sennett}, \citenamefont {Buonanno}, \citenamefont {Kramer},\
  and\ \citenamefont {Wex}}]{Shao:2017gwu}%
  \BibitemOpen
  \bibfield  {author} {\bibinfo {author} {\bibfnamefont {Lijing}\ \bibnamefont
  {Shao}}, \bibinfo {author} {\bibfnamefont {Noah}\ \bibnamefont {Sennett}},
  \bibinfo {author} {\bibfnamefont {Alessandra}\ \bibnamefont {Buonanno}},
  \bibinfo {author} {\bibfnamefont {Michael}\ \bibnamefont {Kramer}}, \ and\
  \bibinfo {author} {\bibfnamefont {Norbert}\ \bibnamefont {Wex}},\ }\bibfield
  {title} {\enquote {\bibinfo {title} {{Constraining nonperturbative
  strong-field effects in scalar-tensor gravity by combining pulsar timing and
  laser-interferometer gravitational-wave detectors}},}\ }\href {\doibase
  10.1103/PhysRevX.7.041025} {\bibfield  {journal} {\bibinfo  {journal} {Phys.
  Rev.}\ }\textbf {\bibinfo {volume} {X7}},\ \bibinfo {pages} {041025}
  (\bibinfo {year} {2017})},\ \Eprint {http://arxiv.org/abs/1704.07561}
  {arXiv:1704.07561 [gr-qc]} \BibitemShut {NoStop}%
%%CITATION = ARXIV:1704.07561;%%
\bibitem [{\citenamefont {He}\ \emph {et~al.}(2015)\citenamefont {He},
  \citenamefont {Fattoyev}, \citenamefont {Li},\ and\ \citenamefont
  {Newton}}]{He:2014yqa}%
  \BibitemOpen
  \bibfield  {author} {\bibinfo {author} {\bibfnamefont {Xiao-Tao}\
  \bibnamefont {He}}, \bibinfo {author} {\bibfnamefont {Farrukh~J.}\
  \bibnamefont {Fattoyev}}, \bibinfo {author} {\bibfnamefont {Bao-An}\
  \bibnamefont {Li}}, \ and\ \bibinfo {author} {\bibfnamefont {William~G.}\
  \bibnamefont {Newton}},\ }\bibfield  {title} {\enquote {\bibinfo {title}
  {{Impact of the equation-of-state-- gravity degeneracy on constraining the
  nuclear symmetry energy from astrophysical observables}},}\ }\href {\doibase
  10.1103/PhysRevC.91.015810} {\bibfield  {journal} {\bibinfo  {journal} {Phys.
  Rev.}\ }\textbf {\bibinfo {volume} {C91}},\ \bibinfo {pages} {015810}
  (\bibinfo {year} {2015})},\ \Eprint {http://arxiv.org/abs/1408.0857}
  {arXiv:1408.0857 [nucl-th]} \BibitemShut {NoStop}%
%%CITATION = ARXIV:1408.0857;%%
\bibitem [{\citenamefont {Horbatsch}\ and\ \citenamefont
  {Burgess}(2012)}]{Horbatsch:2011nh}%
  \BibitemOpen
  \bibfield  {author} {\bibinfo {author} {\bibfnamefont {M.~W.}\ \bibnamefont
  {Horbatsch}}\ and\ \bibinfo {author} {\bibfnamefont {C.~P.}\ \bibnamefont
  {Burgess}},\ }\bibfield  {title} {\enquote {\bibinfo {title}
  {{Model-Independent Comparisons of Pulsar Timings to Scalar-Tensor
  Gravity}},}\ }\href {\doibase 10.1088/0264-9381/29/24/245004} {\bibfield
  {journal} {\bibinfo  {journal} {Class. Quant. Grav.}\ }\textbf {\bibinfo
  {volume} {29}},\ \bibinfo {pages} {245004} (\bibinfo {year} {2012})},\
  \Eprint {http://arxiv.org/abs/1107.3585} {arXiv:1107.3585 [gr-qc]}
  \BibitemShut {NoStop}%
%%CITATION = ARXIV:1107.3585;%%
\bibitem [{\citenamefont {Danby}(1970)}]{danby1970fundamentals}%
  \BibitemOpen
  \bibfield  {author} {\bibinfo {author} {\bibfnamefont {J.M.A.}\ \bibnamefont
  {Danby}},\ }\href {https://books.google.ch/books?id=4TGWnQEACAAJ} {\emph
  {\bibinfo {title} {Fundamentals of Celestial Mechanics}}}\ (\bibinfo
  {publisher} {MacMillan},\ \bibinfo {year} {1970})\BibitemShut {NoStop}%
\bibitem [{\citenamefont {Poisson}\ and\ \citenamefont
  {Will}(2014)}]{poisson2014gravity}%
  \BibitemOpen
  \bibfield  {author} {\bibinfo {author} {\bibfnamefont {E.}~\bibnamefont
  {Poisson}}\ and\ \bibinfo {author} {\bibfnamefont {C.M.}\ \bibnamefont
  {Will}},\ }\href {https://books.google.ch/books?id=PZ5cAwAAQBAJ} {\emph
  {\bibinfo {title} {Gravity: Newtonian, Post-Newtonian, Relativistic}}}\
  (\bibinfo  {publisher} {Cambridge University Press},\ \bibinfo {year}
  {2014})\BibitemShut {NoStop}%
\bibitem [{\citenamefont {Turner}(1979)}]{Turner:1979yn}%
  \BibitemOpen
  \bibfield  {author} {\bibinfo {author} {\bibfnamefont {Michael~S.}\
  \bibnamefont {Turner}},\ }\bibfield  {title} {\enquote {\bibinfo {title}
  {{Influence of a weak gravitational wave on a bound system of two point
  masses}},}\ }\href {\doibase 10.1086/157429} {\bibfield  {journal} {\bibinfo
  {journal} {Astrophys. J.}\ }\textbf {\bibinfo {volume} {233}},\ \bibinfo
  {pages} {685--693} (\bibinfo {year} {1979})}\BibitemShut {NoStop}%
%%CITATION = ASJOA,233,685;%%
\bibitem [{\citenamefont {Nordtvedt}(1968)}]{Nordtvedt:1968qs}%
  \BibitemOpen
  \bibfield  {author} {\bibinfo {author} {\bibfnamefont {Kenneth}\ \bibnamefont
  {Nordtvedt}},\ }\bibfield  {title} {\enquote {\bibinfo {title} {{Equivalence
  Principle for Massive Bodies. 2. Theory}},}\ }\href {\doibase
  10.1103/PhysRev.169.1017} {\bibfield  {journal} {\bibinfo  {journal} {Phys.
  Rev.}\ }\textbf {\bibinfo {volume} {169}},\ \bibinfo {pages} {1017--1025}
  (\bibinfo {year} {1968})}\BibitemShut {NoStop}%
%%CITATION = PHRVA,169,1017;%%
\bibitem [{\citenamefont {Damour}\ and\ \citenamefont
  {Sch{\"a}fer}(1991)}]{Damour:1991rq}%
  \BibitemOpen
  \bibfield  {author} {\bibinfo {author} {\bibfnamefont {Thibault}\
  \bibnamefont {Damour}}\ and\ \bibinfo {author} {\bibfnamefont {Gerhard}\
  \bibnamefont {Sch{\"a}fer}},\ }\bibfield  {title} {\enquote {\bibinfo {title}
  {{New tests of the strong equivalence principle using binary pulsar data}},}\
  }\href {\doibase 10.1103/PhysRevLett.66.2549} {\bibfield  {journal} {\bibinfo
   {journal} {Phys.Rev.Lett.}\ }\textbf {\bibinfo {volume} {66}},\ \bibinfo
  {pages} {2549--2552} (\bibinfo {year} {1991})}\BibitemShut {NoStop}%
%%CITATION = PRLTA,66,2549;%%
\bibitem [{\citenamefont {Liu}\ \emph {et~al.}(2011)\citenamefont {Liu},
  \citenamefont {Verbiest}, \citenamefont {Kramer}, \citenamefont {Stappers},
  \citenamefont {van Straten},\ and\ \citenamefont {Cordes}}]{Kramer11}%
  \BibitemOpen
  \bibfield  {author} {\bibinfo {author} {\bibfnamefont {K.}~\bibnamefont
  {Liu}}, \bibinfo {author} {\bibfnamefont {J.~P.~W.}\ \bibnamefont
  {Verbiest}}, \bibinfo {author} {\bibfnamefont {M.}~\bibnamefont {Kramer}},
  \bibinfo {author} {\bibfnamefont {B.~W.}\ \bibnamefont {Stappers}}, \bibinfo
  {author} {\bibfnamefont {W.}~\bibnamefont {van Straten}}, \ and\ \bibinfo
  {author} {\bibfnamefont {J.~M.}\ \bibnamefont {Cordes}},\ }\bibfield  {title}
  {\enquote {\bibinfo {title} {Prospects for high-precision pulsar timing},}\
  }\href {\doibase 10.1111/j.1365-2966.2011.19452.x} {\bibfield  {journal}
  {\bibinfo  {journal} {Monthly Notices of the Royal Astronomical Society}\
  }\textbf {\bibinfo {volume} {417}},\ \bibinfo {pages} {2916--2926} (\bibinfo
  {year} {2011})},\ \Eprint {http://arxiv.org/abs/arXiv:1107.3086}
  {arXiv:arXiv:1107.3086 [astro-ph]} \BibitemShut {NoStop}%
\bibitem [{\citenamefont {{Damour}}\ and\ \citenamefont
  {{Taylor}}(1991)}]{1991ApJ...366..501D}%
  \BibitemOpen
  \bibfield  {author} {\bibinfo {author} {\bibfnamefont {T.}~\bibnamefont
  {{Damour}}}\ and\ \bibinfo {author} {\bibfnamefont {J.~H.}\ \bibnamefont
  {{Taylor}}},\ }\bibfield  {title} {\enquote {\bibinfo {title} {{On the
  orbital period change of the binary pulsar PSR 1913 + 16}},}\ }\href
  {\doibase 10.1086/169585} {\bibfield  {journal} {\bibinfo  {journal} {\apj}\
  }\textbf {\bibinfo {volume} {366}},\ \bibinfo {pages} {501--511} (\bibinfo
  {year} {1991})}\BibitemShut {NoStop}%
\bibitem [{\citenamefont {{Shklovskii}}(1970)}]{1970SvA....13..562S}%
  \BibitemOpen
  \bibfield  {author} {\bibinfo {author} {\bibfnamefont {I.~S.}\ \bibnamefont
  {{Shklovskii}}},\ }\bibfield  {title} {\enquote {\bibinfo {title} {{Possible
  Causes of the Secular Increase in Pulsar Periods.}}}\ }\href@noop {}
  {\bibfield  {journal} {\bibinfo  {journal} {Soviet Astronomy}\ }\textbf
  {\bibinfo {volume} {13}},\ \bibinfo {pages} {562} (\bibinfo {year}
  {1970})}\BibitemShut {NoStop}%
\bibitem [{\citenamefont {Freire}\ \emph {et~al.}(2012)\citenamefont {Freire},
  \citenamefont {Kramer},\ and\ \citenamefont {Wex}}]{Freire:2012nb}%
  \BibitemOpen
  \bibfield  {author} {\bibinfo {author} {\bibfnamefont {Paulo~C.C.}\
  \bibnamefont {Freire}}, \bibinfo {author} {\bibfnamefont {Michael}\
  \bibnamefont {Kramer}}, \ and\ \bibinfo {author} {\bibfnamefont {Norbert}\
  \bibnamefont {Wex}},\ }\bibfield  {title} {\enquote {\bibinfo {title} {{Tests
  of the universality of free fall for strongly self-gravitating bodies with
  radio pulsars}},}\ }\href {\doibase 10.1088/0264-9381/29/18/184007}
  {\bibfield  {journal} {\bibinfo  {journal} {Class.Quant.Grav.}\ }\textbf
  {\bibinfo {volume} {29}},\ \bibinfo {pages} {184007} (\bibinfo {year}
  {2012})},\ \Eprint {http://arxiv.org/abs/1205.3751} {arXiv:1205.3751 [gr-qc]}
  \BibitemShut {NoStop}%
%%CITATION = ARXIV:1205.3751;%%
\bibitem [{\citenamefont {Kehl}\ \emph {et~al.}(2017)\citenamefont {Kehl},
  \citenamefont {Wex}, \citenamefont {Kramer},\ and\ \citenamefont
  {Liu}}]{Kehl:2016mgp}%
  \BibitemOpen
  \bibfield  {author} {\bibinfo {author} {\bibfnamefont {Marcel~S.}\
  \bibnamefont {Kehl}}, \bibinfo {author} {\bibfnamefont {Norbert}\
  \bibnamefont {Wex}}, \bibinfo {author} {\bibfnamefont {Michael}\ \bibnamefont
  {Kramer}}, \ and\ \bibinfo {author} {\bibfnamefont {Kuo}\ \bibnamefont
  {Liu}},\ }\bibfield  {title} {\enquote {\bibinfo {title} {{Future
  measurements of the Lense-Thirring effect in the Double Pulsar}},}\ }\href
  {\doibase 10.1142/9789813226609_0195} {\bibfield  {journal} {\bibinfo
  {journal} {Proceedings, 14th Marcel Grossmann Meeting on Recent Developments
  in Theoretical and Experimental General Relativity, Astrophysics, and
  Relativistic Field Theories (MG14) (In 4 Volumes): Rome, Italy, July 12-18,
  2015}\ }\textbf {\bibinfo {volume} {2}},\ \bibinfo {pages} {1860--1865}
  (\bibinfo {year} {2017})},\ \Eprint {http://arxiv.org/abs/1605.00408}
  {arXiv:1605.00408 [astro-ph.HE]} \BibitemShut {NoStop}%
%%CITATION = ARXIV:1605.00408;%%
\bibitem [{\citenamefont {Freire}\ \emph {et~al.}(2011)\citenamefont {Freire}
  \emph {et~al.}}]{Freire:2010tf}%
  \BibitemOpen
  \bibfield  {author} {\bibinfo {author} {\bibfnamefont {P.~C.~C.}\
  \bibnamefont {Freire}} \emph {et~al.},\ }\bibfield  {title} {\enquote
  {\bibinfo {title} {{On the nature and evolution of the unique binary pulsar
  J1903+0327}},}\ }\href {\doibase 10.1111/j.1365-2966.2010.18109.x} {\bibfield
   {journal} {\bibinfo  {journal} {Mon. Not. Roy. Astron. Soc.}\ }\textbf
  {\bibinfo {volume} {412}},\ \bibinfo {pages} {2763} (\bibinfo {year}
  {2011})},\ \Eprint {http://arxiv.org/abs/1011.5809} {arXiv:1011.5809
  [astro-ph.GA]} \BibitemShut {NoStop}%
%%CITATION = ARXIV:1011.5809;%%
\bibitem [{\citenamefont {{Foster}}\ \emph {et~al.}(1993)\citenamefont
  {{Foster}}, \citenamefont {{Wolszczan}},\ and\ \citenamefont
  {{Camilo}}}]{1993ApJ...410L..91F}%
  \BibitemOpen
  \bibfield  {author} {\bibinfo {author} {\bibfnamefont {R.~S.}\ \bibnamefont
  {{Foster}}}, \bibinfo {author} {\bibfnamefont {A.}~\bibnamefont
  {{Wolszczan}}}, \ and\ \bibinfo {author} {\bibfnamefont {F.}~\bibnamefont
  {{Camilo}}},\ }\bibfield  {title} {\enquote {\bibinfo {title} {{A new binary
  millisecond pulsar}},}\ }\href {\doibase 10.1086/186887} {\bibfield
  {journal} {\bibinfo  {journal} {ApJ}\ }\textbf {\bibinfo {volume} {410}},\
  \bibinfo {pages} {L91--L94} (\bibinfo {year} {1993})}\BibitemShut {NoStop}%
\bibitem [{\citenamefont {Zhu}\ \emph {et~al.}(2015)\citenamefont {Zhu} \emph
  {et~al.}}]{Zhu:2015mdo}%
  \BibitemOpen
  \bibfield  {author} {\bibinfo {author} {\bibfnamefont {W.~W.}\ \bibnamefont
  {Zhu}} \emph {et~al.},\ }\bibfield  {title} {\enquote {\bibinfo {title}
  {{Testing Theories of Gravitation Using 21-Year Timing of Pulsar Binary
  J1713+0747}},}\ }\href {\doibase 10.1088/0004-637X/809/1/41} {\bibfield
  {journal} {\bibinfo  {journal} {Astrophys. J.}\ }\textbf {\bibinfo {volume}
  {809}},\ \bibinfo {pages} {41} (\bibinfo {year} {2015})},\ \Eprint
  {http://arxiv.org/abs/1504.00662} {arXiv:1504.00662 [astro-ph.SR]}
  \BibitemShut {NoStop}%
%%CITATION = ARXIV:1504.00662;%%
\bibitem [{\citenamefont {Zhu}\ \emph {et~al.}(2018)\citenamefont {Zhu} \emph
  {et~al.}}]{Zhu:2018etc}%
  \BibitemOpen
  \bibfield  {author} {\bibinfo {author} {\bibfnamefont {W.~W.}\ \bibnamefont
  {Zhu}} \emph {et~al.},\ }\bibfield  {title} {\enquote {\bibinfo {title}
  {{Tests of Gravitational Symmetries with Pulsar Binary J1713+0747}},}\
  }\href@noop {} {\  (\bibinfo {year} {2018})},\ \Eprint
  {http://arxiv.org/abs/1802.09206} {arXiv:1802.09206 [astro-ph.HE]}
  \BibitemShut {NoStop}%
%%CITATION = ARXIV:1802.09206;%%
\bibitem [{fre()}]{freire}%
  \BibitemOpen
  \href@noop {} {}\bibinfo {note} {{\it Pulsars in globular clusters}, \url{
  http://www.naic.edu/~pfreire/GCpsr.html}}\BibitemShut {NoStop}%
\bibitem [{\citenamefont {Ridolfi}\ \emph {et~al.}(2016)\citenamefont {Ridolfi}
  \emph {et~al.}}]{Ridolfi:2016fet}%
  \BibitemOpen
  \bibfield  {author} {\bibinfo {author} {\bibfnamefont {A.}~\bibnamefont
  {Ridolfi}} \emph {et~al.},\ }\bibfield  {title} {\enquote {\bibinfo {title}
  {{Long-term observations of the pulsars in 47 Tucanae - I. A study of four
  elusive binary systems}},}\ }\href {\doibase 10.1093/mnras/stw1850}
  {\bibfield  {journal} {\bibinfo  {journal} {Mon. Not. Roy. Astron. Soc.}\
  }\textbf {\bibinfo {volume} {462}},\ \bibinfo {pages} {2918--2933} (\bibinfo
  {year} {2016})},\ \Eprint {http://arxiv.org/abs/1607.07248} {arXiv:1607.07248
  [astro-ph.HE]} \BibitemShut {NoStop}%
%%CITATION = ARXIV:1607.07248;%%
\bibitem [{\citenamefont {Wex}(2000)}]{wex_2000}%
  \BibitemOpen
  \bibfield  {author} {\bibinfo {author} {\bibfnamefont {Norbert}\ \bibnamefont
  {Wex}},\ }\bibfield  {title} {\enquote {\bibinfo {title} {Small-eccentricity
  binary pulsars and relativistic gravity},}\ }\href {\doibase
  10.1017/S0252921100059200} {\bibfield  {journal} {\bibinfo  {journal}
  {International Astronomical Union Colloquium}\ }\textbf {\bibinfo {volume}
  {177}},\ \bibinfo {pages} {113--116} (\bibinfo {year} {2000})}\BibitemShut
  {NoStop}%
\bibitem [{\citenamefont {Stairs}\ \emph {et~al.}(2005)\citenamefont {Stairs},
  \citenamefont {Faulkner}, \citenamefont {Lyne}, \citenamefont {Kramer},
  \citenamefont {Lorimer} \emph {et~al.}}]{Stairs:2005hu}%
  \BibitemOpen
  \bibfield  {author} {\bibinfo {author} {\bibfnamefont {Ingrid~H.}\
  \bibnamefont {Stairs}}, \bibinfo {author} {\bibfnamefont {A.J.}\ \bibnamefont
  {Faulkner}}, \bibinfo {author} {\bibfnamefont {A.G.}\ \bibnamefont {Lyne}},
  \bibinfo {author} {\bibfnamefont {M.}~\bibnamefont {Kramer}}, \bibinfo
  {author} {\bibfnamefont {D.R.}\ \bibnamefont {Lorimer}},  \emph {et~al.},\
  }\bibfield  {title} {\enquote {\bibinfo {title} {{Discovery of three
  wide-orbit binary pulsars: Implications for binary evolution and equivalence
  principles}},}\ }\href {\doibase 10.1086/432526} {\bibfield  {journal}
  {\bibinfo  {journal} {Astrophys.J.}\ }\textbf {\bibinfo {volume} {632}},\
  \bibinfo {pages} {1060--1068} (\bibinfo {year} {2005})},\ \Eprint
  {http://arxiv.org/abs/astro-ph/0506188} {arXiv:astro-ph/0506188 [astro-ph]}
  \BibitemShut {NoStop}%
%%CITATION = ASTRO-PH/0506188;%%
\bibitem [{\citenamefont {Gonzalez}\ \emph {et~al.}(2011)\citenamefont
  {Gonzalez}, \citenamefont {Stairs}, \citenamefont {Ferdman}, \citenamefont
  {Freire}, \citenamefont {Nice} \emph {et~al.}}]{Gonzalez:2011kt}%
  \BibitemOpen
  \bibfield  {author} {\bibinfo {author} {\bibfnamefont {M.E.}\ \bibnamefont
  {Gonzalez}}, \bibinfo {author} {\bibfnamefont {I.H.}\ \bibnamefont {Stairs}},
  \bibinfo {author} {\bibfnamefont {R.D.}\ \bibnamefont {Ferdman}}, \bibinfo
  {author} {\bibfnamefont {P.C.C.}\ \bibnamefont {Freire}}, \bibinfo {author}
  {\bibfnamefont {D.J.}\ \bibnamefont {Nice}},  \emph {et~al.},\ }\bibfield
  {title} {\enquote {\bibinfo {title} {{High-Precision Timing of 5 Millisecond
  Pulsars: Space Velocities, Binary Evolution and Equivalence Principles}},}\
  }\href {\doibase 10.1088/0004-637X/743/2/102} {\bibfield  {journal} {\bibinfo
   {journal} {Astrophys.J.}\ }\textbf {\bibinfo {volume} {743}},\ \bibinfo
  {pages} {102} (\bibinfo {year} {2011})},\ \Eprint
  {http://arxiv.org/abs/1109.5638} {arXiv:1109.5638 [astro-ph.HE]} \BibitemShut
  {NoStop}%
%%CITATION = ARXIV:1109.5638;%%
\bibitem [{\citenamefont {Kramer}\ and\ \citenamefont
  {Stappers}(2015)}]{Kramer:2015bea}%
  \BibitemOpen
  \bibfield  {author} {\bibinfo {author} {\bibfnamefont {Michael}\ \bibnamefont
  {Kramer}}\ and\ \bibinfo {author} {\bibfnamefont {Ben}\ \bibnamefont
  {Stappers}},\ }\href
  {http://inspirehep.net/record/1383201/files/arXiv:1507.04423.pdf} {\enquote
  {\bibinfo {title} {{Pulsar Science with the SKA}},}\ } (\bibinfo {year}
  {2015}),\ \Eprint {http://arxiv.org/abs/1507.04423} {arXiv:1507.04423
  [astro-ph.IM]} \BibitemShut {NoStop}%
%%CITATION = ARXIV:1507.04423;%%
\bibitem [{\citenamefont {Barr}\ \emph {et~al.}(2017)\citenamefont {Barr},
  \citenamefont {Freire}, \citenamefont {Kramer}, \citenamefont {Champion},
  \citenamefont {Berezina}, \citenamefont {Bassa}, \citenamefont {Lyne},\ and\
  \citenamefont {Stappers}}]{Barr:2016vxv}%
  \BibitemOpen
  \bibfield  {author} {\bibinfo {author} {\bibfnamefont {E.~D.}\ \bibnamefont
  {Barr}}, \bibinfo {author} {\bibfnamefont {P.~C.~C.}\ \bibnamefont {Freire}},
  \bibinfo {author} {\bibfnamefont {M.}~\bibnamefont {Kramer}}, \bibinfo
  {author} {\bibfnamefont {D.~J.}\ \bibnamefont {Champion}}, \bibinfo {author}
  {\bibfnamefont {M.}~\bibnamefont {Berezina}}, \bibinfo {author}
  {\bibfnamefont {C.~G.}\ \bibnamefont {Bassa}}, \bibinfo {author}
  {\bibfnamefont {A.~G.}\ \bibnamefont {Lyne}}, \ and\ \bibinfo {author}
  {\bibfnamefont {B.~W.}\ \bibnamefont {Stappers}},\ }\bibfield  {title}
  {\enquote {\bibinfo {title} {{A Massive Millisecond Pulsar in an Eccentric
  Binary}},}\ }\href {\doibase 10.1093/mnras/stw2947} {\bibfield  {journal}
  {\bibinfo  {journal} {Mon. Not. Roy. Astron. Soc.}\ }\textbf {\bibinfo
  {volume} {465}},\ \bibinfo {pages} {1711--1719} (\bibinfo {year} {2017})},\
  \Eprint {http://arxiv.org/abs/1611.03658} {arXiv:1611.03658 [astro-ph.HE]}
  \BibitemShut {NoStop}%
%%CITATION = ARXIV:1611.03658;%%
\bibitem [{\citenamefont {Hui}\ \emph {et~al.}(2018)\citenamefont {Hui},
  \citenamefont {Wu}, \citenamefont {Han}, \citenamefont {Kong},\ and\
  \citenamefont {Tam}}]{Hui:2018mkc}%
  \BibitemOpen
  \bibfield  {author} {\bibinfo {author} {\bibfnamefont {C.~Y.}\ \bibnamefont
  {Hui}}, \bibinfo {author} {\bibfnamefont {Kinwah}\ \bibnamefont {Wu}},
  \bibinfo {author} {\bibfnamefont {Qin}\ \bibnamefont {Han}}, \bibinfo
  {author} {\bibfnamefont {A.~K.~H.}\ \bibnamefont {Kong}}, \ and\ \bibinfo
  {author} {\bibfnamefont {P.~H.~T.}\ \bibnamefont {Tam}},\ }\bibfield  {title}
  {\enquote {\bibinfo {title} {{On the orbital properties of millisecond pulsar
  binaries}},}\ }\href {\doibase 10.3847/1538-4357/aad5ec} {\bibfield
  {journal} {\bibinfo  {journal} {Astrophys. J.}\ }\textbf {\bibinfo {volume}
  {864}},\ \bibinfo {pages} {30} (\bibinfo {year} {2018})},\ \Eprint
  {http://arxiv.org/abs/1807.09001} {arXiv:1807.09001 [astro-ph.HE]}
  \BibitemShut {NoStop}%
%%CITATION = ARXIV:1807.09001;%%
\bibitem [{\citenamefont {Wex}(2014)}]{Wex:2014nva}%
  \BibitemOpen
  \bibfield  {author} {\bibinfo {author} {\bibfnamefont {Norbert}\ \bibnamefont
  {Wex}},\ }\href@noop {} {\enquote {\bibinfo {title} {{Testing Relativistic
  Gravity with Radio Pulsars}},}\ } (\bibinfo {year} {2014}),\ \Eprint
  {http://arxiv.org/abs/1402.5594} {arXiv:1402.5594 [gr-qc]} \BibitemShut
  {NoStop}%
%%CITATION = ARXIV:1402.5594;%%
\bibitem [{\citenamefont {Watson}(1995)}]{watson1995treatise}%
  \BibitemOpen
  \bibfield  {author} {\bibinfo {author} {\bibfnamefont {G.N.}\ \bibnamefont
  {Watson}},\ }\href {https://books.google.ch/books?id=Mlk3FrNoEVoC} {\emph
  {\bibinfo {title} {A Treatise on the Theory of Bessel Functions}}},\
  Cambridge Mathematical Library\ (\bibinfo  {publisher} {Cambridge University
  Press},\ \bibinfo {year} {1995})\BibitemShut {NoStop}%
\end{thebibliography}%

 \end{document}